\pdfoutput=1
\documentclass[oneside,letterpaper,12pt]{book}
\usepackage{wudisser}
\usepackage{amsmath,amssymb,amsfonts}
\usepackage{graphicx}
\usepackage{lscape}
\usepackage{url}
\usepackage[hang,small,bf]{caption}
\usepackage{bm}
\defheaders{\FANCY}
\newcommand{\profs}{Wai-Mo Suen\textrm{,} Chair\\,Quo-Shin Chi\\,Ramanath Cowsik\\,Zohar Nussinov\\,Xiang Tang\\,Clifford M. Will}
\renewcommand{\title}{Properties of Neutron Star Critical Collapses}
\renewcommand{\author}{Mew-Bing Wan}
\renewcommand{\month}{December}
\renewcommand{\year}{2009}
\renewcommand{\date}{\month~\year}
\newcommand{\dept}{Physics}
\newcommand{\chair}{Wai-Mo Suen}

\begin{document}

  \begin{frontmatter}
    \titlepage{\dept}{\profs}{\title}{\author}{\date}

    \lhead{}
\rhead{\textit{Acknowledgements}}
\begin{acknowledgements}

The people who made this dissertation possible is innumerable.
First and foremost, I would like to thank my advisor, Wai-Mo Suen, who gave me the opportunity to pursue my ambition in the field of general relativity at large, 
to be initiated into the field of numerical relativity in his group and who taught me physical intuition and skepticism, and clarity, conciseness and forcefulness in thought.
Without him, I will not be where I am now. 

Special thanks to Ke-Jian Jin for sharing his work of many years on the code 
that has enabled us to gain a deeper understanding of the findings in neutron star critical gravitational collapses; for his comradeship and collaboration.
Not forgetting also my WUGRAV colleagues, Jing Zeng, Hui-Min Zhang, Jian Tao, Han Wang, Randy Wolfmeyer and Dimitris Manolidis 
who have provided precious companionship during those hard times, 
mental and emotional support, invaluable discussions on general relativity and numerical relativity at large and 
various collaborations throughout my entire graduate career at Washington University.

Many thanks to Emanuele Berti, Renato Feres, Jos\'{e} Antonio Font, Carsten Gundlach, Adam Helfer, J\'{e}r\^{o}me Novak, Luciano Rezzolla, Ulrich Sperhake, Xiang Tang and Clifford Will for the various discussions and input relevant to this dissertation.    
Thanks to Ke-Jian Jin, Sai Iyer and Richard Bose for help and input on all things technical.

I thank the Physics Department and the Graduate School for providing the means of financial support for four years and the Dean's Dissertation Fellowship for the fifth year 
and for all the support provided on the academics at large.
I am deeply grateful to Julia Hamilton for all the motherly, kind and genuine support in all things that make a 
graduate student's life especially that of an alien in a foreign country, so much move livable.        
Thanks Sarah Hadley and Alison Verbeck for making all the little things happen, without which academics just would not survive.

I would also like thank the various friends and family for their genuine love, kindness, friendship and practical advice without which I would not be able to make it.
I will be forever indebted to my parents who gave of themselves to me, even throughout this journey of pursuing my dream. 
   
\end{acknowledgements}

    \defheaders{\FANCY}
    \lhead{}
    \tables{1}{1}{1}
    \lhead{}
\rhead{}
\begin{abstract}[1]{\title}{\author}{\dept}{\year}{\chair}

Critical phenomena in gravitational collapse opened a new mathematical vista into the theory of general relativity and may
ultimately entail fundamental physical implication in observations.
However, at present, the dynamics of critical phenomena in gravitational collapse scenarios
are still largely unknown. My thesis seeks to understand the properties of the threshold in the solution space of the Einstein field equations
between the black hole and neutron star phases, understand the properties of the neutron star critical solution
and clarify the implication of these results on realistic astrophysical scenarios.
We develop a new set of neutron star-like initial data to establish the universality of the neutron star critical solution and analyze the structure of neutron star and 
neutron star-like critical collapses via the study of the phase spaces.
We also study the different time scales involved in the neutron star critical solution and analyze the 
properties of the critical index via comparisons between neutron star and neutron star-like initial data.
Finally, we explore the boundary of the attraction basin of the neutron star critical solution and its transition to a known set of non-critical fixed points.

\end{abstract}

  \end{frontmatter}

\defheaders{\FANCY}

  \begin{main}

   \chapter{Introduction}

\section{Background and motivation}

The field of numerical relativity has been a rich interplay of different disciplines throughout its relatively young life.
Besides the advances in numerical methods applied to the theory of general relativity, eg. spectral methods \cite{Grandclement09}, finite-element methods \cite{Aksoylu08},\cite{Korobkin08} and multigrid methods \cite{Natchu07},\cite{Hawley03}, 
diverse fields such as theoretical computational science, eg. adaptive mesh refinement algorithms and algebraic computation; fluid dynamics, eg. magnetohydrodynamical 
flows; the theory of partial differential equations; the theory of the global structure of spacetimes, eg. the study of the structure of singularities and the dynamics of the 
spacetimes surrouding it; theoretical astrophysics, eg. binary pulsars and black holes, 
rotating stars, and supernovae core collapses; cosmology, eg. galactic clusters and collisions; and astronomy, eg. gravitational radiation detection, gamma-ray burst 
modelling; have contributed to its development \cite{Etienne09},\cite{Evans05},\cite{Brown09},\cite{Stephens08},\cite{Andersson05},\cite{Shibata06},\cite{Buonanno07},\cite{SuenNSGC},\cite{Stergioulas95},\cite{Ott07},\cite{Martel94},\cite{Aylott09},\cite{Will06},\cite{Granot99}, and vice versa. 
An example is the application of the Godunov method in dealing with hydrodynamic shocks 
when solving the Einstein field equations on a numerical finite-differenced grid. Godunov first suggested this method in 1959 as a means to solve partial differential equations. 
In the application of his method in numerical relativistic hydrodynamics, the relativistic fluid is cast as a Riemann problem, which consists of a conservation law together 
with a piecewise constant data having a single discontinuity. The use of the Godunov method is now widespread and known to be highly effective in all state-of-the-art numerical 
general relativistic hydrodynamics simulations of astrophysical systems and phenomena \cite{Font00},\cite{Font02}.

In 1993, a phenomenon is found by Choptuik \cite{Choptuik93} in general relativistic simulations, analogous to phase transitions in condensed matter systems, thereby 
earning the name critical phenomenon. Critical phenomena are the first \textit{new} phenomena discovered via numerical simulation in the theory of
general relativity. The phenomena observed describes 
the existence of a threshold between distinct possible end states of a gravitational collapse scenario, and new 
behavior of the gravitating system on this threshold reflective of that undergone by condensed matter systems during phase transitions, eg. liquid-gas transitions and 
ferromagnetic phase transitions. The dominant features of this behavior include fine-tuning, universality and scale-invariance, where a term called \textit{order~parameter} 
is used to describe a physical quantity that follows a power-law behavior, 
and where the phase transitions are categorized according to whether the order parameter changes discretely or continuously. 
The scale-invariant feature of these condensed matter systems, eg. the dimensionless quality of the correlation length which causes the particle clusters to 
exhibit a fractal structure, and the renormalization group transformation performed on the correlation length, is carried over analogously to the gravitational collapse 
behavior on the threshold.
As such, the discovery of critical phenomena in gravitational collapse became another prominent example where a separate field of physics has enriched the field
of numerical relativity. In addition, it has sinced opened a new mathematical vista into the theory of relativity. 
In the first decade after this discovery, research has been focused
on analytically establishing the existence and generality of critical phenomena in general relativity.
Due to the complexity of Einstein's field equations, such analytic analyses of the phenomena have all assumed spherical symmetry, and involved simplified matter or massless 
systems. Overview of these works has been covered comprehensively by Gundlach in his review \cite{Gundlach07}.  
Later works gradually ventured into more relaxed symmetry, ie. 
axisymmetry, and considered more complicated matter models. However, the field of critical gravitational collapse still dwells very much in the theoretical and 
mathematical arena, and has been used only as a tool to understand new and interesting aspects of the theory of general relativity, eg. the cosmic censorship conjecture and 
the emergence of fractality.

In 2003, Noble and Choptuik \cite{Noble08} began looking into how critical phenomena could be triggered in static neutron star models in spherical symmetry. Neutron stars refer 
to stars with masses on the order of 1.5 $M_{\odot}$, radii of about 12 km, and central densities 5 to 10 times the nuclear equilibrium density of about 0.16 $fm^3$, thereby 
making them some of the densest massive objects found in the astrophysical realm \cite{Baym75},\cite{Baym79},\cite{Heiselberg00}. 
They were first predicted by Baade and Zwicky in 1933 \cite{Baade34} in a study on the 
origins of supernovae, and discovered by Bell and Hewish in 1967 as a source of regular radio pulses in the Crab Nebula \cite{Hewish68}. 
Neutron stars are now thought to be formed as an aftermath of the gravitational collapse of the core of a massive star of more than 8 $M_{\odot}$ at the end of its life 
\cite{Lattimer04}. They are also believed to be created from the accretion-induced collapse of massive white dwarfs. 
Their interiors, in particular their outer cores, consist of neutron superfluids with proton superconductors. These interiors lose energy at a rapid rate via neutrino 
emission. A standard neutrino cooling scenario called the Urca process \cite{Pethick92} requires the proton to neutron ratio to exceed $1/8$. 
Each reaction in this process produces a neutrino and antineutrino via alternate beta and inverse-beta decays, losing thermal energy in the process. 
The period whereby this occurs beginning from the explosion is about $3\times 10^5$ years old. In 1974, Hulse and Taylor discovered a binary neutron star system \cite{Hulse75}, a 
system which is also termed as a binary pulsar, due to their emission of radio pulses as first observed by Bell and Zwicky.
Two stars starting from large separations slowly inspiral into each other due to loss of angular momentum and energy. When their orbits shrink beneath the innermost stable 
circular orbit, they enter a coalescence phase, where they begin to plunge and merge. Even though this phase is characterized by very strong and dynamical gravitational and 
hydrodynamical processes, the plunging can be approximately described as two stars colliding head-on into each other. Due to the pervasiveness 
of such systems in galactic clusters as well as the dynamics of their orbital decrease and energy loss, coalescing binary neutron stars 
are also generally considered good candidates as sources of gravitational radiation detectable by both existing ground-based detectors such as LIGO, VIRGO, GEO600, and 
TAMA300 and the proposed space-based detector LISA \cite{Cutler02}.

In current state-of-the-art 
numerical simulations of such processes, full general relativistic hydrodynamic is employed, where the standard Tolman-Oppenheimer-Volkoff models are solved 
together with a polytropic equation of state.  
The study of the time scales involved in gravitational collapses in colliding neutron stars was first considered by Miller, Suen and Tobias \cite{Miller01} using fully 
general relativistic hydrodynamics simulations. In this study, the neutron stars are modelled such that they infall into each other from infinity. As a result, a dividing 
line is found to exist between neutron star masses producing collapses that occur within a 
dynamical time scale, ie. prompt collapses, and those that occur after neutrino cooling settles in, ie. delayed collapses. As the problem is studied using a 3-dimensional code, 
the quest of determining the dividing line between these two scenarios becomes computationally expensive. This drove the need to construct a 2-dimensional code capable of 
high resolutions in the finite-differencing scheme employed in the code (refer to Appendix A), which in turn can be employed to resolve the fine structure behavior at the dividing line.  
In 2007, such an axisymmetric construction was undertaken by Jin and Suen \cite{Jin07} in a study that explores this dividing line in fine detail. They report that critical 
phenomena is observed along this dividing line, characterizing an oscillatory threshold between the black hole and neutron star end states. Furthermore, the same phenomena 
is seen to occur when the equation of state of the neutron star system is made to vary infinitesimally. Although the same equation of state is used in both this study and in 
the study undertaken by Noble and Choptuik earlier, the former does not exhibit scale-invariance at the threshold. With all these considerations, this study poses a 
significant departure from previous works and approaches the realm of realistic astrophysical phenomena. Furthermore, the determination of the gravitational radiation 
signature of the unstable modes of such gravitational collapses of neutron star systems may provide insights to gravitational radiation emission data. 
However, at present, the dynamics of critical phenomena in such astrophysical gravitational collapse scenarios are still largely unknown. 

In addition, the study of non-rotating neutron stars is still a deviation from realistic systems typically formed in the universe. In particular, coalescences 
of binary neutron star inspirals produce hypermassive neutron stars that support themselves against gravitational collapse via \textit{differential} rotation. These 
hypermassive neutron stars undergo diverse astrophysical mechanisms, eg. angular momentum transport, magnetic breaking, as well as gravitational radiation emission, before 
collapsing into black hole states. Rotating neutron stars are also formed from rotational core collapse of supernovae. With this in mind, 
Jin and Suen \cite{Jinpr} recently incorporated angular momentum into the head-on 
colliding neutron star system. The axis of angular momentum is parallel to the axis of collision. In this recent study, they find similar critical phenomena in the 
simultaneously oscillating and rotating merger.  

Given this background, my thesis is motivated by a three-fold objective as follows:\\
(1) understand the properties of neutron star critical collapses, 
(2) understand the properties of the threshold in the solution space of the Einstein field equations between the black hole and neutron star end states, and 
(3) clarify the implication of the results on realistic astrophysical scenarios.

\section{Overview of thesis}

In Chapter 2 of the thesis, the theoretical basis of numerical relativity is presented, namely detailed derivations of the 3+1 ADM formalism in
solving the system of Einsteins field equations in the theory of general relativity.
Chapter 3 presents the various constructs required in the implementation of the 
3+1 formalism in neutron star simulations, where we see the coupling of the spacetime with the hydrodynamical matter equations adapted to the 3+1 formalism. 
It further explores in particular how conformal decomposition and the BSSN formalism is employed in the solution and evolution of 
the system of Einsteins field equations in the theory 
of general relativity coupled with hydrodynamically-described matter, the basis of the GRAstro-3D code \cite{Font00},\cite{Font02}. We also present the main concepts behind the 
set-up and mechanism of the GRAstro-2D code. In Chapter 4, we study the Tolman-Volkoff-Oppeheimer spacetime and the ingredients of stellar perturbation theory as well 
as numerical observations in stellar phase transitions in order to enable us to understand the time scales involved in neutron star critical collapses. 
Chapter 5 in turn presents the basic concepts of critical gravitational collapse as well as some features of its implementation in numerical simulations of neutron stars.
In this chapter, the dynamical systems picture is brought forward prominently as an important tool in understanding the structure of critical solutions.

Chapter 6 contains the bulk of the analysis and simulation results in line with the three-fold objective mentioned above, 
utilizing all the ingredients presented in former chapters.
In Section 6.1, curve fitting is employed in order to establish the time scales and oscillation frequencies
exhibited by the critical solution and their universality across several crucial parameters in the critical
solution. These results are compared with the values obtained from analytic perturbative analysis of
equilibrium TOV configurations as presented in Chapter 4 of the thesis.
The construction of a neutron star-like initial data is carried out and evolution of various sets of this new
initial data is studied and analyzed in Sections 6.2 and 6.3 in order to provide evidence that the neutron
star critical solution is a semi-attractor. The boundary of its attraction basin and the transition to a known set of non-critical fixed points is explored.
In Section 6.4, the evolutions of the neutron star initial data sets
obtained by Jin and Suen \cite{Jin07} are studied in detail in the aspects of its spacetime and matter properties.
Convergence tests and various parameter searches for the construction of a phase diagram are also performed in this section. 
The power of the radiation emitted at the brink of critical collapse is examined using the quadrupole approximation 
in order to shed light on the structure of the semi-attractor, ie. on whether it is a limit cycle or a limit point. 
In Section 6.5, convergence tests are performed for the critical indices obtained
from different sets of the new initial data in order to determine whether the neutron star critical index has
a 1-parameter or 2-parameter dependence. The mass dependence of the critical index is also studied. 

Chapter 7 summarizes the main conclusions drawn from the analyses and simulation results and propose interpretations
of them in the framework of astrophysical relevance. Finally, the Appendix contains explanations of the finite-differencing scheme, of the basic geometric constructs of the pushforward and pullback operators used in the previous chapters, and of the 
derivation of geometric units used in this thesis.

   \chapter{3+1 formalism in numerical relativity}

Throughout the past half century, the theory of general relativity has proven to be extremely successful in the understanding of 
phenomena of massive objects interacting under strongly dynamical gravitational fields. In particular, numerical advances within 
the field have made enormous headway in solving the celebrated two-body problem in geometrodynamics. As a result, we are now able to 
understand to a great extent the various physical phenomena observed in binary neutron star coalescences and binary black hole 
coalescences. This includes the understanding of the broader context of gravitational collapse of compact objects, the various mechanisms 
that trigger their occurence and their effects on the spacetime structure surrounding these systems via the emission of gravitational waves. 

Numerical advances in the theory are characterized by the success of fully coupled general relativistic simulations of the astrophysical 
systems as mentioned above. These simulations employ massive numerical algorithms that solve the Einstein field equations coupled with 
matter. Their theoretical basis is the 3+1 formalism which reduces the Einstein field equations to an initial-value problem 
\cite{ADM62},\cite{York73},\cite{York79}. In this chapter, we present this formalism as it is adapted to the simulations 
performed for the study in this thesis, ie. the study of critical phenomena in gravitational collapse of neutron stars. 

\section{Hypersurfaces and foliations}

\begin{figure}
\begin{center}
\includegraphics[scale=0.8]{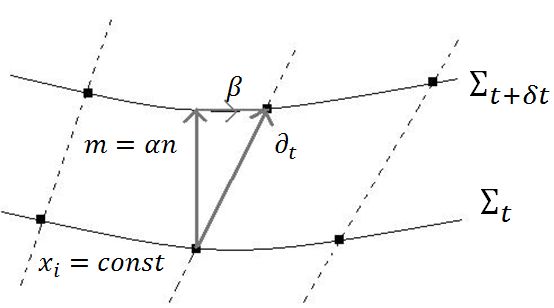}
\caption[Construction of an adapted coordinate system.]{}
\end{center}
\end{figure}

In the 3+1 formalism, the 4-dimensional spacetime is foliated by 3-dimensional spatial hypersurfaces of constant coordinate time. We 
denote the spacetime by $M$ and the hypersurface by $H$. The hypersurface $H$ is obtained by an embedding of a 3-dimensional manifold 
in the 4-dimensional spacetime via a 1-to-1 mapping that ensures that the hypersurface does not intersect itself. The embedding induces 
a push-forward mapping between vectors on $H$ to vectors on $M$, and a pull-back mapping between linear forms on $H$ and those on $M$ (refer to Appendix B). 

As the hypersurfaces are labeled by constant coordinate time, $t$, the gradient 1-form $\mathbf{d}t$ is timelike and can be 
written in component form as $(\mathbf{d}t)_{\mu}=\bigtriangledown_{\mu}t$. We denote the future-directed timelike unit vector normal to 
a hypersurface as $\mathbf{n}$. We also define as Eulerian observers a class of observers whose 4-velocity is collinear with this vector 
and thus whose worldlines are orthogonal to the hypersurface. The vector $\mathbf{n}$ is collinear with the timelike normal evolution 
vector $\mathbf{\bigtriangledown}t$ as follows:
\begin{equation}
\label{eq:geod1}
\mathbf{n}=N^{-1}\bm{\bigtriangledown}t,
\end{equation}
whereas the 1-form, $\mathbf{\bar{n}}$, which is dual to $\mathbf{n}$, is collinear with the gradient 1-form $\mathbf{d}t$, as follows:
\begin{equation}
\label{eq:geod2}
\mathbf{\bar{n}}=-N\mathbf{d}t.
\end{equation}
N is thus the lapse function that ensures the normalization relation:
\begin{equation}
\label{eq:geod3}
\mathbf{n}\cdot\mathbf{n}=\langle\mathbf{\bar{n}},\mathbf{n}\rangle=-1.
\end{equation}

On each hypersurface, we introduce a spatial coordinate system $(x^i)=(x^1,x^2,x^3)$ which constitutes a well-behaved 
4-dimensional coordinate system on $M$, $(x^{\alpha})=(t,x^1,x^2,x^3)$ when varied smoothly between neighboring hypersurfaces. The natural 
basis for this coordinate system is denoted by $(\partial_\alpha)=(\partial_t,\partial_i)$ where:
\begin{eqnarray}
\label{eq:geod4}
\bm{\partial}_t & := & \frac{\partial}{\partial t} \nonumber \\
\bm{\partial}_i & := & \frac{\partial}{\partial x^i}, i\in{1,2,3}.
\end{eqnarray}
$\bm{\partial}_t$ is the time vector tangent to the lines of constant spatial coordinates, whereas $\bm{\partial}_i, 
i\in{1,2,3}$ are vectors tangent to the hypersurface. We define as "coordinate observers" a class of observers whose 4-velocity is collinear 
with the time vector.

The difference between the time vector $\bm{\partial}_t$ and the timelike normal evolution vector $\bm{\bigtriangledown}t$ is 
the shift vector $\bm{\beta}$, and the relation is as follows:
\begin{equation}
\label{eq:geod5}
\bm{\partial}_t=\bm{\bigtriangledown}t+\bm{\beta}=N\mathbf{n}+\bm{\beta}.
\end{equation}
The shift vector thus enables the freedom to choose how the spatial coordinate system changes from hypersurface to hypersurface. Fig. 2.1 shows 
the construction of an adapted coordinate system. 

The 4-metric $\mathbf{g}$ can be written in terms of components $g_{\alpha\beta}$ with respect to the coordinates $(x^{\alpha})$ as 
follows:
\begin{equation}
\label{eq:geod6}
g_{\alpha\beta}=\mathbf{g}(\bm{\partial}_{\alpha},\bm{\partial}_{\beta}).
\end{equation}
Using Eq.~\eqref{eq:geod5}, the $00$-component of the 4-metric is thus given by:
\begin{equation}
\label{eq:geod7}
g_{00}=\mathbf{g}(\bm{\partial}_t,\bm{\partial}_t)=\bm{\partial}_t\cdot\bm{\partial}_t 
=-N^2(\mathbf{n}\cdot\mathbf{n})+\bm{\beta}\cdot\bm{\beta}=-N^2+\beta_i\beta^i,
\end{equation} 
whilst the $0i$-components are given by:
\begin{equation}
\label{eq:geod8}
g_{0i}=\mathbf{g}(\bm{\partial}_t,\bm{\partial}_i)=(N\mathbf{n}+\bm{\beta})\cdot\bm{\partial}_i=\beta_i,
\end{equation}
as the vector $\bm{\partial}_i$ is tangent to the hypersurface. The 3-metric $\bm{\gamma}$ is thus induced on each hypersurface 
with respect to the adapted coordinate system via this relation:
\begin{equation}
\label{eq:geod9}
g_{ij}=\mathbf{g}(\bm{\partial}_i,\bm{\partial}_j)=\bm{\gamma}(\bm{\partial}_i,\bm{\partial}_j)=\gamma_{ij}.
\end{equation}

The orthogonal projector that maps the 4-dimensional metric $\mathbf{g}$ onto the 3-dimensional hypersurfaces
is given in terms of components with respect to the coordinates $(x^{\alpha})$ by:
\begin{equation}
\label{eq:geod10}
\gamma^{\alpha}_{\beta}=\delta^{\alpha}_{\beta}+n^{\alpha}n_{\beta}.
\end{equation}
The operator acts on the normal timelike unit vector, as follows:
\begin{equation}
\label{eq:geod11}
\gamma^{\alpha}_{\beta}n^{\beta}=\delta^{\alpha}_{\beta}n^{\beta}+n^{\alpha}n_{\beta}n^{\beta}
=n^{\alpha}-n^{\alpha}=0,
\end{equation}
and on any vector tangent to the hypersurfaces, as follows:
\begin{equation}
\label{eq:geod12}
\gamma^{\alpha}_{\beta}\partial_{\alpha}=\delta^{\alpha}_{\beta}\partial_{\alpha}+n^{\alpha}n_{\beta}\partial_{\alpha} 
=\partial_{\beta}.
\end{equation}

\section{The covariant and Lie derivatives}

The 4-dimensional spacetime $M$ with the metric $\mathbf{g}$ possesses an associated
connection, $\bm{\bigtriangledown}$, denoted as the affine connection or
covariant derivative, which enables the comparison between vectors evaluated at 2 different points along the
congruence of curves generated by a vector field. We define the vector
evaluated at a point $P$, with coordinates $x^{\alpha}$ along the
congruence, as $v^{\alpha}_P$ and the vector evaluated at a point Q, with
coordinates $x^{\alpha}+\delta x^{\alpha}$, as $v^{\alpha}_Q$. This vector
at point Q is given by a Taylor's expansion as follows:
\begin{equation}
\label{eq:geod13}
v^{\alpha}(x+\delta x)=v^{\alpha}(x)+\delta x^{\beta}\partial_{\beta}v^{\alpha}.
\end{equation}
We further introduce another vector at point Q which is 'parallel' to the
vector at point P, and denote it as $v^{\alpha}_P+\-{\delta}v^{\alpha}$.
$\-{\delta}v^{\alpha}$ is collinear with $v^{\alpha}_P$ and $\delta x^{\alpha}$
and thus can be written as:
\begin{equation}
\label{eq:geod14}
\-{\delta}v^{\alpha}(x)=-\Gamma^{\alpha}_{\mu\beta}v^{\mu}(x)\delta x^{\beta}.
\end{equation} 
The connection or covariant derivative then evaluates the difference between $v^{\alpha}_P$
and $v^{\alpha}_Q$ as follows:
\begin{equation}
\label{eq:geod15}
\bm{\bigtriangledown}_{\alpha}v^{\beta}=\lim_{\delta x^{\alpha}\rightarrow 0} 
\frac{1}{\delta x^{\alpha}}[v^{\alpha}_Q-v^{\alpha}_P]
=v^{\beta}_{,\alpha}+\Gamma^{\beta}_{\mu\alpha}v^{\mu},
\end{equation}
where $\Gamma^{\beta}_{\mu\alpha}$ are thus denoted as the connection
coefficients, also known as Christoffel symbols. The covariant derivative can be generalized to
apply to the differentiation of a tensor $\mathbf{T}$ of any rank in the 4-dimensional spacetime
$M$, along a congruence generated by any vector field $\mathbf{u}$. The 3-dimensional covariant derivative, $\mathbf{DT}$, 
is thus obtained from the projection of the 4-dimensional covariant derivative, $\bm{\bigtriangledown T}$, onto
the 3-dimensional hypersurfaces, as follows:
\begin{equation}
\label{eq:geod16}
\mathbf{DT}=\bm{\gamma\bigtriangledown T}.
\end{equation}

The connection coefficients components can be given in terms of the 4-metric components as follows:
\begin{equation}
\label{eq:geod17}
\Gamma^{\gamma}_{\alpha\beta}=\frac{1}{2}g^{\gamma\mu}[g_{\alpha\mu,\beta}+g_{\beta\mu,\alpha}-g_{\alpha\beta,\mu}]. 
\end{equation}
This can be written analogously for the 3-dimensional connection
coefficients in terms of the 3-metric components, $\gamma_{ij}$.

The 4-dimensional Lie derivative for $M$ measures the distortion of the 4-dimensional coordinate
system. As it is based solely on coordinate bases, the 4-dimensional Lie
derivative is considered a more fundamental construct compared to the affine
connection. The Lie derivative is given by the difference between the vector evaluated at point Q, $v^{\alpha}_Q$, 
and the vector evaluated at point P that is 'dragged along' to point Q, given by:
\begin{eqnarray} 
\label{eq:geod18}
v^{\alpha}_P(x+\delta x) & = & v'^{\alpha}(x+\delta x) \nonumber \\
& = & v'^{\alpha}(x+\delta s u^{\alpha}) \nonumber \\
& = & \frac{\partial x'^{\alpha}}{\partial x^{\beta}}v^{\beta} \nonumber \\
& = & (\delta^{\alpha}_{\beta}+\delta s\partial_{\beta}u^{\alpha})v^{\beta}(x) \nonumber \\
& = & v^{\alpha}(x)+\delta s(\partial_{\beta}u^{\alpha}v^{\beta}(x),
\end{eqnarray}
where $s$ is the affine parameter along the congruence of curves generated by the vector field $\mathbf{u}$.
Therefore, the derivative is given by:
\begin{eqnarray}
\label{eq:geod19}
\mathcal{L}_u v^{\alpha} & = & \lim_{\delta s\rightarrow 0}\frac{1}{\delta s}
[v^{\alpha}(x+\delta x)-v'^{\alpha}(x+\delta x)] \nonumber \\
& = & \lim_{\delta s\rightarrow 0}\frac{1}{\delta s}
[v^{\alpha}(x)+\delta s u^{\mu}\partial_{\mu}v^{\alpha}-v^{\alpha}(x)-\delta s
\partial_{\mu}u^{\alpha}v^{\mu}(x)] \nonumber \\
& = & u^{\mu}\partial_{\mu}v^{\alpha}-\partial_{\mu}u^{\alpha}v^{\mu}. 
\end{eqnarray}
The 3-dimensional Lie derivative acts on vectors tangent to the hypersurfaces in the same way.

\section{The intrinsic and extrinsic curvatures}

The intrinsic curvature of the 4-dimensional spacetime $M$ is described by the
non-vanishing of the commutator of any vector, $\mathbf{v}$ in $M$,
$(\bigtriangledown_{\alpha}\bigtriangledown_{\beta}-\bigtriangledown_{\beta}\bigtriangledown_{\alpha})v^{\mu}$
as follows:
\begin{equation}
\label{eq:geod20}
(\bigtriangledown_{\alpha}\bigtriangledown_{\beta}-\bigtriangledown_{\beta}\bigtriangledown_{\alpha})v^{\mu}
=\bigtriangledown_{[\alpha}\bigtriangledown_{\beta]}v^{\mu}=R^{\alpha}_{\gamma\alpha\beta}v^{\gamma}. 
\end{equation}
Again, this can be generalized to a tensor of any rank in $M$. The
contraction of the intrinsic curvature, also known as the Riemann tensor $R^{\alpha}_{\gamma\alpha\beta}$, 
gives the Ricci tensor $R_{\gamma\beta}$. The contraction of the Ricci tensor in turn gives the Ricci scalar
$R=g^{\gamma\beta}R_{\gamma\beta}$. This 4-dimensional Ricci scalar is independent of the
ambient coordinate system of $M$. The expressions for the 3-dimensional
Riemann tensor, Ricci tensor and Ricci scalar take on analogous forms. We
shall denote them as $^{(3)}R^l_{ijk}$, $^{(3)}R_{ij}$ and $^{(3)}R$ respectively.
Similarly, the 3-dimensional Ricci scalar is independent of how the hypersurfaces are embedded in $M$.

Conversely, the extrinsic curvature describes how the hypersurfaces are embedded in
$M$, or more specifically, it measures the curvature of the hypersurfaces in the embedding. 
It is given by the orthogonal projection of the covariant derivative of the timelike normal unit vector $\mathbf{n}$
along any vector $\mathbf{u}$ tangent to the hypersurfaces, as follows:
\begin{equation}
\label{eq:geod21}
K=-\bm{\gamma}\bm{\bigtriangledown}_{\mathbf{u}}\mathbf{n}.
\end{equation}
By invoking the vanishing of the covariant derivative of the 4-metric as a
direct result of Eq.~\eqref{eq:geod17}, the covariant expression of the projection
operator in Eq.~\eqref{eq:geod10}, and the $n^{\sigma}\bigtriangledown_{\nu}n_{\sigma}=0$
identity, the extrinsic curvature can also be expressed 
as the Lie derivative of the 3-metric along $\mathbf{n}$ as follows:
\begin{eqnarray}
\label{eq:geod22}
K_{\alpha\beta} & = & (\bigtriangledown_{\mu}n_{\alpha})\gamma^{\mu}_{\beta} \nonumber \\
& = & \bigtriangledown_{\mu}(N\bigtriangledown_{\alpha}t)\gamma^{\mu}_{\beta} \nonumber \\
& = & ((\bigtriangledown_{\mu}N)\bigtriangledown_{\alpha}t+N\bigtriangledown_{\mu}\bigtriangledown_{\alpha}t)\gamma^{\mu}_{\beta} \nonumber \\
& = & D_{\beta}N\bigtriangledown_{\alpha}t+N\bigtriangledown_{\alpha}(-N^{-1}n_{\mu})\gamma^{\mu}_{\beta} \nonumber \\
& = & -N^{-1}n_{\alpha}D_{\beta}N-\bigtriangledown_{\alpha}(N^{-1}(Nn_{\mu}\gamma^{\mu}_{\beta})
-\bigtriangledown_{\alpha}n_{\mu}\gamma^{\mu}_{\beta} \nonumber \\
& = & -n_{\alpha}D_{\beta}\ln N-(\bigtriangledown_{\alpha}n_{\mu})\delta^{\mu}_{\beta}
-(\bigtriangledown_({\alpha}n_{\mu})n^{\mu}n_{\beta} \nonumber \\
& = & -n_{\alpha}D_{\beta}\ln N-\bigtriangledown_{\alpha}n_{\beta} \nonumber \\
& = & -n_{\alpha}n^{\mu}\bigtriangledown_{\mu}n_{\beta}-\bigtriangledown_{\alpha}n_{\beta} \nonumber \\
& = & -\frac{1}{2}[n_{\alpha}n^{\mu}\bigtriangledown_{\mu}n_{\beta}+n_{\beta}n^{\mu}\bigtriangledown_{\mu}n_{\alpha}
+\bigtriangledown_{\alpha}n_{\beta}+\bigtriangledown_{\beta}n_{\alpha}] \nonumber \\
& = & -\frac{1}{2}[n_{\alpha}n^{\mu}\bigtriangledown_{\mu}n_{\beta}+n_{\beta}n^{\mu}\bigtriangledown_{\mu}n_{\alpha}
-n_{\alpha}n^{\mu}\bigtriangledown_{\beta}n_{\mu}-n_{\beta}n^{\mu}\bigtriangledown_{\alpha}n_{\mu}
+\bigtriangledown_{\alpha}(g_{\beta\mu}n^{\mu})+ \nonumber \\ 
& & \bigtriangledown_{\beta}(g_{\alpha\mu}n^{\mu})] \nonumber \\
& = & -\frac{1}{2}[n^{\mu}\bigtriangledown_{\mu}(n_{\alpha}n_{\beta})
-n_{\alpha}n^{\mu}\bigtriangledown_{\beta}n_{\mu}-n_{\beta}n^{\mu}\bigtriangledown_{\alpha}n_{\mu}
+g_{\beta\mu}\bigtriangledown_{\alpha}n^{\mu}+g_{\alpha\mu}\bigtriangledown_{\beta}n^{\mu}] \nonumber \\
& = & -\frac{1}{2}[n^{\mu}\bigtriangledown_{\mu}(g_{\alpha\beta}+n_{\alpha}n_{\beta})
+(g_{\beta\mu}+n_{\beta}n_{\mu})\bigtriangledown_{\alpha}n^{\mu}
+(g_{\alpha\mu}+n_{\alpha}n_{\mu})\bigtriangledown_{\beta}n^{\mu}] \nonumber \\
& = & -\frac{1}{2}[n^{\mu}\bigtriangledown_{\mu}\gamma_{\alpha\beta}
+\gamma_{\beta\mu}\bigtriangledown_{\alpha}n^{\mu}+\gamma_{\alpha\mu}\bigtriangledown_{\beta}n^{\mu}] \nonumber \\
& = & -\frac{1}{2}\mathcal{L}_{\mathbf{n}}\bm{\gamma}. 
\end{eqnarray}

\section{The Gauss-Codazzi relations}

The various contractions of the 4-dimensional Riemann tensor $^{(4)}\mathbf{R}$ along the 3-dimensional hypersurfaces 
and the timelike unit normal vector $\mathbf{n}$ are essential for the 3+1 formulation of the Einstein field equations.
Solutions to the field equations can be obtained when the equations are cast
in an initial value problem involving constraint equations on the initial
hypersurface and evolution equations on subsequent hypersurfaces. Due to the fact that
these contractions of $^{(4)}\mathbf{R}$ do not involve any timelike derivatives of the metric tensor $\mathbf{g}$, 
they will be used to construct the constraint equations. In this section, we present how these contractions are obtained.
Covariant derivatives of vectors and tensors will hereby be denoted by the subscripted
semicolon, and the vectors and tensors along the hypersurface will use indices denoted with the alphabet
instead of Greek letters.

We begin by considering the full projection of $^{(4)}\mathbf{R}$
onto a hypersurface. Using Eq.~\eqref{eq:geod10} and Eq.~\eqref{eq:geod21}, we first calculate the commutator of the
double covariant derivative of the projection tensor $\mathbf{\gamma}$ as follows:
\begin{eqnarray}
\label{eq:geod23}
\gamma^{\alpha}_{a;\beta\gamma}\gamma^{\beta}_b\gamma^{\gamma}_c-\gamma^{\alpha}_{a;\gamma\beta}\gamma^{\gamma}_c\gamma^{\beta}_b 
& = & -\Gamma^d_{bc}\Gamma^e_{ad}\gamma^{\alpha}_e+\Gamma^d_{cb}\Gamma^e_{ad}\gamma^{\alpha}_e
+\Gamma^d_{bc}K_{ad}n^{\alpha}-\Gamma^d_{cb}K_{ad}n^{\alpha}+ \nonumber \\
& & K_{bc}\gamma^{\alpha}_{a;\beta}n^{\beta}-K_{cb}\gamma^{\alpha}_{a;\beta}n^{\beta} 
+\Gamma^d_{ab,c}\gamma^{\alpha}_d-\Gamma^d_{ac,b}\gamma^{\alpha}_d+ \nonumber \\
& & \Gamma^d_{ab}(\Gamma^e_{dc}\gamma^{\alpha}_e)-\Gamma^d_{ac}(\Gamma^e_{db}\gamma^{\alpha}_e)
-\Gamma^d_{ab}K_{dc}n^{\alpha}+\Gamma^d_{ac}K_{db}n^{\alpha}- \nonumber \\
& & K_{ab,c}n^{\alpha}+K_{ac,b}n^{\alpha}-K_{ab}n^{\alpha}_{;\gamma}\gamma^{\gamma}_c+K_{ac}n^{\alpha}_{;\beta}\gamma^{\beta}_b \nonumber \\
& = & (\Gamma^e_{ab,c}-\Gamma^e_{ac,b}+\Gamma^d_{ab}\Gamma^e_{dc}-\Gamma^d_{ac}\Gamma^e_{db})\gamma^{\alpha}_e- \nonumber \\
& & (K_{ab,c}-\Gamma^d_{ac}K_{db})n^{\alpha}+(K_{ac,b}-\Gamma^d_{ac}K_{dc})n^{\alpha}- \nonumber \\
& & K_{ab}n^{\alpha}_{;\gamma}\gamma^{\gamma}_c+K_{ac}n^{\alpha}_{;\beta}\gamma^{\beta}_b \nonumber \\
& = & -R^m_{abc}\gamma^{\mu}_m-(K_{ab|c}-K_{ac|b})n^{\mu}- \nonumber \\
& & K_{ab}n^{\mu}_{;\gamma}\gamma^{\gamma}_c+K_{ac}n^{\mu}_{;\beta}\gamma^{\beta}_b. 
\end{eqnarray}
Denoting analogously to $^{(4)}R^{\mu}_{\alpha\beta\gamma}$, the 3-dimensional intrinsic curvature
$^{(3)}R^i_{ljk}$, in terms of vectors $v^i$ tangent to the hypersurface as follows:
\begin{equation}
\label{eq:geod24}
-^{(3)}R^i_{ljk}v^l=v^i_{;jk}-v^i_{;kj},
\end{equation}
the previous equation can be written as:
\begin{equation}
\label{eq:geod25}
^{(4)}R^{\mu}_{\alpha\beta\gamma}\gamma^{\alpha}_a\gamma^{\beta}_b\gamma^{\gamma}_c 
=^{(3)}R^m_{abc}\gamma^{\mu}_m+(K_{ab;c}-K_{ac;b})n^{\mu}+K_{ab}n^{\mu}_{;\gamma}\gamma^{\gamma}_c
-K_{ac}n^{\mu}_{;\beta}\gamma^{\beta}_b.
\end{equation}
A further projection of this equation along the hypersurface yields the
Gauss relation as follows:
\begin{equation}
\label{eq:geod26}
^{(4)}R^{\mu}_{\alpha\beta\gamma}\gamma^{\alpha}_a\gamma^{\beta}_b\gamma^{\gamma}_c\gamma^d_{\mu}
=^{(3)}R^d_{abc}+(K^d_b K_{ac}-K^d_c K_{ab}).
\end{equation}
The Gauss relation can be contracted in the $\mu$ and $\beta$ indices using
again Eq.~\eqref{eq:geod10} to yield the following:
\begin{equation}
\label{eq:geod27}
\gamma^{\alpha}_a\gamma^{\gamma}_c{}^{(4)}R_{\alpha\gamma}+\gamma_{c\mu}n^{\alpha}\gamma^{\beta}_a
n^{\gamma}{}^{(4)}R^{\mu}_{\alpha\beta\gamma}=^{(3)}R_{ac}+KK_{ac}-K^b_c K_{ab}.
\end{equation}
Further contraction in the $a$ and $c$ indices will yield the contracted
Gauss relation as follows:
\begin{equation}
\label{eq:geod28}
^{(4)}R+2^{(4)}R_{\mu\alpha}n^{\mu}n^{\alpha}=^{(3)}R+K^2-K_{ab}K^{ab},
\end{equation}
which will be used in the 3+1 formulation of the Einstein field equations.

We now consider the projection of Eq.~\eqref{eq:geod26} along the timelike
normal unit vector $\mathbf{n}$ which results in the Codazzi relation:
\begin{equation}
\label{eq:geod29}
^{(4)}R^{\mu}_{\alpha\beta\gamma}n_{\mu}\gamma^{\alpha}_a\gamma^{\beta}_b\gamma^{\gamma}_c=K_{ab;c}-K_{ac;b}.
\end{equation}
The Codazzi relation can be similarly contracted in the $a$ and $b$ indices
to yield the contracted Codazzi relation:
\begin{equation}
\label{eq:geod30}
^{(4)}R^{\mu}_{\gamma}n_{\mu}\gamma^{\gamma}_c=K_{;c}-K^b_{c;b}.
\end{equation}

\section{The Ricci relation}

The projection of $^{(4)}\mathbf{R}$ twice along the hypersurface and
twice along the timelike unit normal vector $\mathbf{n}$ yields the Ricci
relation which will be used to construct the evolution equations in the 3+1
formulation of the Einstein field equations. We obtain this by considering 
the projection of the commutator of the double covariant derivative of the
timelike unit normal vector itself, and using the expression for $K_{\alpha\beta}$
in Eq.~\eqref{eq:geod22} and the component form of Eq.~\eqref{eq:geod21}, as well as the
projection onto the hypersurface of the Lie derivative of $K_{\alpha\beta}$ with respect to the
timelike normal evolution vector $\mathbf{m}$, as follows:
\begin{eqnarray}
\label{eq:geod31}
\gamma_{\alpha\mu}n^{\sigma}\gamma^{\nu}_{\beta}{}^{(4)}R^{\mu}_{\rho\nu\sigma}n^{\rho}
& = & \gamma_{\alpha\mu}n^{\sigma}\gamma^{\nu}_{\beta}(n^{\mu}_{;\nu\sigma}-n^{\mu}_{;\sigma\nu}) \nonumber \\
& = & \gamma_{\alpha\mu}n^{\sigma}\gamma^{\nu}_{\beta}
[-(K^{\mu}_{\sigma}+D^{\mu}ln Nn_{\sigma})_{;\nu}+(K^{\mu}_{\nu}+D^{\mu}ln Nn_{\nu})_{;\sigma}] \nonumber \\
& = & \gamma_{\alpha\mu}n^{\sigma}\gamma^{\nu}_{\beta}
[-K^{\mu}_{\sigma;\nu}-n_{\sigma;\nu}D^{\mu}ln N-n_{\sigma}(D^{\mu}ln N)_{;\nu}-K^{\mu}_{\nu;\sigma}- \nonumber \\
& & n_{\nu;\sigma}D^{\mu}ln N-n_{\nu}(D^{\mu}ln N)_{;\sigma}] \nonumber \\
& = & \gamma_{\alpha\mu}\gamma^{\nu}_{\beta}[-n^{\sigma}K^{\mu}_{\sigma;\nu}+(D^{\mu}ln N)_{;\nu}
+n^{\sigma}K^{\mu}_{\nu;\sigma}+n^{\sigma}n_{\nu;\sigma}D^{\mu}ln N] \nonumber \\
& = & \gamma_{\alpha\mu}\gamma^{\nu}_{\beta}[K^{\mu}_{\sigma}n^{\sigma}_{;\nu}+(D^{\mu}ln N)_{;\nu}
+n^{\sigma}K^{\mu}_{\nu;\sigma}+D_{\nu}ln N D^{\mu}ln N] \nonumber \\
& = & -K_{\alpha\sigma}K^{\sigma}_{\beta}+D_{\beta}D_{\alpha}ln N
+\gamma^{\mu}_{\alpha}\gamma^{\nu}_{\beta}n^{\sigma}K_{\mu\nu;\sigma}+D_{\alpha}ln N D_{\beta}ln N] \nonumber \\
& = & -K_{\alpha\sigma}K^{\sigma}_{\beta}+\frac{1}{N}D_{\beta}D_{\alpha}N
+\gamma^{\mu}_{\alpha}\gamma^{\nu}_{\beta}n^{\sigma}K_{\mu\nu;\sigma} \nonumber \\
& = & -K_{\alpha\sigma}K^{\sigma}_{\beta}+\frac{1}{N}D_{\beta}D_{\alpha}N
+\frac{1}{N}(\mathcal{L}_{\mathbf{m}}K_{\alpha\beta}+2K_{\alpha\mu}K^{\mu}_{\beta}) \nonumber \\
& = & \frac{1}{N}D_{\beta}D_{\alpha}N+\frac{1}{N}\mathcal{L}_{\mathbf{m}}K_{\alpha\beta}+K_{\alpha\mu}K^{\mu}_{\beta}. 
\end{eqnarray}
 
\section{Projections of the stress-energy tensor}

The various projections of the stress-energy tensor onto the 3-dimensional
hypersurface and along the timelike unit normal vector $\mathbf{n}$ are needed to construct the
matter sources for the 3+1 formulation of the Einstein field equations. 

We first present the full projection of this tensor along $\mathbf{n}$. We recall
from Section 2.1 that the 4-velocity of the Eulerian observers is
definitionally the timelike normal unit vector $\mathbf{n}$. Therefore, the
projection of the stress-energy tensor along $\mathbf{n}$ is the matter
energy density, which is a scalar measured by these Eulerian observers. We denote this matter
energy density as $E$. 

The mixed projection of the stress-energy tensor is called the matter
momentum density, $\mathbf{p}$, which is a linear form tangent to the
hypersurface. Its component form is given by:
\begin{equation}
\label{eq:geod32}
p_{\alpha}=-T_{\mu\nu}n^{\mu}n^{\nu}.
\end{equation}

The full projection of the stress-energy tensor along the hypersurface is
called the matter stress tensor, $\mathbf{S}$, which is a bilinear form
tangent to the hypersurface. Its componenet form is given by:
\begin{equation}
\label{eq:geod33}
S_{\alpha\beta}=T_{\mu\nu}\gamma^{\mu}_{\alpha}\gamma^{\nu}_{\beta}.
\end{equation}

Using Eq.~\eqref{eq:geod10} in this component form of the matter stress tensor and
taking the trace, we obtain the following relation:
\begin{equation}
\label{eq:geod34}
T=S-E.
\end{equation} 

\section{3+1 decomposition of the Einstein field equations}

Armed with the various projections of the intrinsic curvature tensor
$\bm{^{(4)}R}$ and the stress-energy tensor $\mathbf{S}$, we are now ready to
present the full 3+1 decomposition of the Einstein field equations. In the
decomposition, we will utilize two forms of the field equations, namely:
\begin{equation}
\label{eq:geod35}
\bm{^{(4)}R}-\frac{1}{2}^{(4)}R\mathbf{g}=8\pi\mathbf{T},
\end{equation}
and its equivalent:
\begin{equation}
\label{eq:geod36}
\bm{^{(4)}R}=8\pi(\mathbf{T}-\frac{1}{2}T\mathbf{g}),
\end{equation}
where $T:=g^{\mu\nu}T_{\mu\nu}$ is the trace of the stress-energy tensor
$\mathbf{T}$.

To construct the Hamiltonian constraint components of the Einstein field
equations, we apply the twice-contracted Gauss relation obtained in Section
2.4, into the twice-contracted Eq.~\eqref{eq:geod35} along the timelike normal
unit vector $\mathbf{n}$, as follows:
\begin{eqnarray}
\label{eq:geod37}
R_{\mu\nu}n^{\mu}n^{\nu}-\frac{1}{2}^{(4)}R g_{\mu\nu}n^{\mu}n^{\nu} & = & 8\pi T_{\mu\nu}n^{\mu}n^{\nu} \nonumber \\
2R_{\mu\nu}n^{\mu}n^{\nu}+^{(4)}R & = & 16\pi E \nonumber \\
^{(3)}R+K^2-K_{ab}K^{ab} & = & 16\pi E. 
\end{eqnarray}

The momentum constraint components of the field equations however will make
use of the contracted Codazzi relation as obtained in Section 2.4, in the
mixed projection of Eq.~\eqref{eq:geod35} once along $\mathbf{n}$ and once along
the hypersurface, as follows:
\begin{eqnarray}
\label{eq:geod38}
R_{\mu\nu}n^{\nu}\gamma^{\mu}_{\alpha}-\frac{1}{2}^{(4)}R g_{\mu\nu}n^{\nu}\gamma^{\mu}_{\alpha}
& = & 8\pi T_{\mu\nu}n^{\nu}\gamma^{\mu}_{\alpha} \nonumber \\
K_{;c}-K^b_{c;b} & = & 8\pi p_c. 
\end{eqnarray}

To construct the evolution components of the field equations, we first
combine the Ricci relation (Section 2.5) with the once-contracted Gauss relation (Section 2.4):
\begin{equation}
\label{eq:geod39}
\gamma^{\mu}_{\alpha}\gamma^{\nu}_{\beta}{}^{(4)}R_{\mu\nu} 
=^{(3)}R_{\alpha\beta}+KK_{\alpha\beta}-\frac{1}{N}D_{\beta}D_{\alpha}N
-\frac{1}{N}\mathcal{L}_{\mathbf{m}}K_{\alpha\beta}-2K_{\alpha\mu}K^{\mu}_{\beta}. 
\end{equation}
We then substitute this into the equivalent form of the field equations Eq.~\eqref{eq:geod36}
in this construction:
\begin{eqnarray}
\label{eq:geod40}
\gamma^{\mu}_{\alpha}\gamma^{\nu}_{\beta}{}^{(4)}R_{\mu\nu}
& = & 8\pi(\gamma^{\mu}_{\alpha}\gamma^{\nu}_{\beta}T_{\mu\nu}-\frac{1}{2}T\gamma^{\mu}_{\alpha}\gamma^{\nu}_{\beta}g_{\mu\nu}) \nonumber \\  
^{(3)}R_{\alpha\beta}+KK_{\alpha\beta}-\frac{1}{N}D_{\beta}D_{\alpha}N- \nonumber \\
\frac{1}{N}\mathcal{L}_{\mathbf{m}}K_{\alpha\beta}-2K_{\alpha\mu}K^{\mu}_{\beta} 
& = & 8\pi[S_{\alpha\beta}-\frac{1}{2}(S-E)\gamma_{\alpha\beta}] \nonumber \\
\mathcal{L}_{\mathbf{m}}K_{\alpha\beta} & = & -D_{\beta}D_{\alpha}N+N\{^{(3)}R_{\alpha\beta}+KK_{\alpha\beta}- \nonumber \\
& & 2K_{\alpha\mu}K^{\mu}_{\beta}+4\pi[(S-E)\gamma_{\alpha\beta}-2S_{\alpha\beta}]\}. 
\end{eqnarray}

   \chapter{Application of the 3+1 formalism in neutron star simulations}

\section{3+1 decomposition of general relativistic hydrodynamics}

The conservation of the stress-energy tensor $\mathbf{T}$ ensures that, as
long as they are satisfied on the initial hypersurface, the
Einstein field equations are satisfied for all time. The conservation is
given as:
\begin{equation}
\label{eq:bssn1}
T^{\mu\nu}_{;\mu}=0.
\end{equation}

In neutron star simulations, the neutron star matter is characterized by a 
perfect fluid, where there is no presence of heat conduction and other stresses besides 
pressure. The stress-energy tensor for a perfect fluid is purely
defined by the matter energy density and the pressure. In component form, the
stress-energy tensor is given by:
\begin{equation}
\label{eq:bssn2}
T_{\mu\nu}=\rho hu_{\mu}u^{\nu}+pg_{\mu\nu}.
\end{equation}

To incorporate these equations into the 3+1 formalism, we decompose them
into the 3+1 form by first writing them in a 1st order flux conservative
form:
\begin{equation}
\label{eq:bssn3}
\partial_t\vec{U}+\partial_i\vec{F}^i=\vec{S},
\end{equation}
where $\vec{U}$, $\vec{F}^i$ and $\vec{S}$ are the evolved state vector, 
the flux vector and source vector respectively. The evolved state vector 
$\vec{U}$ is written in terms of the primitive variables, 
ie. the matter density $\rho$, fluid velocity vector $v^i$, and specific internal energy
density $\epsilon$ as follows:
\begin{equation}
\label{eq:bssn4}
\vec{U}=\begin{bmatrix}
D \\ 
S_j \\ 
\tau 
\end{bmatrix}=\begin{bmatrix}
\sqrt{\gamma}W\rho \\
\sqrt{\gamma}\rho hW^2 v_j \\
\sqrt{\gamma}(\rho hW^2-P-W\rho)
\end{bmatrix},
\end{equation}
where the fluid velocity vector is a 3-velocity related to the 4-velocity
$u^i$ as:
\begin{equation}
\label{eq:bssn5}
\{u^\mu\}=\frac{W}{N}\{1,Nv^i-\beta^i\},
\end{equation}
with $W$ representing the Lorentz factor $W=1/\sqrt{1-\gamma_{ij}v^i v^j}$, 
and $h$ is the specific enthalpy which can be written in terms of the specific internal energy density as follows:
\begin{equation}
\label{eq:bssn6}
h=1+\epsilon+P/\rho.
\end{equation}
The flux vector $\vec{F}^i$ is defined as:
\begin{equation}
\label{eq:bssn7}
\vec{F}^i=\begin{bmatrix}
N(v^i-\beta^i/N)D \\ 
N((v^i-\beta^i/N)S_j+\sqrt{\gamma}P\delta^i_j) \\
N((v^i-\beta^i/N)\tau+\sqrt{\gamma}v^i P)  
\end{bmatrix},
\end{equation}
whilst the source vector $\vec{S}$ is defined as:
\begin{equation}
\label{eq:bssn8}
\vec{S}=\begin{bmatrix}
0 \\   
N\sqrt{\gamma}T^{\mu\nu}g_{\nu\sigma}\Gamma^\sigma_{\mu j} \\
N\sqrt{\gamma}(T^{\mu t}\partial_\mu N-NT^{\mu\nu}\Gamma^t_{\mu\nu})  
\end{bmatrix}.
\end{equation}

\section{Conformal decomposition} 

Conformal decomposition was introduced by Lichnerowicz in 1944 \cite{Gourgoulhon07} to facilitate a more efficient resolution of the constraint equations
in obtaining valid initial data for the initial value problem. In the
decomposition, the 3-metric $\bm{\gamma}$ is written in terms of a
conformal factor $\Psi$, which is a positive scalar field, and a conformal 3-metric $\bm{\tilde{\gamma}}$
as follows:
\begin{equation}
\label{eq:bssn9}
\bm{\gamma}=\Psi^4\bm{\tilde{\gamma}},
\end{equation}
where $\det\bm{\tilde{\gamma}}=\det f=1$ when Cartesian coordinates are used.
The extrinsic curvature $\mathbf{K}$ of the 3-dimensional hypersurface is
decomposed into its trace $K:=K^i_i=\gamma^{ij}K_{ij}$ and a traceless form $\mathbf{A}$ as follows:
\begin{equation}
\label{eq:bssn10}
\mathbf{A}:=\mathbf{K}-\frac{1}{3}K\gamma,
\end{equation}
where $\gamma^{ij}A_{ij}=0$. We recall that the expression of the extrinsic
curvature can be written in terms of the Lie-derivative of the 3-metric,
as shown in Eq.~\eqref{eq:geod22}.

We substitute the new form of $\mathbf{K}$ as written in Eq.~\eqref{eq:bssn10} and $\bm{\gamma}$ 
into this to obtain the following:
\begin{equation}
\label{eq:bssn11}
\mathcal{L}_{\mathbf{m}}(\Psi^4\tilde{\gamma_{ij}})=-2NA_{ij}-\frac{2}{3}NK\gamma_{ij}.
\end{equation} 
Taking the trace of this equation with respect to $\bm{\tilde{\gamma}}$
and applying the general law of variation of the determinant of an invertible
matrix twice, we obtain the evolution equation for the conformal factor $\Psi$
as follows:
\begin{equation}
\label{eq:bssn12}
(\frac{\partial}{\partial t}-\mathcal{L}_{\bm{\beta}})\ln\Psi=\frac{1}{6}(\tilde{D}_i\beta^i-NK).
\end{equation}

Using the general law of variation of any invertible matrix, 
Eq.~\eqref{eq:bssn11} also yields the evolution equation for the conformal
metric $\bm{\tilde{\gamma}}$ as follows:
\begin{equation}
\label{eq:bssn13}
(\frac{\partial}{\partial t}-\mathcal{L}_{\bm{\beta}})\tilde{\gamma}_{ij}
=-2N\Psi^{-4}A_{ij}-\frac{2}{3}\tilde{D}_k\beta^k\tilde{\gamma}_{ij}.
\end{equation}

In order to obtain the conformally-decomposed evolution equation for the extrinsic curvature tensor
$\mathbf{K}$, we recall the evolution component of the Einstein field
equations in the 3+1 formalism as obtained in Chapter 2, ie. Eq.~\eqref{eq:geod40},
and substitute Eq.~\eqref{eq:bssn10} into it. Also using Eq.~\eqref{eq:geod22}, we obtain:
\begin{equation}
\label{eq:bssn14}
\mathcal{L}_{\mathbf{m}}K_{ij}=\mathcal{L}_{\mathbf{m}}A_{ij}+\frac{1}{3}\mathcal{L}_{\mathbf{m}}K\gamma_{ij}-\frac{2}{3}NKK_{ij}.
\end{equation}
From the identity:
\begin{eqnarray}
\label{eq:bssn15}
\mathcal{L}_{\mathbf{m}}K & = & \gamma^{ij}\mathcal{L}_{\mathbf{m}}K_{ij}+K_{ij}\mathcal{L}_{\mathbf{m}}\gamma^{ij} \nonumber \\
& = & \gamma^{ij}\mathcal{L}_{\mathbf{m}}K_{ij}+2NK_{ij}K^{ij},
\end{eqnarray}
we thus obtain:
\begin{eqnarray}
\label{eq:bssn16}
\mathcal{L}_{\mathbf{m}}K & = & -D_i D^i N+N[R+K^2+4\pi(S-3E)] \nonumber \\
& = & -D_i D^i N+N[4\pi(E+S)+K_{ij}^{ij}].
\end{eqnarray}
We now substitute this and Eq.~\eqref{eq:geod40} back into Eq.~\eqref{eq:bssn14}, employing
again Eq.~\eqref{eq:bssn10}, to obtain:
\begin{eqnarray}
\label{eq:bssn17}
\mathcal{L}_{\mathbf{m}} A_{ij} & = & -D_i D_j N+N[R_{ij}+\frac{1}{3}KA_{ij}-2A_{ik}A^k_j-8\pi(S_{ij}-\frac{1}{3}S\gamma_{ij})]+ \nonumber \\
& & \frac{1}{3}(D_k D^k N-NR)\gamma_{ij}.
\end{eqnarray} 
Eq.s~\eqref{eq:bssn16} and \eqref{eq:bssn17} represent the trace part and the traceless part of
the evolution equation for $\mathbf{K}$. We now conformally decompose the
trace part, ie. Eq.~\eqref{eq:bssn16} by subtituting Eq.~\eqref{eq:bssn10} and $\tilde{A}^{ij}=\Psi^4 A^{ij}$ into it
to obtain the following:
\begin{equation}
\label{eq:bssn18}
\mathcal{L}_{\mathbf{m}}K=-\Psi^4(\tilde{D}_i\tilde{D}^i N+2\tilde{D}_i
\ln\Psi\tilde{D}^i N)+N[4\pi(E+S)+\tilde{A}_{ij}\tilde{A}^{ij}+\frac{K^2}{3}]. 
\end{equation}
Similarly, employing Eq.~\eqref{eq:bssn12}, the conformal connection
$C^k_{ij}=\Gamma^k_{ij}-\frac{1}{2}\tilde{\gamma}^{kl}(\frac{\partial\tilde{\gamma}_{lj}}{\partial
x^i}+\frac{\partial\tilde{\gamma}_{il}}{\partial
x^j}-\frac{\partial\tilde{\gamma}_{ij}}{\partial x^l})$ and the resulting
conformal Ricci tensor and conformal Ricci scalar, we conformally decompose the traceless part into:
\begin{eqnarray}
\label{eq:bssn19}
\mathcal{L}_{\mathbf{m}}\tilde{A}_{ij} & = & -\frac{2}{3}\tilde{D}_k\beta^k\tilde{A}_{ij}
+N[K\tilde{A}_{ij}-2\tilde{\gamma}^{kl}\tilde{A}_{ik}\tilde{A}_{jl}-8\pi(\Psi^{-4}S_{ij}-\frac{1}{3}S\tilde{\gamma}_{ij})]+ \nonumber \\
& & \Psi^{-4}\{-\tilde{D}_i\tilde{D}_j N+2\tilde{D}_i \ln\Psi\tilde{D}_j N+2\tilde{D}_j \ln\Psi\tilde{D}_i N+ \nonumber \\
& & \frac{1}{3}(\tilde{D}_k\tilde{D}^k N-4\tilde{D}_k \ln\Psi\tilde{D}^k N)\tilde{\gamma}_{ij}+ \nonumber \\
& & N[\tilde{R}_{ij}-\frac{1}{3}\tilde{R}\tilde{\gamma}_{ij}-2\tilde{D}_i\tilde{D}_j \ln\Psi+4\tilde{D}_i \ln\Psi\tilde{D}_j \ln\Psi+ \nonumber \\
& & \frac{2}{3}(\tilde{D}_k\tilde{D}^k \ln\Psi-2\tilde{D}_k \ln\Psi\tilde{D}^k \ln\Psi)\tilde{\gamma}_{ij}]\}. 
\end{eqnarray}

We now present the conformal decomposition of the constraint part of the
Einstein field equations in the 3+1 formalism as obtained in Chapter 2, ie.
Eq.s~\eqref{eq:geod37} and  \eqref{eq:geod38}. We substitute Eq.~\eqref{eq:bssn10} and the conformal Ricci scalar,
$R=\Psi^{-4}\tilde{R}-8\Psi^{-5}\tilde{D}_i\tilde{D}^i\Psi$, into the
3+1 Hamiltonian constraint equation, ie. Eq.~\eqref{eq:geod37}, to yield:
\begin{equation}
\label{eq:bssn20}
\tilde{D}_i\tilde{D}^i\Psi-\frac{1}{8}\tilde{R}\Psi+(\frac{1}{8}\tilde{A}_{ij}\tilde{A}^{ij}-\frac{1}{12}K^2+2\pi
E)\Psi^5=0.
\end{equation}
Similarly, the conformal decomposition of the momentum constraint equation,
ie. Eq.~\eqref{eq:geod38}, yields:
\begin{equation}
\label{eq:bssn21}
\tilde{D}_j\tilde{A}^{ij}\tilde{D}_j \ln\Psi-\frac{2}{3}\tilde{D}^i K=8\pi\Psi^4 p^i.
\end{equation}

These six equations, ie. Eq.s~\eqref{eq:bssn12},~\eqref{eq:bssn13},~\eqref{eq:bssn16},~\eqref{eq:bssn19},~\eqref{eq:bssn20} and \eqref{eq:bssn21}, constitute the
conformal 3+1 Einstein field equations. This set of equations is solved 
for the conformal 3-metric $\tilde{\gamma}_{ij}$, the conformal traceless
part of the extrinsic curvature $\tilde{A}_{ij}$, the conformal factor
$\Psi$ and the trace of the extrinsic curvature $K$. We recover the
physical 3-metric $\bm{\gamma}$ and the physical extrinsic curvature
$\mathbf{K}$ via the following:
\begin{eqnarray}
\label{eq:bssn22}
\gamma_{ij} & = & \Psi^4\tilde{\gamma}_{ij} \nonumber \\
K_{ij} & = & \Psi^4(\tilde{A}_{ij}+\frac{1}{3}K\tilde{\gamma}_{ij}).
\end{eqnarray}

However, in our neutron star simulations, we employ a modified form of the above
conformal 3+1 Einstein field equations. This modified formulation is called
the Baumgarte-Shapiro-Shibata-Nakamura (BSSN) formulation. In this
formulation, a vector is introduced by Shibata and Nakamura as well as
Baumgarte and Shapiro to restore the Laplacian nature of the conformal Ricci tensor
that is written in terms of the conformal metric. 

In order to do this, the expression of the conformal Ricci
tensor in terms of the conformal metric, $\bm{\tilde{\gamma}}$, is considered as follows:
\begin{equation}
\label{eq:bssn23}
\tilde{R}_{ij}=\frac{\partial}{\partial
x^k}\tilde{\Gamma}^k_{ij}-\frac{\partial}{\partial
x^j}\tilde{\Gamma}^k_{ik}+\tilde{\Gamma}^k_{ij}\tilde{\Gamma}^l_{kl}-\tilde{\Gamma}^k_{il}\tilde{\Gamma}^l_{kj}.
\end{equation}
We introduce a new tensor field $\bm{\Delta}$ as follows:
\begin{equation}
\label{eq:bssn24}
\Delta^k_{ij}:=\tilde{\Gamma}^k_{ij}-\bar{\Gamma}^k_{ij},
\end{equation}
where $\bar{\Gamma}^k_{ij}$ are the connection coefficients for the flat
metric with respect to the coordinates $(x^i)$. The components of this
tensor field can also be written as:
\begin{equation}
\label{eq:bssn25}
\Delta^k_{ij}=\frac{1}{2}\tilde{\gamma}^{kl}(\mathcal{D}_i\tilde{\gamma}_{lj}+\mathcal{D}_j\tilde{\gamma}_{il}-\mathcal{D}_l\tilde{\gamma}_{ij}),
\end{equation}
where $\mathcal{D}_i$ represents the covariant derivative associated with
the flat metric. 

Using the expression~\eqref{eq:bssn24} together with the fact that the Ricci tensor
vanishes for a flat metric and the identity $\Delta^k_{ik}=0$, we obtain:
\begin{eqnarray}
\label{eq:bssn26}
\tilde{R}_{ij} & = & \frac{\partial}{\partial x^k}\Delta^k_{ij}-\frac{\partial}{\partial
x^i}\Delta^k_{ik}+\Delta^k_{ij}\Delta^l_{kl}+\bar{\Gamma}^l_{kl}\Delta^k_{ij}+\bar{\Gamma}^k_{ij}\Delta^l_{kl}-\Delta^k_{il}\Delta^l_{kj}- \nonumber \\
& & \bar{\Gamma}^l_{kj}\Delta^k_{il}-\bar{\Gamma}^k_{il}\Delta^l_{kj} \nonumber \\
& = & \frac{\partial}{\partial
x^k}\Delta^k_{ij}+\bar{\Gamma}^l_{kl}\Delta^k_{ij}-\bar{\Gamma}^l_{kj}\Delta^k_{il}-\bar{\Gamma}^k_{il}\Delta^l_{kj}-\Delta^k_{il}\Delta^l_{kj} \nonumber \\
& = & \frac{\partial}{\partial
x^k}\Delta^k_{ij}+\bar{\Gamma}^k_{kl}\Delta^l_{ij}-\bar{\Gamma}^l_{ki}\Delta^k_{lj}-\bar{\Gamma}^l_{kj}\Delta^k_{il}-\Delta^k_{il}\Delta^l_{kj} \nonumber \\
& = & \mathcal{D}_k\Delta^k_{ij}-\Delta^k_{il}\Delta^l_{kj}. 
\end{eqnarray}
 
Using expression~\eqref{eq:bssn25} on the previous result yields:
\begin{eqnarray}
\label{eq:bssn27}
\tilde{R}_{ij} & = & \frac{1}{2}\mathcal{D}_k[\tilde{\gamma}^{kl}(\mathcal{D}_i\tilde{\gamma}_{lj}+\mathcal{D}_j\tilde{\gamma}_{il}-\mathcal{D}_l\tilde{\gamma}_{ij})]
-\Delta^k_{il}\Delta^l_{kj} \nonumber \\
& = & \frac{1}{2} \{\mathcal{D}_k[\mathcal{D}_i(\tilde{\gamma}^{kl}\tilde{\gamma}_{lj})-\tilde{\gamma}_{lj}\mathcal{D}_i\tilde{\gamma}^{kl}
+\mathcal{D}_j(\tilde{\gamma}^{kl}\tilde{\gamma}_{il})-\tilde{\gamma}_{il}\mathcal{D}_j\tilde{\gamma}^{kl}]- \nonumber \\
& & \mathcal{D}_k\tilde{\gamma}^{kl}\mathcal{D}_l\tilde{\gamma}_{ij}-\tilde{\gamma}^{kl}\mathcal{D}_k\mathcal{D}_l\tilde{\gamma}_{ij}\}-\Delta^k_{il}\Delta^l_{kj} \nonumber \\
& = & \frac{1}{2}(-\mathcal{D}_k\tilde{\gamma}_{lj}\mathcal{D}_i\tilde{\gamma}^{kl}-\tilde{\gamma}_{lj}\mathcal{D}_k\mathcal{D}_i\tilde{\gamma}^{kl}
-\mathcal{D}_k\tilde{\gamma}_{il}\mathcal{D}_j\tilde{\gamma}^{kl}-\tilde{\gamma}_{il}\mathcal{D}_k\mathcal{D}_j\tilde{\gamma}^{kl}- \nonumber \\
& & \mathcal{D}_k\tilde{\gamma}^{kl}\mathcal{D}_l\tilde{\gamma}_{ij}-\tilde{\gamma}^{kl}\mathcal{D}_k\mathcal{D}_l\tilde{\gamma}^{ij})-\Delta^k_{il}\Delta^l_{kj} \nonumber \\
& = & -\frac{1}{2}(\tilde{\gamma}^{kl}\mathcal{D}_k\mathcal{D}_l\tilde{\gamma}_{ij}+\tilde{\gamma}_{ik}\mathcal{D}_j\mathcal{D}_l\tilde{\gamma}^{kl}
+\tilde{\gamma}_{jk}\mathcal{D}_i\mathcal{D}_l\tilde{\gamma}^{kl})+\mathcal{O}_{ij}(\bm{\tilde{\gamma}},\bm{\mathcal{D}\tilde{\gamma}}),
\end{eqnarray}
where
$\mathcal{O}_{ij}(\bm{\tilde{\gamma}},\bm{\mathcal{D}\tilde{\gamma}}):=
-\frac{1}{2}(\mathcal{D}_k\tilde{\gamma}_{lj}\mathcal{D}_i\tilde{\gamma}^{kl}+\mathcal{D}_k\tilde{\gamma}_{il}\mathcal{D}_j\tilde{\gamma}^{kl}
+\mathcal{D}_k\tilde{\gamma}^{kl}\mathcal{D}_l\tilde{\gamma}_{ij})-\Delta^k_{il}\Delta^l_{kj}.$

The Ricci tensor as rendered in the previous expression can be viewed as a
Laplace operator acting on the conformal metric
$\bm{\tilde{\gamma}}$ yielding second-derivatives on the right hand
side. However, the second and third terms on the previous expression spoils
the elliptic character of the Ricci tensor operator. It is here that Baumgarte
and Shapiro introduce the vector $\tilde{\Gamma}^i=-\mathcal{D}_j\tilde{\gamma}^{ij}$, which turns the
previous expression into the following:
\begin{equation}
\label{eq:bssn28}
\tilde{R}_{ij}=\frac{1}{2}(-\tilde{\gamma}^{kl}\mathcal{D}_k\mathcal{D}_l\tilde{\gamma}_{ij}+\tilde{\gamma}_{ik}\mathcal{D}_j\tilde{\Gamma}^k
+\tilde{\gamma}_{jk}\mathcal{D}_i\tilde{\Gamma}^k)+\mathcal{O}_{ij}(\bm{\tilde{\gamma}},\bm{\mathcal{D}\tilde{\gamma}}).
\end{equation}
Taking the trace of this expression of the conformal Ricci tensor, as well
as recalling the identity $\tilde{\gamma}^{ij}\mathcal{D}_l\tilde{\gamma}_{ij}=2\Delta^k_{lk}=0$, 
the conformal Ricci scalar is thus:
\begin{equation}
\label{eq:bssn29}
\tilde{R}=\mathcal{D}_k\tilde{\Gamma}^k+\mathcal{O}(\bm{\tilde{\gamma}},\bm{\mathcal{D}\tilde{\gamma}}),
\end{equation}
where
$\mathcal{O}(\bm{\tilde{\gamma}},\bm{\mathcal{D}\tilde{\gamma}})
:=\frac{1}{2}\tilde{\gamma}^{ij}\mathcal{D}_k\tilde{\gamma}^{ij}\mathcal{D}_l\tilde{\gamma}_{ij}
+\tilde{\gamma}^{ij}\mathcal{O}_{ij}(\bm{\tilde{\gamma}},\bm{\mathcal{D}\tilde{\gamma}})$,
which is a term that does not contain any second derivatives of $\bm{\tilde{\gamma}}$
and is quadratic in the first derivatives.

Earlier, Shibata and Nakamura introduced the covector $F_i=\mathcal{D}^j\tilde{\gamma}_{ij}$
instead of the vector $\tilde{\Gamma}^i$, which is related to the latter as follows:
\begin{equation}
\label{eq:bssn30}
F_i=\tilde{\gamma}_{ij}\tilde{\Gamma}^j-(\tilde{\gamma}^{jk}-f^{jk})\mathcal{D}_k\tilde{\gamma}_{ij}.
\end{equation}
However, the vector $\tilde{\Gamma}^i$ has an edge over the covector $F_i$
because it covers all the second derivatives of the conformal metric that
do not contribute to the Laplacian operator.  

\section{Gauge choices}

Looking at the set of conformally decomposed Einstein field equations
obtained in the previous section, we note that there are no derivatives of either the lapse function $N$ or shift
vector $\bm{\beta}$. This indicates that in the solution of the field
equations, the lapse function and the shift vector can be freely chosen
while yielding the same physical solution. We recall from
Chapter 2 that the lapse function determines how the spacetime is sliced and
the shift vector determines the choice of coordinates on these spacetime
slices. The choice of the lapse and the shift thus changes the
form of the Einstein field equations to be solved, making it either more
hyperbolic or more elliptic in nature. As the success of numerical simulations
in modeling neutron star systems depends crucially on the well-behavedness
or non-singularity of the coordinate functions, the freedom of the lapse and
shift choice gives us the privilege to adjust the hyperbolicity of the field
equations based on the nature of the physical system to be studied. 
We will discuss in this section the common choices made particularly in neutron star simulations. 

The simplest lapse choice is called the geodesic slicing, which sets $N=1$.
Since the acceleration co-vector can be given by:
\begin{equation}
\label{eq:bssn31}
a_{\alpha}=D_{\alpha}\ln N,
\end{equation}
setting $N=1$ renders zero acceleration for the Eulerian observers, ie. they
travel along geodesics of the spacetime, hence the name of this lapse choice.
This choice permits only limited evolution of the spacetime due to its
focusing property that results in coordinate singularities. 

A more popular choice is the maximal slicing which sets the extrinsic curvature scalar $K=0$.
This lapse choice maximizes the volume of the hypersurface. The volume
enclosed within a closed 2-dimensional surface lying on a hypersurface is
given by:
\begin{equation}
\label{eq:bssn32}
V=\int_{\mathcal{V}}\sqrt{\gamma}d^3 x,
\end{equation}
with $\gamma$ is the determinant of the metric $\bm{\gamma}$ with
respect to the coordinates $(x^i)$ on the hypersurface. The change of
hypervolume can thus be written:
\begin{equation}
\label{eq:bssn33}
\frac{dV}{dt}=\int_{\mathcal{V}_t}\frac{\partial\sqrt{\gamma}}{\partial t}d^3 x,
\end{equation}
where $\mathcal{V}_t$ is the domain at time $t$. Using the identity
$\frac{1}{\sqrt{\gamma}}\frac{\partial\sqrt{\gamma}}{\partial t}=-NK+D_i\beta^i$
that is derived from the evolution equation of the 3-metric, the variation
of the hypervolume in the previous equation can be written as:
\begin{eqnarray}
\label{eq:bssn34}
\frac{dV}{dt} & = & \int_{\mathcal{V}_t}[-NK+D_i\beta^i]\sqrt{\gamma}d^3 x \nonumber \\
& = & -\int_{\mathcal{V}_t}NK\sqrt{\gamma}d^3 x+\oint_{\mathcal{S}}\beta^i s_i\sqrt{q}d^2 y \nonumber \\
& = & -\int_{\mathcal{V}_t}NK\sqrt{\gamma}d^3 x,
\end{eqnarray}
where $\mathbf{s}$ is the unit normal to $\mathcal{S}$ lying in the
hypersurface, $\mathbf{q}$ is the induced metric on $\mathcal{S}$ with $q=q_{ab}$ and
$(y^a)$ are the coordinates on $\mathcal{S}$. Therefore, setting $K=0$ on
the hypersurface renders the hypervolume extremal with respect to the
variations in the domain bounded by $\mathcal{S}$. With a metric of a
Lorentzian signature, this extremum is a maximum, hence the name of this
lapse choice. Combining the maximal slicing condition with the evolution
equation for the extrinsic curvature scalar $K$, Eq.~\eqref{eq:bssn16}, as derived in
the Section 3.2, we obtain:
\begin{equation}
\label{eq:bssn35}
D_i D^i N=N[4\pi(E+S)+K_{ij}K^{ij}],
\end{equation}
an elliptic equation which imposes the condition on all subsequent hypersurfaces in the spacetime.
Maximal slicing avoids the formation of coordinate singularities in the
evolution of a physical system. As we recall from Chapter 2, the extrinsic
curvature scalar is defined as the divergence of the unit normal vector
$\mathbf{n}$, alternatively called the 4-velocity field of the Eulerian
observers. Fixing $K=0$ thus results in $\bm{\bigtriangledown}\cdot\mathbf{n}=0$, 
which prevents the Eulerian observers from converging towards any coordinate singularity that forms due to the focusing effect of gravity. 
This happens when the lapse function and thus the proper time between two adjacent hypersurfaces tends
to zero as the coordinate time tends to infinity.

The next important category of lapse choice is called the $1+log$ slicing,
which is a generalization of the harmonic slicing introduced by Bona,
Mass\'{o}, Seidel and Stela \cite{Bona95}. Harmonic slicing was introduced by
Choquet-Bruhat and Ruggeri \cite{Choquet83} in an attempt to write the 3+1
Einstein field equations in a hyperbolic form, and sets
$\Box_{\mathbf{g}}t=0$, a harmonic condition for the time coordinate.   
Using the relation $\sqrt{-g}=N\sqrt{\gamma}$, the expression for the 4-metric
$g^{\alpha\beta}$ delineated in Chapter 2 and again the identity
$\frac{1}{\sqrt{\gamma}}\frac{\partial\sqrt{\gamma}}{\partial t}=-NK+D_i\beta^i$, this d'Alembertian becomes:
\begin{equation}
\label{eq:bssn36}
(\frac{\partial}{\partial t}-\mathcal{L}_{\bm{\beta}})N=-KN^2,
\end{equation} 
which is an evolution equation for the lapse function.
The $1+log$ slicing thus generalizes this equation as follows:
\begin{equation}
\label{eq:bssn37}
(\frac{\partial}{\partial t}-\mathcal{L}_{\bm{\beta}})N=-KN^2 f(N),
\end{equation}
where $f$ is an arbitrary function with $f(N)=1$ corresponding to the
harmonic slicing and $f(N)=0$ corresponding to the geodesic slicing.
Using the identity $\frac{1}{\sqrt{\gamma}}\frac{\partial\sqrt{\gamma}}{\partial t}=-NK+D_i\beta^i$
and employing normal coordinates where $\bm{\beta}=0$, this equation becomes:
\begin{equation}
\label{eq:bssn38}
\frac{\partial N}{\partial t}=\frac{\partial}{\partial t}\ln\gamma,
\end{equation}
which yields $N=1+\ln\gamma$ as one of its solutions, hence the name of
$1+log$ slicing. The $1+log$ slicing produces foliations very similar the
to maximal slicing and thus have strong singularity avoidance, its biggest
advantage, making it another popular choice for neutron star simulations.

We now move on to the gauge choices commonly made in the shift vector
for neutron star simulations. Again, the simplest of these choices are the
normal coordinates, which set $\bm{\beta}=0$, where the 4-velocity field
$\mathbf{n}$ or the unit normal vector field lines are parallel to the constant spatial coordinate field lines,
hence the name normal coordinates. An advantage of this choice is its
incapability in introducing any pathologies of its own. However, in rotating
star spacetimes, the employment of this shift choice can result in coordinate
shears due to the fact that the unit normal vector field lines are not
parallel to the stationary Killing vector field $\xi$ lines.   

The minimal distortion shift was introduced by Smarr and York in 1978
\cite{Smarr78} in an attempt to minimize the time derivative of the conformal
3-metric, $\bm{\dot{\tilde{\gamma}}}:=\bm{\mathcal{L}_{\partial_t}\tilde{\gamma}}$.
The components of this time derivative are:
\begin{equation}
\label{eq:bssn39}
\dot{\tilde{\gamma_{ij}}}=\frac{\partial\tilde{\gamma}_{ij}}{\partial t},
\end{equation}
which are related to the distortion tensor components $Q_{ij}$ as follows:
\begin{equation}
\label{eq:bssn40}
Q_{ij}=\Psi^4\frac{\partial\tilde{\gamma}_{ij}}{\partial t},
\end{equation}
which has 5 degrees of freedom, ie. 6 freely chosen components minus 1
constraint requiring $\det\dot{\tilde{\gamma}}_{ij}=0$. The distortion tensor $\mathbf{Q}$ 
is a measure of the change of the shape of spatial domain $\mathcal{V}$ within a 
fixed coordinate boundary from one hypersurface to the next.
The minimal distortion shift hence seeks to minimize the change in the shape of
the spatial domain $\mathcal{V}$. This can be done by choosing coordinates
$(x^i)$ that set the distortion tensor $\mathbf{Q}$ identically equal to zero.
Taking into account that the 3 degrees of freedom for the coordinate choice
are insufficient to constrain the 5 degrees of freedom for the distortion
tensor $\mathbf{Q}$, we decompose the distortion tensor $\mathbf{Q}$ into a
longitudinal part and a transverse-traceless part as follows:
\begin{equation}
\label{eq:bssn41}
Q_{ij}=(\mathcal{L}X)_{ij}+Q^{TT}_{ij},
\end{equation}
where $\mathcal{L}X$ is the conformal Killing operator associated with the physical
metric $\bm{\gamma}$ acting on some vector field $\mathbf{X}$. Taking
the divergence of this relation, we obtain:
\begin{equation}
\label{eq:bssn42}
D^j Q_{ij}=D^j(\mathcal{L}X)_{ij},
\end{equation}
which possesses 3 degrees of freedom that can be constrained by the
coordinate choice. Hence, the minimum distortion shift conditon becomes:
\begin{equation}
\label{eq:bssn43}
D^j Q_{ij}=0.
\end{equation}
Using the evolution equation for the 3-metric $\bm{\gamma}$, 
namely, $\mathcal{L}_{\mathbf{m}} \gamma_{ij}=-2NK_{ij}$,
the identity $\frac{1}{\sqrt{\gamma}}\frac{\partial\sqrt{\gamma}}{\partial t}=-NK+D_i\beta^i$, 
and the expression for the traceless part of the extrinsic curvature tensor
$\mathbf{A}$, Eq.~\eqref{eq:bssn10}, this condition yields:
\begin{equation}
\label{eq:bssn44}
-2ND_j A^{ij}-2A^{ij}D_j N+D_i\beta_j+D^j(D_j\beta_i-2D_k\beta^k\gamma_{ij})=0.
\end{equation}
Furthermore, we can employ the 3+1 momentum constraint equation, Eq. (2.38),
to obtain the following elliptic equation of shift evolution resulting from the
minimal distortion condition:
\begin{equation}
\label{eq:bssn45}
D_j D^j\beta^i+\frac{1}{3}D^i D_j\beta^j+R^i_j\beta^j=16\pi Np^i+\frac{4}{3}ND^i K+2A^{ij}D_j N.
\end{equation}

A more computationally-feasible shift choice based on the minimal distortion shift
is the $\Gamma$-freezing shift, introduced by Alcubierre and Br\"{u}gmann \cite{Alcubierre01}.
This shift choice sets $\mathcal{D}_j\dot{\tilde{\gamma}}^{ij}
=\frac{\partial}{\partial t}(\mathcal{D}_j\tilde{\gamma}^{ij})=0$.
The covariant derivative $\mathcal{D}_j$ can be written in terms of the
connection coefficients for the flat metric $\bar{\Gamma}^i_{jk}$ with
respect to the coordinates $(x^i)$ as follows:
\begin{equation}
\label{eq:bssn46}
\mathcal{D}_j\tilde{\gamma}^{ij}=\tilde{\gamma}^{jk}(\bar{\Gamma}^i_{jk}-\tilde{\Gamma}^i_{jk})=-\tilde{\Gamma}^i.
\end{equation}
Hence, the $\Gamma$-freezing shift condition sets $\frac{\partial\tilde{\Gamma}^i}{\partial t}=0$.
To apply this condition on the shift evolution, we write the Lie derivative
of the conformal 3-metric $\tilde{\gamma}_{ij}$ in terms of the covariant derivative
as follows:
\begin{equation}
\label{eq:bssn47}
\dot{\tilde{\gamma}}^{ij}=2N\tilde{A}^{ij}+\beta^k\mathcal{D}_k\tilde{\gamma}^{ij}-\tilde{\gamma}^{kj}\mathcal{D}_k\beta^k
-\tilde{\gamma}^{ik}\mathcal{D}_k\beta^j+\frac{2}{3}\mathcal{D}_k\beta^k\tilde{\gamma}^{ij},
\end{equation}
the divergence of which, via the expression $\mathcal{D}_j\tilde{\gamma}^{ij}=-\tilde{\Gamma}^i$, 
the conformal 3+1 momentum constraint equation, and the relationship between
the conformal and flat covariant derivatives, becomes:
\begin{eqnarray}
\label{eq:bssn48}
\frac{\partial\tilde{\Gamma}^i}{\partial t} = -2N\mathcal{D}_j\tilde{A}^{ij}-2\tilde{A}^{ij}\mathcal{D}_j N+\beta^k\mathcal{D}_k\tilde{\Gamma}^i
-\tilde{\Gamma}^k\mathcal{D}_k\beta^i+\frac{2}{3}\tilde{\Gamma}^i\mathcal{D}_k\beta^k+ \nonumber \\ 
\tilde{\gamma}^{jk}\mathcal{D}_j\mathcal{D}_k\beta^i+\frac{1}{3}\tilde{\gamma}^{ij}\mathcal{D}_j\mathcal{D}_k\beta^k, \nonumber \\ 
\tilde{\gamma}^{jk}\mathcal{D}_j\mathcal{D}_k\beta^i+\frac{1}{3}\tilde{\gamma}^{ij}\mathcal{D}_j\mathcal{D}_k\beta^k
+\frac{2}{3}\tilde{\Gamma}^i\mathcal{D}_k\beta^k-\tilde{\Gamma}^k\mathcal{D}_k\beta^i+\beta^k\mathcal{D}_k\tilde{\Gamma}^i \nonumber \\
= 2N[8\pi\Psi^4 p^i-\tilde{A}^{jk}(\tilde{\Gamma}^i_{jk}-\bar{\Gamma}^i_{jk})-6\tilde{A}^{ij}\mathcal{D}_j \ln\Psi
+\frac{2}{3}\tilde{\gamma}^{ij}\mathcal{D}_j K]+ \nonumber \\
2\tilde{A}^{ij}\mathcal{D}_j N.
\end{eqnarray}
The $\Gamma$-freezing shift condition thus yields an elliptic equation for the shift.
Alcubierre and Br\"{u}gmann \cite{Alcubierre01} turned it into a parabolic equation
using the relation $\frac{\partial\beta^i}{\partial t}=k\frac{\partial\tilde{\Gamma}^i}{\partial t}$
with $k$ being a positive constant. A modification of this is used in the
neutron star simulations presented in this thesis, namely:
\begin{equation}
\label{eq:bssn49}
\frac{\partial\beta^i}{\partial t}=C_1\tilde{\Gamma}^i-C_2\beta^i,
\end{equation}
where we set $C_1=C_2=0$.
   
\section{The initial value problem}

In Chapter 2, we see that the Einstein field equations are separated into
the constraint part and the evolution part, such that their resolution
amounts to solving an initial value problem using the constraint part to obtain initial data that will
be propagated forward in time using the evolution part of the field equations.
As the initial data is constrained, it is a non-trivial astrophysical problem
common particularly in neutron star simulations, to ascertain that the solution to the constraint part of the field equations
yields the physical system to be studied. 

In this section, we discuss the conformal transverse traceless method
proposed by York \cite{York73},\cite{York74},\cite{York79}, a method that has been employed in our general
relativistic hydrodynamics simulations of neutron star and neutron star-like systems.  
We recall that in Section 3.2, we have obtained the conformally decomposed constraint part of the Einstein field equations. 
In 1973 and 1979, York solved this system of equations by further decomposing the trace part of the
conformal extrinsic curvature $\hat{A}^{ij}$ into a longitudinal part and a
transverse part, as follows:
\begin{equation}
\label{eq:bssn50}
\hat{A}^{ij}=(\tilde{\mathcal{L}}X)^{ij}+\hat{A}^{ij}_{TT},
\end{equation}
where $\hat{A}^{ij}_{TT}$ is both traceless and transverse with respect to the
conformal metric, ie. $\tilde{\gamma}_{ij}\hat{A}^{ij}_{TT}=0$ and
$\mathcal{D}_j\hat{A}^{ij}_{TT}=0$ respectively, and $(\tilde{\mathcal{L}}X)^{ij}$ is the conformal
Killing operator associated with the conformal metric $\bm{\tilde{\gamma}}$
that acts on the vector field $\mathbf{X}$ as follows:
\begin{equation}
\label{eq:bssn51}
(\tilde{\mathcal{L}}X)^{ij}:=\tilde{D}^i X^j+\tilde{D}^j X^i-\frac{2}{3}\tilde{D}_k X^k\tilde{\gamma}^{ij},
\end{equation}
and is also traceless, ie. $\tilde{\gamma}_{ij}(\tilde{\mathcal{L}}X)^{ij}=0$. 

We incorporate this longitudinal/transverse decomposition and the identity
$\tilde{\Delta}_{\mathcal{L}}X^i=\tilde{D}_j \hat{A}^{ij}$ into the
conformally decomposed constraint part of the Einstein field equations, Eq.s
~\eqref{eq:bssn20} and ~\eqref{eq:bssn21} and obtain the following:
\begin{equation}
\label{eq:bssn52}
\tilde{D}_i\tilde{D}^i\Psi-\frac{1}{8}\tilde{R}\Psi+\frac{1}{8}[(\tilde{\mathcal{L}}X)_{ij}+\hat{A}^{TT}_{ij}][(\tilde{\mathcal{L}}X)^{ij}+\hat{A}^{ij}_{TT}]\Psi^{-7}+ 
2\pi\tilde{E}\Psi^{-3}-\frac{1}{12}K^2\Psi^5=0, 
\end{equation}
\begin{equation}
\label{eq:bssn53}
\tilde{\Delta}_{\mathcal{L}}X^i-\frac{2}{3}\Psi^6\tilde{D}^i K=8\pi\tilde{p}^i,
\end{equation}
where $(\tilde{\mathcal{L}}X)_{ij}:=\tilde{\gamma}_{ik}\tilde{\gamma}_{jl}(\tilde{\mathcal{L}}X)^{kl}$
and $\hat{A}^{TT}_{ij}=\tilde{\gamma}_{ik}\tilde{\gamma}_{jl}\hat{A}^{kl}_{TT}$.

Eq.s~\eqref{eq:bssn52} and~\eqref{eq:bssn53} above show that in solving this system of
equations, the conformal metric $\bm{\tilde{\gamma}}$, the symmetric traceless and transverse tensor
$\hat{A}^{ij}_{TT}$, the extrinsic curvature scalar $K$, and the conformal
matter variables $(\tilde{E},\tilde{p}^i)$, can be freely chosen, whilst the
conformal factor $\Psi$ and the vector $\mathbf{X}$ will be determined as
results of the initial value solve. We then construct the physical metric
as $\gamma_{ij}=\Psi^4\tilde{\gamma}_{ij}$, the physical extrinsic curvature tensor as
$K^{ij}=\Psi^{-10}((\tilde{\mathcal{L}}X)^{ij}+\hat{A}^{ij}_{TT})+\frac{1}{3}\Psi^{-4}K\tilde{\gamma}^{ij}$,
the physical matter energy density as $E=\Psi^{-8}\tilde{E}$ and the
physical matter momentum density vector as $p^i=\Psi^{-10}\tilde{p}^i$,
which form a set of initial data $(\bm{\gamma},\mathbf{K},E,\mathbf{p})$ that
satisfies the constraint part of the Einstein field equations, Eq.s~\eqref{eq:geod37} and \eqref{eq:geod38}.

\section{GRAstro-2D as an axisymmetric general relativistic hydrodynamics code}

The GRAstro-2D code is based on the Cactus Computational Toolkit \cite{Cactus}
and the GRAstro-3D code \cite{Font00},\cite{Font02}. Thus, similar to the GRAstro-3D code, 
it solves the full 3+1 Einstein field equations as presented in Chapter 2,
using the BSSN scheme as presented earlier in Section 3.2. 
The evolution of the code is similarly unconstrained, ie. the constraint equations are only solved at the initial time and \textit{not} throughout the 
evolution. As such, violation of the constraint equations are monitored throughout the evolution to determine convergence of the code.
The code also employs the same initial value problem solver and finite-differencing scheme as used in GRAstro-3D.  
In this section, we present the basic concepts behind the modifications made by \cite{Jin07} on the GRAstro-3D code for the
purpose of adapting the former to perform numerical simulations of axisymmetric systems.
In both GRAstro-3D and GRAstro-2D, we use geometric units where we set $G=c=1$ (refer to Appendix C).

The basic concepts are based on the Cartoon technique introduced by
Alcubierre, Brandt, Br\"{u}gmann, Holz, Seidel, Takahashi and Thornbug in
2005 \cite{Alcubierre01a}. This technique is able to avoid the problems
normally encountered in other axisymmetric techniques, eg. techniques that
employ a cylindrical $(\rho,z,\theta)$ coordinate grids, where physically
non-singular variables may become indeterminate $0/0$ forms along the
$z$-axis. Although these indeterminate forms can be regularized by
L'Hospital's rule, their evaluation becomes inaccurate in finite
differencing schemes where the grid spacing is finite when its limit is
required to go to zero to be consistent with its analytic counterpart. In
addition to this problem, some of the variables in the 3+1 Einstein field equations
are obtained in finite differencing schemes using a summation of terms that
may include these indeterminate forms. Regularization of such variables
require detailed analysis of the entire system of the 3+1 Einstein field
equations in the axisymmetric coordinate system near the axis of symmetry, 
a difficult although not impossible undertaking.

\begin{figure}
\begin{center}
\includegraphics[scale=0.8]{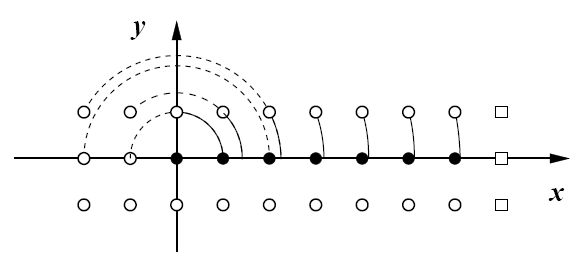}
\caption[The Cartoon technique.]{Figure taken from \cite{Alcubierre01a}.}
\end{center}
\end{figure}

However, Cartesian coordinate grids do not introduce any such pathology, as
the coordinate system is completely regular even at the grid origin. The
Cartoon technique employs a 3-dimensional Cartesian $(x,y,z)$ coordinate grid that is restricted to the $y=0$ plane by only one
finite-diffence-length. It uses continuous rotational symmetry to provide
the boundary conditions for this thin 3-dimensional slab. Fig. 3.1 shows     
the construction of this slab, where the $z$ axis is taken as the axis of
symmetry. The solid dots represent grid points where
variables take on the values calculated on the original 3-dimensional
coordinate grid. The white dots represent grid points where the variables
are calculated via the Cartoon boundary conditions. The white squares represent grid
points where the variables are calculated by imposing a physical boundary. 
The Cartoon boundary condition is described by a rotational coordinate transformation $R$. If we
consider arbitrary tensor fields $T$ on the 3-dimensional coordinate grid, a
rotation of such a tensor field by an angle $-\theta_0$ about the axis of
symmetry is equivalent to the rotation of the coordinate system by an angle
$\theta_0$. The coordinate transformation $R$ can thus be written as:
\begin{equation}
\label{eq:bssn54}
(R(\theta_0)^i_j)=(\frac{\partial x'^{i}}{\partial x^j})=
\left[\begin{array}{ccc} \cos\theta_0 & -\sin\theta_0 & 0 \\ 
                         \sin\theta_0 & \cos\theta_0  & 0 \\ 
                         0 & 0 & 0 \end{array} \right ],
\end{equation}
with $\cos\theta_0=x/\rho$, $\sin\theta_0=y/\rho$, and $\rho=\sqrt{x^2+y^2}$. It operates on the tensor field $T$ as follows:
\begin{equation}
\label{eq:bssn55}
T^{i_1,i_2,...}_{j_1,j_2,...}(x,y,z)=R^{i_1}_{k_1}R^{i_2}_{k_2}...(R^{-1})^{l_1}_{j_1}(R^{-1})^{l_2}_{j_2}...T^{k_1,k_2,...}_{l_1,l_2,...}(\rho,0,z).
\end{equation}
This tranformation law applies to partial derivatives of tensors in the same way.
A 1-dimensional interpolation will be needed to calculate fields at the
points $(\rho,0,z)$ that are not part of the original 3-dimensional coordinate grid.
This interpolation may require points in the $x<0$ range, where the values
are calculated via the same coordinate transformation Eq.~\eqref{eq:bssn53} as shown by
the dashed lines in Fig. 3.1.

   \chapter{Stellar perturbations and phase transitions}

\section{The TOV spacetime}

The TOV (Tolman-Oppenheimer-Volkoff) spacetime is an exact solution in general relativity 
modelling fluid balls such as isolated stars. It is the only known general relativistic 
static equilibrium configuration for spherical objects with matter. The line element of this spacetime is:
\begin{equation}
\label{eq:tov1} 
ds^2=-e^\nu dt^2+e^\lambda dr^2+r^2(d\theta^2+\sin^2\theta d\phi^2),
\end{equation}
which contains no non-diagonal components. 
$r$ in this line element represents the coordinate radius of the fluid ball in question, 
which we will henceforth call the Schwarzschild radius, whereas 
$\nu$ and $\lambda$ are purely functions of $r$, indicating the time-independence of this spacetime.
This line element is used to model neutron stars in this thesis. 
The connection coefficients, Ricci tensor and scalar that result from this line element are as follows:
\begin{eqnarray}
\label{eq:tov2}
G^0_0=R^0_0-\frac{1}{2}g^0_0 R=-\frac{1}{r^2}+e^{-\lambda}(\frac{1}{r^2}-\frac{\lambda'}{r}) \nonumber \\
G^1_1=R^1_1-\frac{1}{2}g^1_1 R=-\frac{1}{r^2}+e^{-\lambda}(\frac{1}{r^2}+\frac{\nu'}{r}) \nonumber \\
G^2_2=G^3_3=R^2_2-\frac{1}{2}g^2_2 R=\frac{1}{4}e^{-\lambda}((\nu')^2-\nu'\lambda'+2\nu''+2(\frac{\nu'}{r}-\frac{\lambda'}{r})),
\end{eqnarray}
with all other components of the Einstein tensor $G^i_j$ being zero. 

Within the interior of the neutron star, the momentum-energy tensor $T^i_j$ is given by:
\begin{equation}
\label{eq:tov3}
T^i_j=(\rho_e+p)u^i u_j+p\delta^i_j=diag(-\rho_e,p,p,p),
\end{equation}
where $\rho_e$ here is the matter-energy density. 
Using GRTensorII on Mathematica, the Einstein field equations are computed to be as follows:\\
$G^0_0$:
\begin{eqnarray}
\label{eq:tov4}
\frac{1}{r^2}-e^{-\lambda}(\frac{1}{r^2}-\frac{\lambda'}{r})=8\pi\rho_e \nonumber \\
\lambda'=\frac{1-e^{\lambda}}{r}+8\pi re^{\lambda}\rho_e
\end{eqnarray}
$G^1_1$:
\begin{eqnarray}
\label{eq:tov5}
\frac{1}{r^2}-e^{-\lambda}(\frac{1}{r^2}+\frac{\nu'}{r})=-8\pi p \nonumber \\
\nu'=-\lambda'+8\pi re^{\lambda}(\rho_e+p)
\end{eqnarray}
$G^2_2=G^3_3$:
\begin{equation}
\label{eq:tovG22}
-(\nu'-\lambda')e^{-\lambda}(\frac{1}{2r}+\frac{\nu'}{4})-\frac{1}{2}e^{-\lambda}\nu''=-8\pi p.
\end{equation}
Eq.~\eqref{eq:tov4} can be solved for the function $\lambda$ when it is written as an integral as follows:
\begin{eqnarray}
\label{eq:tov6}
(re^{-\lambda})' & = & 1-8\pi\rho_e r^2 \nonumber \\
e^{-\lambda} & = & 1-\frac{2m(r)}{r},
\end{eqnarray}
where 
\begin{equation}
\label{eq:tov7}
m(r)\equiv 4\pi\int^r_0\rho_e(r')r'^2 dr'.
\end{equation}
Employing Eq.s~\eqref{eq:tov4} and ~\eqref{eq:tov6} in Eq.~\eqref{eq:tov5} gives us:
\begin{equation}
\label{eq:tov8}
\nu'=\frac{2m+8\pi r^3 p}{r(r-2m)}.
\end{equation}
We then use the former two equations together with the 
derivative of Eq.~\eqref{eq:tov5} to write Eq.~\eqref{eq:tovG22} entirely in terms of $\nu'$ and $p'$ as follows:
\begin{eqnarray}
\label{eq:tov9}
(-2\nu'+8\pi re^{\lambda}(\rho_e+p))e^{-\lambda}(\frac{1}{2r}+\frac{\nu'}{4})- \nonumber \\
\frac{e^{-\lambda}}{2r}[(r\nu'+1)\lambda'+(16\pi rp+8\pi r^2 p')e^{\lambda}-\nu']=-8\pi p \nonumber \\
p'=-\frac{1}{2}(\rho_e+p)\nu'.
\end{eqnarray}
Eq.s~\eqref{eq:tov7},~\eqref{eq:tov8} and ~\eqref{eq:tov9} form a set of ordinary differential equations that
can be solved to yield our theoretical stellar models. In our simulations, we choose a value for $\rho_e(r=0)$ with $m(r=0)=0$, 
employ an equation of state giving $p$ as a function of $\rho_e$, and integrate the former equations
from $r=0$ to a radius where the pressure $p$ vanishes. This radius defines the surface of the neutron star, which we will denote henceforth as $R$. 
We employ an equation of state popularly used in neutron star simulations
called the polytropic equation of state coupled with a $\Gamma$ law 
which sets $p=\kappa\rho^{\Gamma}=\rho_e\epsilon(\Gamma-1)$ for the initial data with $\kappa$ and $\Gamma$ as arbitrary constants. 
These solutions give us models of the interior of the neutron stars.

For the exterior region of the neutron star, $p=0$ and $m=M$ where $M$ is the gravitational mass of the star as evaluated at spatial infinity, 
which we will denote henceforth as the ADM (Arnowitt-Deser-Misner) mass.
Therefore, Eq.~\eqref{eq:tov6} becomes:
\begin{equation}
\label{eq:tov10}
e^{-\lambda}=1-\frac{2M}{r},
\end{equation}
and Eq.~\eqref{eq:tov8} yields:
\begin{equation}
\label{eq:tov11}
\nu'=\frac{2M}{r(r-2M)}.
\end{equation}
The line element for the exterior region thus becomes:
\begin{equation}
\label{eq:tov12}
ds^2=-(1-\frac{2M}{r})dt^2+(1-\frac{2M}{r})^{-1}dr^2+r^2(d\theta^2+\sin^2\theta d\phi^2),
\end{equation}
which is the standard Schwarzschild line element.

Since we perform these calculations within a 3-dimensional Cartesian grid framework, 
we use an alternative form of the Schwarzschild line element as follows:
\begin{eqnarray}
\label{eq:tov13}
ds^2 & = & -(1-\frac{2M}{r})dt^2+(1-\frac{2M}{r})^{-1}dr^2+r^2(d\theta^2+\sin^2\theta d\phi^2) \nonumber \\
& = & -Adt^2+B(d{r_{iso}}^2+{r_{iso}}^2(d\theta^2+\sin^2\theta d\phi^2) \nonumber \\
& = & -Adt^2+B(dx^2+dy^2+dz^2),
\end{eqnarray}
where it can be computed that:
\begin{eqnarray} 
\label{eq:tov14}
A & = & (\frac{2r_{iso}-M}{2r_{iso}+M})^2 \nonumber \\
B & = & (1+\frac{M}{2r_{iso}})^4,
\end{eqnarray}
and:
\begin{eqnarray}
\label{eq:tov15}
r & = & r_{iso}(1+\frac{M}{2r_{iso}})^2 \nonumber \\
& = & \frac{r_{iso}(2r_{iso}+M)^2}{(2r_{iso})^2} \nonumber \\
dr & = & \frac{\sqrt{r(r-2M)}}{r_{iso}}dr_{iso}.
\end{eqnarray}
Due to Eq.~\eqref{eq:tov10}, the differential for the Schwarzschild radius in Eq.~\eqref{eq:tov15} for the stellar interior
can be written analogously as:
\begin{equation}
\label{eq:tov16}
dr=\frac{\sqrt{r(r-2m)}}{r_{iso}}dr_{iso}.
\end{equation}
The set of ordinary differential equations for the stellar interior thus becomes:
\begin{eqnarray}
\label{eq:tov17}
\frac{dm}{dr_{iso}}=\frac{4\pi\rho_e r^2\sqrt{r(r-2m)}}{r_{iso}} \nonumber \\
\frac{d\nu}{dr_{iso}}=\frac{2(m+4\pi pr^3)}{\sqrt{r(r-2m)}r_{iso}} \nonumber \\
\frac{dp}{dr_{iso}}=-(\rho_e+p)\frac{(m+4\pi pr^3)}{\sqrt{r(r-2m)}r_{iso}}.
\end{eqnarray} 
The solution in the stellar interior is matched to the solution in the exterior to obtain a global solution. 
These solutions were first considered by Tolman, Oppenheimer and Volkoff in 1939 \cite{Tolman39},\cite{Oppenheimer39} to model isolated stars, 
henceforth the name TOV spacetime.

\section{Perturbations on the TOV spacetime}

In this section, we present small amplitude perturbations to the TOV spacetime 
as first introduced by Thorne and Campolattaro in 1967 \cite{Thorne67}.
These perturbations are viewed as vector displacements of the fluid elements 
with respect to the coordinate system of the spacetime. In terms of the line element,
the perturbation is introduced as follows:
\begin{equation}
\label{eq:pert1}
d{s'}^2=ds^2+h_{\mu\nu}dx^{\mu}dx^{\nu},
\end{equation}
where $d{s'}^2$ is the perturbed line element, $ds^2$ is the original line element of the TOV spacetime
as described in Eq.s~\eqref{eq:tov1} and \eqref{eq:tov12} and $h_{\mu\nu}$ are the components
of the perturbation metric. The components of the perturbation metric and the fluid element displacement vector   
can be categorized into quantities that transform as scalar fields, vectors or tensors under a rotational group.
The quantities that transform as scalar fields are constants under a rotation group. They 
can be expanded in terms of scalar spherical harmonics $Y^l_m(\theta,\phi)$
and possess even parity. Those that transform as vectors, eg. $A^{\mu}=\frac{\partial x^{\mu}}{\partial x^{\nu}}A^{\nu}$,
are expanded in terms of vector spherical harmonics $\Psi^l_{mj}=\partial_j Y^l_m(\theta,\phi)$ 
in the even parity and $\Phi^l_{mj}=\epsilon^k_j\partial_k Y^l_m(\theta,\phi)$ in the odd parity 
where $\epsilon^3_2=-\frac{1}{\sin\theta}$, $\epsilon^2_3=\sin\theta$ and $\epsilon^2_2=\epsilon^3_3=0$, 
whereas those that transform as tensors, eg. $A^{\mu\nu}=\frac{\partial x^{\mu}}{\partial x^m}\frac{\partial x^{\nu}}{\partial x^n}A^{mn}$, 
are expanded in terms of tensor spherical harmonics $\Psi^l_{mjk}=Y^l_{m;jk}$,
a covariant derivative with respect to the 3-metric $\gamma_{ij}$ for the TOV spacetime,
$\Phi^l_{mjk}=\gamma_{jk}Y^l_m$ in the even parity and $\xi^l_{mjk}=\frac{1}{2}(\epsilon^n_j\Psi^l_{mnk}+\epsilon^n_k\Psi^l_{mnj})$ 
in the odd parity. $\xi_r$, $h_{00}$, $h_{0r}$ and $h_{rr}$ are quantities that transform as scalars; whereas
$(\xi_{\theta},\xi_{\phi})$, $(h_{0\theta},h_{0\phi})$, $(h_{r\theta},h_{r\phi})$ transform as vectors, and 
$h_{jk}$ with $(j,k=\theta,\phi)$ transform as tensors. Therefore, we can write the overall odd-parity harmonics as follows:
\begin{equation*}
\xi_r=0, \xi_{\theta}=U(r,t)\Phi^l_{m2}, \xi_{\phi}=U(r,t)\Phi^l_{m3} 
\end{equation*}
\begin{equation*}
h_{00}=h_{0r}=h_{rr}=0 
\end{equation*}
\begin{equation}
\label{eq:pert2}
h_{0j}=h_0(r,t)\Phi^l_{mj}, h_{1j}=h_1(r,t)\Phi^l_{mj}, h_{jk}=h_2(r,t)\chi^l_{mjk}.
\end{equation}
The overall even-parity harmonics can be written as follows:
\begin{equation*}
\xi_r=X(r,t)Y^l_m, \xi_{\theta}=V(r,t)\Psi^l_{m2}, \xi_{\phi}=V(r,t)\Psi^l_{m3} 
\end{equation*}
\begin{equation*}
h_{00}=e^{\nu}H_0(r,t)Y^l_m, h_{0r}=H_1(r,t)Y^l_m, h_{rr}=e^{\nu}H_2(r,t)Y^l_m 
\end{equation*}
\begin{equation}
\label{eq:pert3}
h_{0j}=h_0(r,t)\Psi^l_{mj}, h_{1j}=h_1(r,t)\Psi^l_{mj}, h_{jk}=r^2 K(r,t)\Psi^l_{mjk}+r^2 G(r,t)\Phi^l_{mjk}.
\end{equation}
Eq.s~\eqref{eq:pert2} and ~\eqref{eq:pert3} can be simplified via small coordinate tranformations 
$x^{\mu'}=x^{\mu}+\eta^{\nu}(x)$ on the perturbation metric components as follows:
\begin{equation}
\label{eq:pert4}
h'_{\mu\nu}=h_{\mu\nu}+\eta_{\mu;\nu}+\eta_{\nu;\mu},
\end{equation}
where $\eta_0=\eta_r=0$ and $\eta_j=\Lambda(r,t)\Phi^l_{mj}$ for the odd-parity harmonics, and
$\eta_0=M_0(r,t)Y^l_m$, $\eta_1=M(r,t)Y^l_m$ and $\eta_j=M_2(r,t)\Psi^l_{mj}$ for the even-parity harmonics.
For the odd-parity harmonics, the simplification is performed by setting $\Lambda$ so as to annul 
the function $h_2$, whereas for the even-parity harmonics, $M_0$, $M_1$ and $M_2$ are chosen so as 
to annul $h_0$, $h_1$ and $G$. With these gauge simplications, the perturbation metric 
for the odd-parity harmonic with $m=0$ becomes:
\begin{equation}
\label{eq:pert5}
h_{\mu\nu}=\begin{bmatrix}  
0 & 0 & 0 & h_0\sin\theta Y^l_{0,2} \\
0 & 0 & 0 & h_1\sin\theta Y^l_{0,2} \\
0 & 0 & 0 & 0 \\
h_0\sin\theta Y^l_{0,2} & h_1\sin\theta Y^l_{0,2} & 0 & 0
\end{bmatrix},   
\end{equation}
with the corresponding fluid element displacement vector as follows:
\begin{equation}
\label{eq:pert6}
\xi_{\alpha}=\begin{bmatrix}
0 \\
0  \\
U\sin\theta Y^l_{0,2} 
\end{bmatrix},
\alpha=(1,2,3).
\end{equation}
The perturbation metric for the even-parity harmonics for $m=0$ becomes:
\begin{equation}
\label{eq:pert7}
h_{\mu\nu}=\begin{bmatrix}
H_0 e^{\nu} & H_1 & 0 & 0 \\
H_1 & H_2 e^{\lambda} & 0 & 0 \\
0 & 0 & r^2 K & 0 \\
0 & 0 & 0 & r^2\sin^2\theta K 
\end{bmatrix} 
Y^l_0,
\end{equation}
with the corresponding fluid element displacement vector as follows:
\begin{equation}
\label{eq:pert8}
\xi_{\alpha}=\begin{bmatrix}
XY^l_0 \\
V\Psi^l_{02} \\
0
\end{bmatrix}.
\end{equation}
These general forms indicate that the odd-parity harmonics are described by
differential rotations that do not change the profile of the star's density, pressure and shape,
whereas the even-parity harmonics are described by oscillations of the star's density and pressure.

Armed with the previous general forms, we now consider the most basic perturbation mode, ie. the $l=0$ mode. 
The odd-parity harmonics for this mode are:
\begin{equation}
\label{eq:pert9}
\Phi^0_{0j}=0; \chi^0_{0jk}=0,
\end{equation}
and the even-parity harmonics are:
\begin{equation*}
\xi_r=XY^0_0; \Psi^0_{0j}=0; \Psi^0_{0jk}=0
\end{equation*}
\begin{equation}
\label{eq:pert10} 
\Phi^0_{0jk}=\gamma_{jk}Y^0_0.
\end{equation}
Eq.~\eqref{eq:pert9} indicates that the $l=0$ mode contains no odd-parity harmonics, 
and thus does not possess any differential rotations.
However, the perturbation metric for even-parity harmonics for the $l=0$ mode becomes:
\begin{equation}
\label{eq:pert11}
h_{\mu\nu}=\begin{bmatrix}
H_0 e^{\nu} & H_1 & 0 & 0 \\
H_1 & H_2 e^{\lambda} & 0 & 0 \\
0 & 0 & 0 & 0 \\
0 & 0 & 0 & 0
\end{bmatrix}
Y^0_0,
\end{equation}
and the fluid element displacement vector becomes:
\begin{equation}
\label{eq:pert12}
\xi^{\alpha}=\begin{bmatrix}
-r^2 e^{-\alpha/2}XY^0_0 \\
0 \\
0 
\end{bmatrix}.
\end{equation}  
We can thus write the perturbed line element for the $l=0,m=0$ mode as follows:
\begin{equation}
\label{eq:pert13}
ds^2=e^{\nu}(H_0Y^0_0-1)dt^2+2H_1 Y^0_0 dtdr+e^{\lambda}(H_2 Y^0_0+1)dr^2+r^2(d\theta^2+\sin^2\theta d\phi^2).
\end{equation}
The 4-velocity of the fluid elements corresponding to the aforementioned displacement vector is:
\begin{equation}
\label{eq:pert14}
u^{\beta}=\begin{bmatrix}
e^{-\nu/2}(1+\frac{H_0 Y^0_0}{2}) \\   
-r^{-2}e^{-(\nu+\lambda)/2}X_{,0}Y^0_0 \\
0 \\
0
\end{bmatrix},
\beta=(0,1,2,3).
\end{equation}
From the displacement vector, we also compute the Lagrangian change in the number density of baryons 
contained in a certain stellar volume, as follows:
\begin{eqnarray}
\label{eq:pert15}
\frac{\Delta n}{n} & = & -(\xi^k_{;k}+\frac{1}{2}\frac{\delta[{}^{(3)}g]}{{}^{(3)}g} \nonumber \\
& = & r^{-2}e^{-\lambda/2}X_{,1}Y^0_0-\frac{H_2 Y^0_0}{2},
\end{eqnarray}
where the subscripted colon indicates a covariant derivative with respect to the perturbed metric
to first order in the perturbation functions. The corresponding Eulerian changes 
in the density and pressure of mass-energy of the star hence become:
\begin{eqnarray}
\label{eq:pert16}
\delta\rho_e & = & (\rho_e+p)\frac{\Delta n}{n}+\rho_{e,1}r^{-2}e^{-\lambda/2}XY^0_0 \nonumber \\
\delta p & = & \gamma p\frac{\Delta n}{n}+p_{,1}r^{-2}e^{-\lambda/2}XY^0_0,
\end{eqnarray}
where $\gamma=\frac{\rho_e+p}{p}\frac{p_{,1}}{\rho_{e,1}}$. We note that Eq.s~\eqref{eq:pert13} to ~\eqref{eq:pert16}
constitute spacetime and fluid perturbations that do not violate the spherical symmetry of the equilibrium TOV background. 
Substituting these expressions into Einstein's field equations and simplifying by neglecting higher orders of the perturbation functions, 
we now write the equations of motion for the $l=0$ mode perturbation as follows:\\
$\delta G^1_1=8\pi\delta T^1_1$:
\begin{eqnarray}
\label{eq:pert17}
2r^{-1}e^{-(\lambda+\nu)}H_{,0}-r^{-1}e^{-\lambda}H_{0,1}-r^{-2}e^{-\lambda}(1+r\nu')H_2- \nonumber \\
8\pi\gamma p(r^{-2}e^{-\lambda/2}X_{,1}-{H_2}/2)-8\pi r^{-2}e^{-\lambda/2}p_{,1}X=0
\end{eqnarray}
$\delta G^0_0=8\pi\delta T^0_0$:
\begin{eqnarray}
\label{eq:pert18}
-r^{-1}e^{-\lambda}H_{2,1}+r^{-2}e^{-\lambda}(-1+r\lambda_{,1})H_2+ \nonumber \\
8\pi(\rho_e+p)(r^{-2}e^{-\lambda/2}X_{,1}-{H_2}/2)+8\pi r^{-2}e^{-\lambda/2}\rho_{e,1}X=0
\end{eqnarray}
$\delta G^2_2=8\pi\delta T^2_2$, $\delta G^3_3=8\pi\delta T^3_3$::
\begin{eqnarray}
\label{eq:pert19}
-e^{-\nu}H_{2,00}/2+e^{-(\lambda+\nu)}H_{1,01}+r^{-1}e^{-\lambda+\nu)}(2-r\lambda_{,1})H_{1,0}/2- \nonumber \\
e^{-\lambda}H_{0,11}/2-r^{-4}e^{-\lambda}(2-r\lambda_{,1}+2r\nu_{,1})H_{0,1}/4-r^{-1}e^{-\lambda}(2+r\nu_{,1})H_{2,1}/4+ \nonumber \\
4r^{-1}e^{-\lambda}((\lambda_{,1}-\nu_{,1})(2-r\nu_{,1})-2r\nu_{,11})H_2- \nonumber \\
8\pi\gamma p(r^{-2}e^{-\lambda/2}X_{,1}-{H_2}/2)-8\pi r^{-2}e^{-\lambda/2}p_{,1}X=0
\end{eqnarray}
$\delta G^1_0=8\pi\delta T^1_0$:
\begin{equation}
\label{eq:pert20}
r^{-1}e^{-\lambda}H_{2,0}-8\pi(\rho_e+p)r^{-2}e^{-\lambda/2}X_{,0}=0
\end{equation}
$\delta G^0_1=8\pi\delta T^0_1$:
\begin{equation}
\label{eq:pert21}
r^{-1}e^{-(\lambda+\nu)}(\lambda_{,1}+\nu_{,1})H_1-r^{-1}e^{-\nu}H_{2,0}+8\pi(\rho_e+p)r^{-2}e^{-\nu+\lambda/2}X_{,0}=0
\end{equation}
The Euler equation for this perturbation mode, which is the projection of $\delta T^{\mu}_{1;\nu}=0$ 
along the direction orthogonal to the 4-velocity of the fluid $\mathbf{u}$,
is as follows:
\begin{equation}
\label{eq:pert22}
(\rho_e+p)\xi_{,00}e^{\lambda+\nu}-\frac{(\rho_e+p)}{2}H_{0,1}Y^0_0+(\delta\rho_e+\delta p)\frac{\nu_{,1}}{2}=-\delta p_{,1}.
\end{equation}
Eq.s~\eqref{eq:pert17} to ~\eqref{eq:pert22} can be solved to yield a first order ordinary differential equation for the perturbation function $X$.
To do this, we write the perturbation function in terms of a radial dependence and a harmonic time dependence, $X(r,t)=X(r)e^{i\omega t}$.
We then substitute Eq.s~\eqref{eq:pert16},~\eqref{eq:pert17} and ~\eqref{eq:pert20} into ~\eqref{eq:pert22} to obtain the following:
\begin{eqnarray}
\label{eq:pert23}
(\frac{p_{,1}}{\rho_{e,1}})(r^{-2}e^{-\lambda/2}X_{,11}+((\frac{p_{,1}}{\rho_{e,1}})_{,1}-Z+4\pi r\gamma pe^{\lambda}-\frac{\nu_{,1}}{2})X_{,1}+ \nonumber \\
(\frac{(\nu_{,1})^2}{2}+\frac{2m}{r^3}e^{\lambda}-Z_{,1}-4\pi(\rho_e+p)Zre^{\lambda}+\omega^2 e^{\lambda-\nu})X=0.
\end{eqnarray}

We now consider the $l=1$ mode perturbations. The odd-parity harmonics for this mode are:
\begin{equation*}
\Phi^1_{02}=0; \Phi^1_{\pm 12}=i(\frac{3}{8\pi}^{1/2})e^{\pm i\phi}
\end{equation*}
\begin{equation*}
\Phi^1_{03}=-(\frac{3}{4\pi})^{1/2}\sin^2\theta; \Phi^1_{\pm 13}=\mp(\frac{3}{8\pi})^{1/2}e^{\pm i\phi}\sin\theta\cos\theta
\end{equation*}
\begin{equation}
\label{eq:pert24}
\chi^1_{mjk}=0, j,k=2,3,
\end{equation}
whereas the even-parity harmonics are:
\begin{equation*}
Y^1_0=(\frac{3}{4\pi})^{1/2}\cos\theta; Y^1_{\pm 1}=\mp(\frac{3}{8\pi}^{1/2})e^{\pm\phi}\sin\theta
\end{equation*}
\begin{equation*}
\Psi^1_{02}=-(\frac{3}{4\pi})^{1/2}\sin\theta; \Psi^1_{\pm 12}=\mp(\frac{3}{8\pi})^{1/2}e^{\pm i\phi}\cos\theta
\end{equation*}
\begin{equation*}
\Psi^1_{03}=0; \Psi^1_{\pm 13}=-i(\frac{3}{8\pi})^{1/2}e^{\pm i\phi}\sin\theta
\end{equation*}
\begin{equation}
\label{eq:pert25}
\Phi^1_{m22}=-\Psi^1_{m22}=Y^1_m; \Phi^1_{m33}=-\Psi^1_{m33}=\sin^2\theta Y^1_m; \Phi_m{}^1_{23}=-\Psi^1_{23}=0.
\end{equation}
In the spirit of Eq.~\eqref{eq:pert2} and performing a simplification via small coordinate transformations previously done for the $l=0$ mode,
which sets $G-K\equiv 0$, we can then write the metric for the odd-parity $l=1$ perturbation mode as follows:
\begin{equation}
\label{eq:pert26}
h_{\mu\nu}=\begin{bmatrix}
H_0 e^{\nu} & i\omega H_1 & 0 & 0 \\
i\omega H_1 & H_2 e^{\lambda} & 0 & 0 \\
0 & 0 & 0 & 0 \\
0 & 0 & 0 & 0 
\end{bmatrix}
Y^1_m,
\end{equation}
where we include a harmonic term in the $10-$ and $01-$ components in order to simplify the resulting system of ordinary differential equations
shown further on. Correspondingly, the fluid element displacement vector is:
\begin{equation}
\label{eq:pert27}
\xi^{\alpha}=\begin{bmatrix}
-r^{-2}e^{-\lambda/2}WY^1_m \\   
\frac{V}{r^2}\Psi^1_m{}^2 \\
\frac{V}{r^2}\Psi^1_m{}^3 
\end{bmatrix},
\end{equation}
where we see that $W$ is the perturbation function describing radial fluid oscillations whereas $V$ describes azimuthal fluid displacements.
The perturbed line element for this perturbation mode is thus:
\begin{equation}
\label{eq:pert28}
ds^2=e^{\nu}(H_0 Y^1_m-1)dt^2+2i\omega H_1 Y^1_m dtdr+e^{\lambda}(H_2 Y^1_m+1)dr^2+r^2(d\theta^2+\sin^2\theta d\phi^2).
\end{equation}
With this line element, the Lagrangian change in the baryonic number density becomes:
\begin{equation}
\label{eq:pert29}
\frac{\Delta n}{n}=(r^{-2}e^{-\lambda/2}W_{,1}+2r^{-2}V-\frac{H_2}{2})Y^1_m,
\end{equation}
which results in the Eulerian changes in the pressure and matter-energy density as follows:
\begin{equation*}
\delta\rho_e=(\frac{(\rho_e+p)}{\gamma p}e^{-\nu/2}S_1 Y^1_m+\rho_{e,1}r^{-2}e^{-\lambda/2}WY^1_m
\end{equation*}
\begin{equation}
\label{eq:pert29a}
\delta p=e^{-\nu/2}S_1 Y^1_m+p_{,1}r^{-2}e^{-\lambda/2}WY^1_m,
\end{equation}
where $S_1=\gamma pe^{\nu/2}(r^{-2}e^{-\lambda/2}W_{,1}+2r^{-2}V-\frac{H_2}{2})$. 
To first order in the perturbation functions, again using GRTensorII on Mathematica, we obtain the Einstein field equations as follows:\\
$\delta G^1_1=8\pi\delta T^1_1$:
\begin{eqnarray}
\label{eq:pert30}
(2r^{-1}e^{-\lambda-\nu}H_{1,0}-r^{-1}e^{-\lambda}H_{0,1}-r^{-2}e^{-\lambda}H_2(1+r\nu_{,1})+H_0 r^{-2})Y^1_m- \nonumber \\
8\pi e^{-\nu/2}S_1-8\pi p_{,1}r^{-2}e^{-\lambda/2}WY^1_m=0
\end{eqnarray}
$\delta G^2_2=8\pi\delta T^2_2$:
\begin{eqnarray}
\label{eq:pert31}
[-\frac{r^{-1}e^{-\lambda-\nu}}{2}(-2+r\lambda_{,1})H_{1,0}-\frac{e^{-\nu}}{2}H_{2,00}+\frac{r^{-1}e^{-\lambda}}{4}(-2+r\lambda_{,1}-2r\nu_{,1})H_{0,1}- \nonumber \\
\frac{r^{-1}e^{-\lambda}}{4}(2+r\nu_{,1})H_{2,1}+e^{-\lambda-\nu}H_{1,01}-\frac{e^{-\lambda}}{2}H_{0,11}+ \nonumber \\
\frac{r^{-2}H_2}{4}(e^{-\lambda}r((\lambda_{,1}-\nu_{,1})(2+r\nu_{,1})-2r\nu_{,11})-2)+\frac{r^{-2}H_0}{2}]Y^1_m- \nonumber \\
8\pi e^{-\nu/2}S_1-8\pi p_{,1}r^{-2}e^{-\lambda/2}WY^1_m=0
\end{eqnarray}
$\delta G^0_0=8\pi\delta T^0_0$:
\begin{eqnarray}
\label{eq:pert32}
[-r^{-1}e^{-\lambda}H_{2,1}+\frac{r^{-2}e^{-\lambda}H_2}{2}(2(-1+r\lambda_{,1})-2e^{\lambda})]Y^1_m+ \nonumber \\
8\pi\frac{(\rho_e+p)}{\gamma p}e^{-\nu/2}S_1 Y^1_m+8\pi\rho_{e,1}r^{-2}e^{-\lambda/2}WY^1_m=0
\end{eqnarray}
$\delta G^1_0=8\pi\delta T^1_0$:
\begin{equation}
\label{eq:pert33}
r^{-1}e^{-\lambda}H_{2,1}+r^{-2}e^{-\lambda}H_1-8\pi(\rho_e+p)r^{-2}e^{-\lambda/2}W_{,0}=0
\end{equation}
$\delta G^j_0=8\pi\delta T^j_0$:
\begin{equation}
\label{eq:pert34}
-\frac{r^{-2}}{2}(H_{2,0}-e^{-\lambda}H_{1,1}+\frac{e^{-\lambda}}{2}(\lambda_{,1}-\nu_{,1})H_1-8\pi(\rho_e+p)r^{-2}V_{,0}=0
\end{equation}
$\delta G^j_1=8\pi\delta T^j_1$:
\begin{equation}
\label{eq:pert35}
H_{1,0}=e^{\nu}H_{0,1}-e^{\nu}(r^{-1}-\frac{\nu_{,1}}{2})H_0+e^{\nu}(r^{-1}+\frac{\nu_{,1}}{2})H_2.
\end{equation}
We write the perturbation functions in terms of a radial dependence and a harmonic time dependence, ie. $H_0(r,t)=H_0(r)e^{i\omega t}$,
$H_1(r,t)=H_1(r)e^{i\omega t}$, $H_2(r,t)=H_2(r)e^{i\omega t}$, $V(r,t)=V(r)e^{i\omega t}$ and $W(r,t)=W(r)e^{i\omega t}$.
We then decouple Eq.~\eqref{eq:pert32} into two equations for $H_1$ and $H_2$ as follows:
\begin{equation}
\label{eq:pert36}
H_2=-r^{-1}H_1+r^{-1}8\pi(\rho_e+p)e^{\lambda/2}W
\end{equation}
\begin{equation}
\label{eq:pert37}
H_{1,1}=r^{-1}(\frac{r}{2}(\lambda_{,1}-\nu{,1})-e^{\lambda})H_1+r^{-1}8\pi(\rho_e+p)e^{3\lambda/2}W-16\pi(\rho_e+p)e^{\lambda}V.
\end{equation}
Eq.~\eqref{eq:pert35} however can be simplified to yield:
\begin{equation}
\label{eq:pert38}
rH_{0,1}=-(\frac{r\nu_{,1}}{2}-1)H_0-(\frac{r\nu_{,1}}{2}+1)H_2+i\omega re^{-\nu}H_2.
\end{equation} 
We also consider the perturbation of the energy-momentum conservation as follows:\\
$\delta(T^{\nu}_{j;\nu})=0$
\begin{eqnarray}
\label{eq:pert39}
-(\rho_e+p)e^{-\nu}\omega^2 V & = & -p_{,1}r^{-2}e^{-\lambda/2}W-p\gamma r^{-2}e^{-\lambda/2}W_{,1}+ \nonumber \\
& & \frac{\gamma p}{2}H_2+\frac{(\rho_e+p)}{2}H_0-2\gamma p r^{-2}V,
\end{eqnarray}
which can be decoupled into two equations for $V$ and $W$ as follows:
\begin{eqnarray}
\label{eq:pert40}
16\pi r^2\omega^2(\rho_e+p)e^{\lambda-\nu}V & = & -8\pi(\rho_e+p)\nu_{,1}e^{\nu/2}W+(\nu_{,1}-2r\omega^2 e^{-\nu})H_1+ \nonumber \\
& & (16\pi r^2\rho_e e^{\lambda}-3r\lambda_{,1})H_0
\end{eqnarray}
\begin{eqnarray}
\label{eq:pert41}
8\pi p\gamma e^{\lambda/2}W_{,1} & = & -r\omega^2 e^{-\nu}H_1-(\frac{r\nu_{,1}}{2}-4\pi p\gamma r^2 e^{\lambda})H_2- \nonumber \\
& & (1-e^{\lambda}-\frac{r\nu_{,1}}{2})H_0-8\pi p_{,1}e^{\lambda/2}W-16\pi p\gamma e^{\lambda}V.
\end{eqnarray}
Eq.s~\eqref{eq:pert36},~\eqref{eq:pert37},~\eqref{eq:pert38},~\eqref{eq:pert40} and ~\eqref{eq:pert41} form a system of third order ordinary differential equations
for the even parity $l=1$ perturbation mode and can be solved to obtain the mode frequency $\omega$.
To consider the effect of this perturbation on the exterior TOV background spacetime, we solve Eq.s~\eqref{eq:pert32},~\eqref{eq:pert35} and ~\eqref{eq:pert39}
in the vacuum surrounding the star. Using Eq.s~\eqref{eq:tov10} to ~\eqref{eq:tov12}, we obtain, as in \cite{Campolattaro70}:
\begin{equation}
\label{eq:pertSS}
H_0=\frac{1}{3}\frac{\varsigma^3\beta+8M^2\beta_{,00}}{\varsigma(1-\varsigma)^2}; H_1=-\frac{2M\varsigma}{(1-\varsigma)^2}\beta_{,0}; 
H_2={\varsigma^2}{(1-\varsigma)^2}\beta,
\end{equation}
where $\varsigma\equiv\frac{2M}{r}$ and $\beta=$arbitrary function of the coordinate time. Considering infinitesimal coordinate transformations, we use
Eq.~\eqref{eq:pert4} to obtain the following for the even-parity $l=1$ mode:
\begin{eqnarray}
\label{eq:pertSS1}
h_{0'0'} & = & [e^{\nu}H_0-2M_{0,0}+\nu_{,1}e^{\nu-\lambda}M_1]Y^l_m \nonumber \\
h_{0'1'} & = & [H_1-M_{0,1}+\nu_{,1}M_0-M_{1,0}]Y^l_m \nonumber \\
h_{1'1'} & = & [e^{\nu}H_2-2M_{1,1}+\lambda_{,1}M_1]Y^l_m \nonumber \\
h_{0'j'} & = & -[M_{2,0}+M_0]\Psi^l_{mj} \nonumber \\
h_{1'j'} & = & -[M_{2,1}-2r^{-1}M_2+M_1]\Psi^l_{mj} \nonumber \\
h_{j'k'} & = & -2[re^{-\lambda}M_1+M_2]\Psi^l_{mjk}.
\end{eqnarray}
Eq.s~\eqref{eq:pertSS1} show that for the even-parity $l=1$ mode perturbation metric to preserve the same form, the following will have to hold true:
\begin{equation}
\label{eq:pertSS2}
h_{0'j'}=h_{1'j'}=h_{j'k'}=0,
\end{equation}
which entails:
\begin{equation}
\label{eq:pertSS3}
M_{2,0}+M_0=0, M_{2,1}-2r^{-1}M_2+M_1=0, 2re^{-\lambda}M_1-2M_2=0.
\end{equation}
Eq.s~\eqref{eq:pertSS3} can be solved to yield the following:
\begin{equation}
\label{eq:pertSS4}
M_0=-a_{,0}f, M_1=ar^{-1}e^{\lambda}f, M_2=af,
\end{equation}
with $f=r\exp[\int^{\infty}_r r^{-1}(1-e^{\lambda})dr]$ and $a=$arbitrary function of the coordinate time. Substituting Eq.~\eqref{eq:pertSS4}
back into the first three equations in ~\eqref{eq:pertSS1}, we obtain:
\begin{eqnarray}
\label{eq:pertSS5}
H'_0 & = & H_0+(2a_{,00}e^{-\nu}+ar^{-1}\nu_{,1})f \nonumber \\
H'_1 & = & H_1+a_{,0}[2r^{-1}(1-e^{\lambda})-\nu_{,1}]f \nonumber \\
H'_2 & = & H_2-ar^{-1}[2r^{-1}(1-e^{\lambda})+\lambda_{,1}]f \nonumber \\ 
W' & = & W-are^{\lambda/2}f \nonumber \\
V' & = & V+af \nonumber \\
S' & = & S.
\end{eqnarray}
Therefore, the perturbation functions $H_0$, $H_1$, $H_2$, $V$ and $W$ are only unique up to the transformations shown in Eq.~\eqref{eq:pertSS5}.
Given Eq.s~\eqref{eq:pertSS5}, the perturbations in the exterior background spacetime Eq.s~\eqref{eq:pertSS} can be set to zero which then preserves
its spherical symmetry. When $H'_0=0$, we obtain $a(t)=-\frac{2}{3}M\beta{t}$. Using this gauge not only guarantees
that the exterior background spacetime is spherically symmetric but as a unique gauge, 
it also can be used to remove the gauge arbitrariness in the former perturbation functions.

\section{Non-radiative pulsations of neutron stars and their frequency modes}

In the previous section, we have delved into non-radiative perturbations on the TOV spacetime, ie. the $l=0$ and even parity $l=1$ modes.
In this section, we will apply the formalism in our neutron star models to obtain their non-radial pulsation modes. 
By considering boundary conditions at the center and at the surface of the star, we employ a shooting method 
to solve the systems of ordinary differential equations in both the $l=0$ and even parity $l=1$ modes obtained in the previous section.

For the $l=0$ mode, we follow Misner \textit{et~al} \cite{MTW} as well as Kokkotas and Ruoff \cite{Kokkotas01} in re-writing Eq.~\eqref{eq:pert23} as:
\begin{equation}
\label{eq:pert42}
\frac{d}{dr}(P\frac{d\zeta}{dr})+(Q+\omega^2 W)\zeta=0,
\end{equation}
where:
\begin{equation}
\label{eq:pert43}
\zeta=r^2 e^{-\nu}X,
\end{equation}
\begin{equation*}
r^2 W=(\rho_e+p)e^{(3\lambda+\nu)/2}
\end{equation*}
\begin{equation*}
r^2 P=\gamma pe^{(\lambda+3\nu)/2}
\end{equation*}
\begin{equation}
\label{eq:pert44}
r^2 Q=e^{(\lambda+3\nu)/2}(\rho_e+p)(\frac{(\nu_{,1})^2}{4}+2\frac{\nu_{,1}}{r}-8\pi e^{2\lambda}p).
\end{equation}
Eq.~\eqref{eq:pert42} can itself be decoupled into a system of two ordinary differential equations as follows:
\begin{equation*}
\frac{d\zeta}{dr}=\frac{\eta}{P}
\end{equation*}
\begin{equation}
\label{eq:pert45}
\frac{d\eta}{dr}=-(\omega^2 W+Q)\zeta.
\end{equation}
We now consider the boundary conditions for neutron star models. At the center of the star, the radial displacement
of the fluid element vanishes, whilst at the surface of the star, the Eulerian change in the pressure of the star vanishes.
These conditions translate respectively as:
\begin{equation}
\label{eq:pert46}
\zeta(r=0)=0,
\end{equation}
\begin{equation}
\label{eq:pert47}
(r^{-2}e^{-\lambda/2}W)_{,1}=(\gamma R)^{-1}(4+(\frac{M}{R})e^{\lambda}W+\omega^2(\frac{R^3}{M})e^{-\nu})(r^{-2}e^{-\lambda/2}W),
\end{equation}
where $R$ denotes the radius and $M$ the ADM mass of the star \cite{Bardeen66}. \cite{Kokkotas01} found that via a Taylor expansion, $\zeta(r)\sim\zeta_0 r^3+\mathcal{O}(r^5)$, 
and $\eta(r)\sim\eta_0+\mathcal{O}(r^2)$, hence we obtain $\zeta_{,1}\sim 0$ and $\eta_{,1}\sim 0$ as $r\rightarrow 0$. 
We then employ a simple shoot and match method to solve the boundary value problem Eq.s~\eqref{eq:pert45} to ~\eqref{eq:pert47} to obtain the
normal modes of pulsation for the neutron star. We input a range of test $\omega^2$ values and for each of these values
integrate Eq.~\eqref{eq:pert45} from $r=0$ to $r=R$ using a simple finite-differencing scheme and 
observe the difference between the integrated value and the value imposed by the boundary condition 
Eq.~\eqref{eq:pert47} at $r=R$. The $\omega$ values that give us zero difference are the normal mode frequencies for the neutron star. 
We note that using this method, convergence is achieved very rapidly. We denote the lowest of these 
frequencies as the fundamental mode. 

For the even-parity $l=1$ mode, we similarly consider the boundary conditions at both the center and at the surface of the star for 
Eq.s~\eqref{eq:pert37},~\eqref{eq:pert38} and ~\eqref{eq:pert41}. We note that at the center of the star, the perturbation functions $H_0$, $H_1$ and $W$ must be finite 
whilst at the surface of the star, they must match with the values that preserve the spherical symmetry of the exterior background spacetime. These conditions thus 
translate as follows: 
\begin{equation} 
\label{eq:pert48} 
H_0\sim hr; H_1\sim -8\pi(\rho_ec+p_c)wr^2; W\sim wr^2 
\end{equation} 
where $h$ and $w$ are arbitrary constants, and $\rho_ec$ and $p_c$ are respectively the matter-energy density and pressure at the center of the star, and: 
\begin{equation} 
\label{eq:pert49} 
H_0(R)=0; H_1(R)=0. 
\end{equation} 
This boundary value problem Eq.s~\eqref{eq:pert37} to ~\eqref{eq:pert41}, and Eq.s~\eqref{eq:pert48} to ~\eqref{eq:pert49} can then be similarly
solved using a shoot and match method where test $\omega^2$ values are chosen and Eq.s~\eqref{eq:pert37} to ~\eqref{eq:pert41} are integrated
from the center to the surface of the star until a match occurs with the boundary conditions at the surface of the star.
In choosing the test $\omega^2$ values, we follow \cite{Lindblom89} in using a variational principle for the $l=1$ mode frequencies as obtained by Detweiler 
in 1975 \cite{Detweiler75}:
\begin{eqnarray}
\label{eq:pert50}
\omega^2\int^R_0 e^{-\nu/2}[\rho_e+p)e^{\lambda/2}(\frac{W^2}{r^2}+2V^2)-e^{-\lambda/2}\frac{(H_1)^2}{8\pi}]dr= \nonumber \\
\int^R_0 e^{\nu/2}\{r^2 e^{\lambda/2}p\gamma[\frac{\delta\rho_e}{\rho_e+p}]^2-\frac{1}{16\pi}(1+r\nu_{,1})e^{-\lambda/2}(H_2)^2\}dr+ \nonumber \\
\frac{\rho_e{R}MW^2(R)}{R^4}.
\end{eqnarray} 
We employ test functions $H_1=-8\pi(\rho_{ec}+p_c)r^2[1-(r/R)^2]$ and $W=r^2[1-(2r/R)^6]$ \cite{Lindblom89} and 
substitute Eq.~\eqref{eq:pert29}, Eq.~\eqref{eq:pert36} and ~\eqref{eq:pert37} into Eq.~\eqref{eq:pert50} to obtain the test $\omega^2$ values
for the neutron star. We note that the variational principle approach is able to give us a good estimate of the mode frequencies even without 
an exact knowledge of the perturbation functions. 

For these perturbation modes, when the fundamental $\omega^2$ obtained is negative, the star's oscillation increases without bound 
and the star is said to be unstable. When the fundamental mode is zero, the star is at a critical point
between a stable branch and unstable branch. The star will remain at this point unless perturbed. We shall delve into this in the following section. 

\section{Equation of state change of neutron stars and its instability time scale}

In the previous sections, we present the formalism and the methods by which we obtain the time scales associated with 
non-radiative pulsation modes for non-rotating neutron stars. In this section, we consider the time scales involved when 
a non-rotating neutron star undergoes phase transitions induced by a change in its equation of state. Since an analytic result has not been established
governing how the time scale of collapse of a neutron star varies with the speed of its phase transition, in this section,
we shall consider numerical results. 

We consider isolated static neutron star models with a polytropic equation of state $p=\kappa\rho^{\Gamma}$ as mentioned in Section 4.1, with $\kappa=80$.
The rest mass of these neutron stars are set to be at the maximum allowed for a static equilibrium star with the adiabatic index $\Gamma=1.9$. 
According to Harrison \textit{et~al} \cite{Harrison65}, the adiabatic index has the upper limit of $2$ imposed by the causality condition. 
It is a well-known Newtonian result that the speed of sound has the ultrarelativistic limit 
$\beta_{sound}=(\frac{dp}{d\rho_e})^{1/2}=[\epsilon(\Gamma-1)]^{1/2}\rightarrow(\Gamma-1)^{1/2}$.
Therefore, we set $\Gamma_0=1.895$ in order to allow a change of equation of state with $\Gamma$ increasing but not exceeding the upper limit of $2$. 
\begin{figure}
\begin{center}
\includegraphics[scale=1.0]{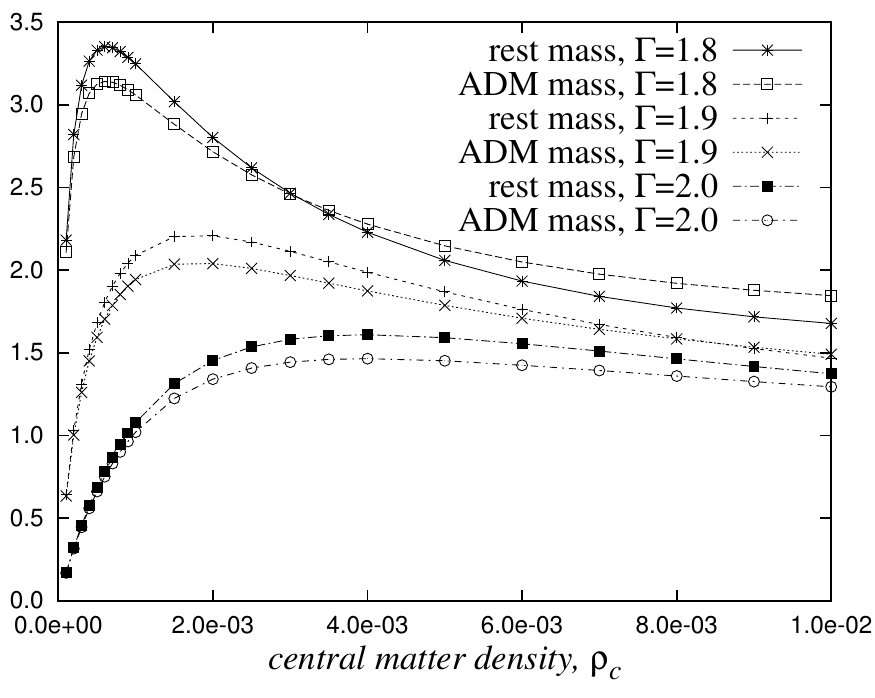}
\caption[Variation of rest mass and ADM mass with respect to central matter density for stars with different equations of state.]{}
\label{fig:4a}
\end{center}
\end{figure}
Fig.~\ref{fig:4a} shows the variation of the rest mass and ADM mass with respect to the central matter density of stars 
modelled with equations of state with different adiabatic indices. As mentioned in the previous section,
there is a maximum for each of the curves, which indicates a critical point between the stable branch on the left and the unstable branch on the right. 
We see that as $\Gamma$ increases, the curve moves further to the right with decreasing maximum. 
As the central matter density is less than 1 in the geometric units used, the pressure of the star 
as well as the number of baryons packed in the star decreases as $\Gamma$ increases.

\begin{figure}
\begin{center}
\includegraphics[scale=1.0]{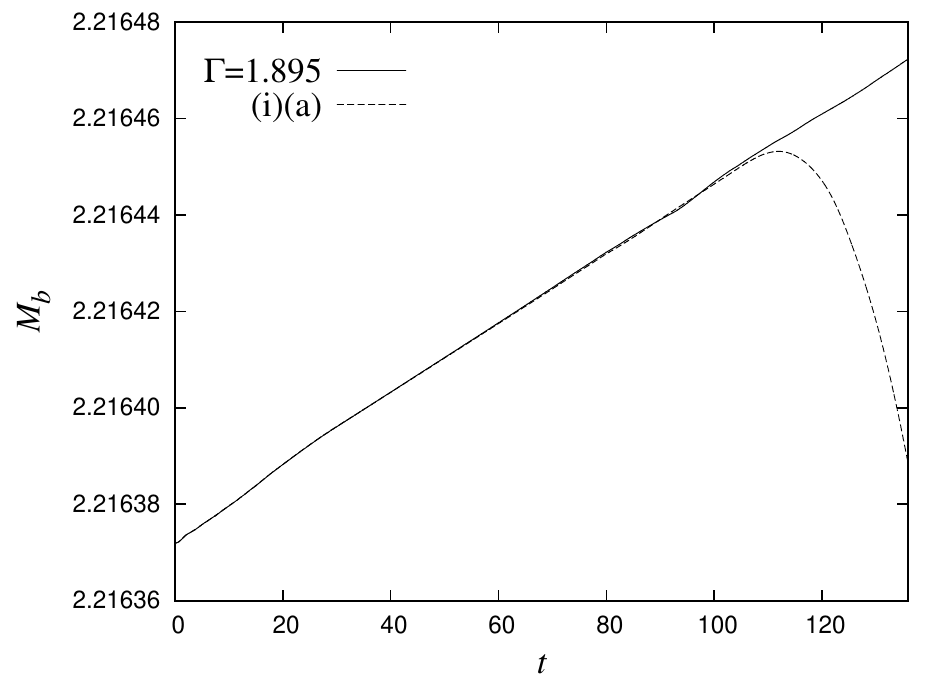}
\caption[Conservation of rest mass for star undergoing phase transition.]{}
\label{fig:mass}
\end{center}
\end{figure}  
We set the rest mass of the stars to be equal to the maximum rest mass of a star with $\Gamma=1.9$, ie. at $M_b=2.2135$. 
With this setup, the stars are thus set on the stable branch of the $\Gamma=1.895$ curve, with their phases allowed to transition with time to the maximum point of the 
$\Gamma=1.9$ curve. We utilize the GRAstro-2D code and use a finite-differencing resolution of $dx=0.025$. Convergence of the results 
with respect to resolution is verified by performing simulations using other resolutions, eg. $dx=0.05,0.075$. A 
slow change is imposed on the equation of state by slowly increasing $\Gamma$ at a constant rate $\frac{d\Gamma}{dt}$, 
simultaneously conserving baryonic mass, which is checked to be conserved to the level $0.00074\%$, shown in Fig.~\ref{fig:mass}. 
Two categories of cases are considered, namely:
\begin{equation*}
(a) \Gamma=\Gamma_0+\frac{d\Gamma}{dt}t, t\leq t_{f1} (b) \Gamma=\Gamma_0+\frac{d\Gamma}{dt}t, t\leq t_{f2}.
\end{equation*}
In (a), $\Gamma$ increases to $\Gamma=1.9$, reaching the critical point of the rest mass-matter density curve for $\Gamma=1.9$ at $t=t_{f1}$, thus 
becoming unstable for $t>t_{f1}$. In (b), $\Gamma$ increases to $\Gamma=1.905$, overshooting the critical point beyond which there is no
equilibrium configuration with the given rest mass, for $\Gamma>1.9$. Within these two categories, neutron stars undergoing three rates of change of 
$\frac{d\Gamma}{dt}$, are investigated, namely:
\begin{equation*}
(i) \frac{d\Gamma}{dt}=0.00005, (ii) \frac{d\Gamma}{dt}=0.000075, (iii) \frac{d\Gamma}{dt}=0.0001.
\end{equation*}

We determine the time scale of the collapse of these neutron stars as they cross the critical point by measuring 
$\delta\rho=\rho-\rho_i$, where $i$ is the central matter density of the star at the critical point of the rest
mass-matter density equilibrium curve for $\Gamma=1.9$, and by measuring the rate of change of $\delta\rho$ and its acceleration.

\begin{figure}
\begin{center}
\includegraphics[scale=1.0]{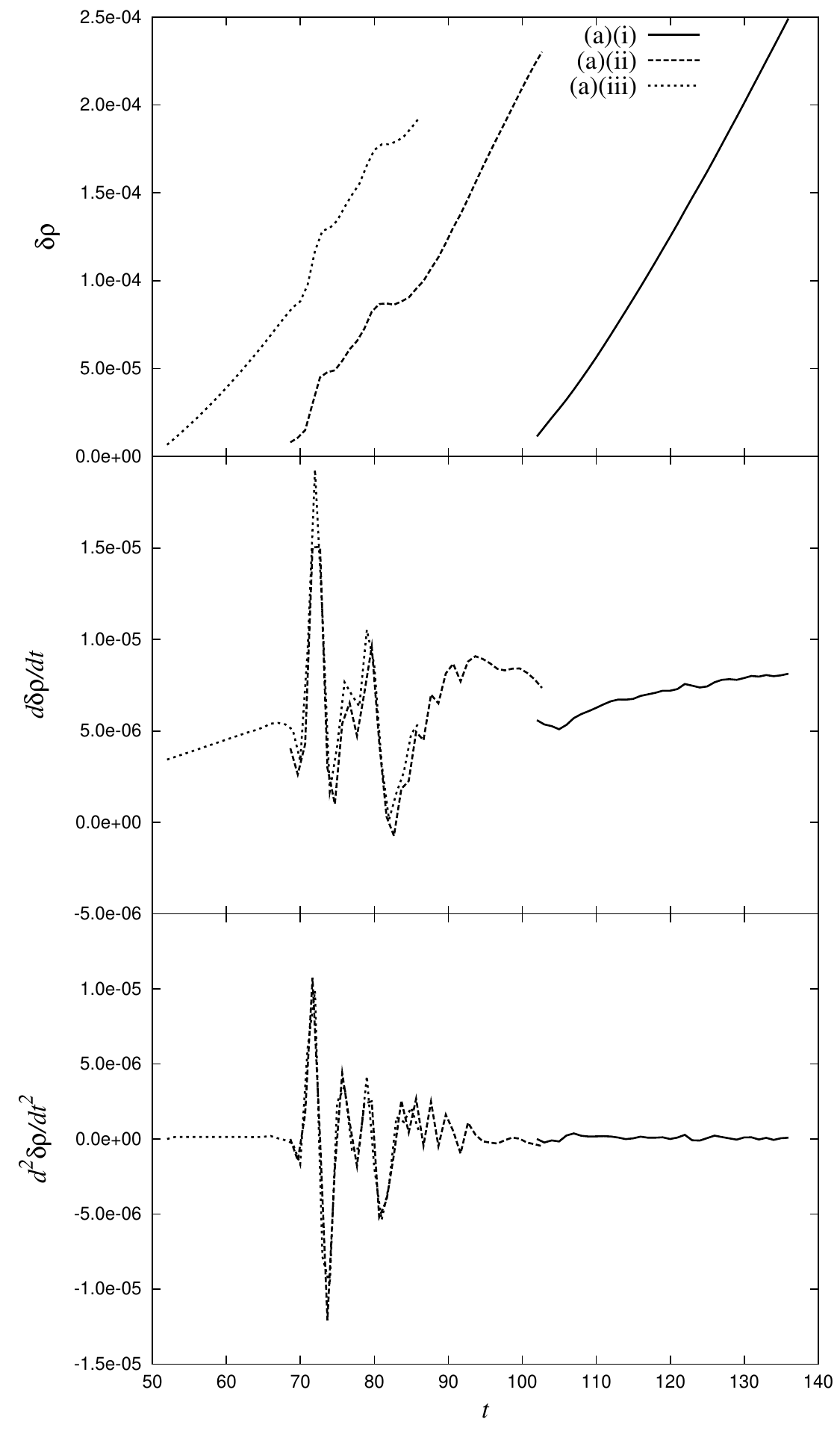}
\caption[Change of $\delta\rho$, $\frac{d\delta\rho}{dt}$ and $\frac{d^2\delta\rho}{dt^2}$ with respect to $t$ for different transition speeds.]{}
\label{fig:4b}
\end{center}
\end{figure}
\begin{figure}
  \begin{tabular}{cc}
     \resizebox{75mm}{!}{\includegraphics{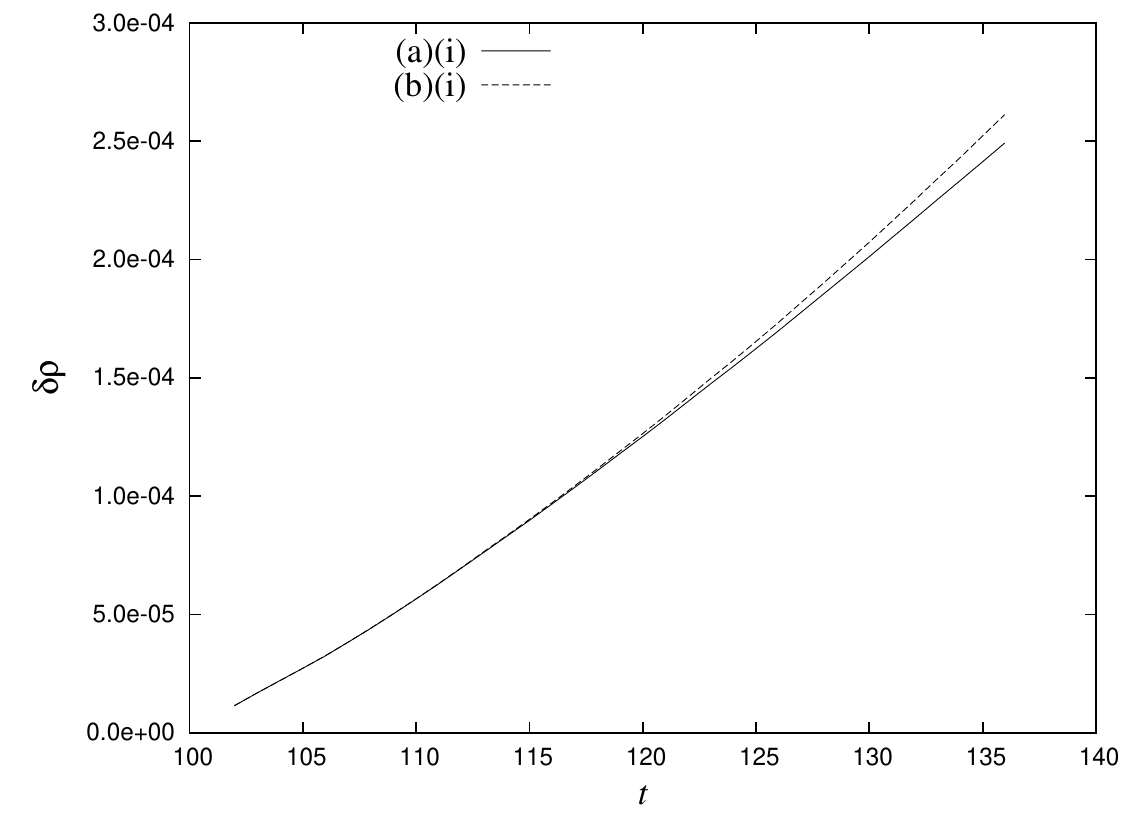}}
     \resizebox{75mm}{!}{\includegraphics{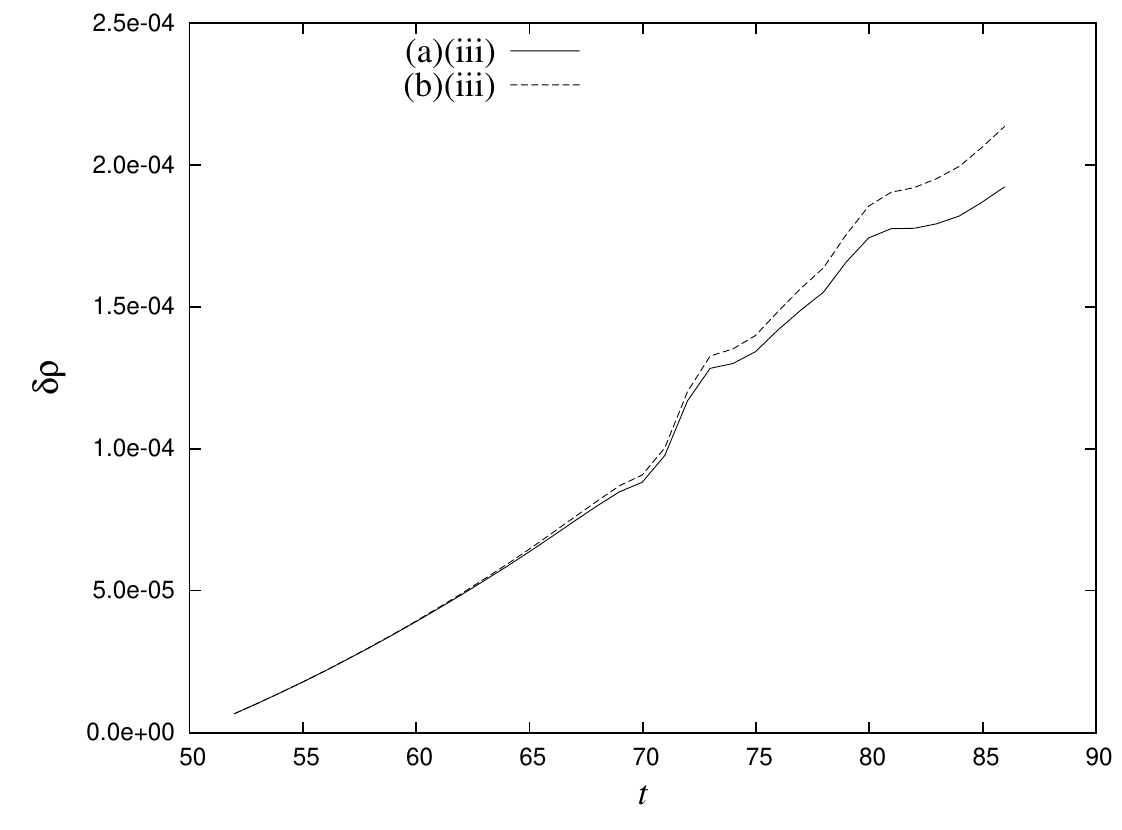}}
  \end{tabular}
\caption[Change of $\delta\rho$ with respect to $t$ for two transition scenarios.]{}
\label{fig:4c}
\end{figure}
\begin{figure}
  \begin{tabular}{cc}
     \resizebox{75mm}{!}{\includegraphics{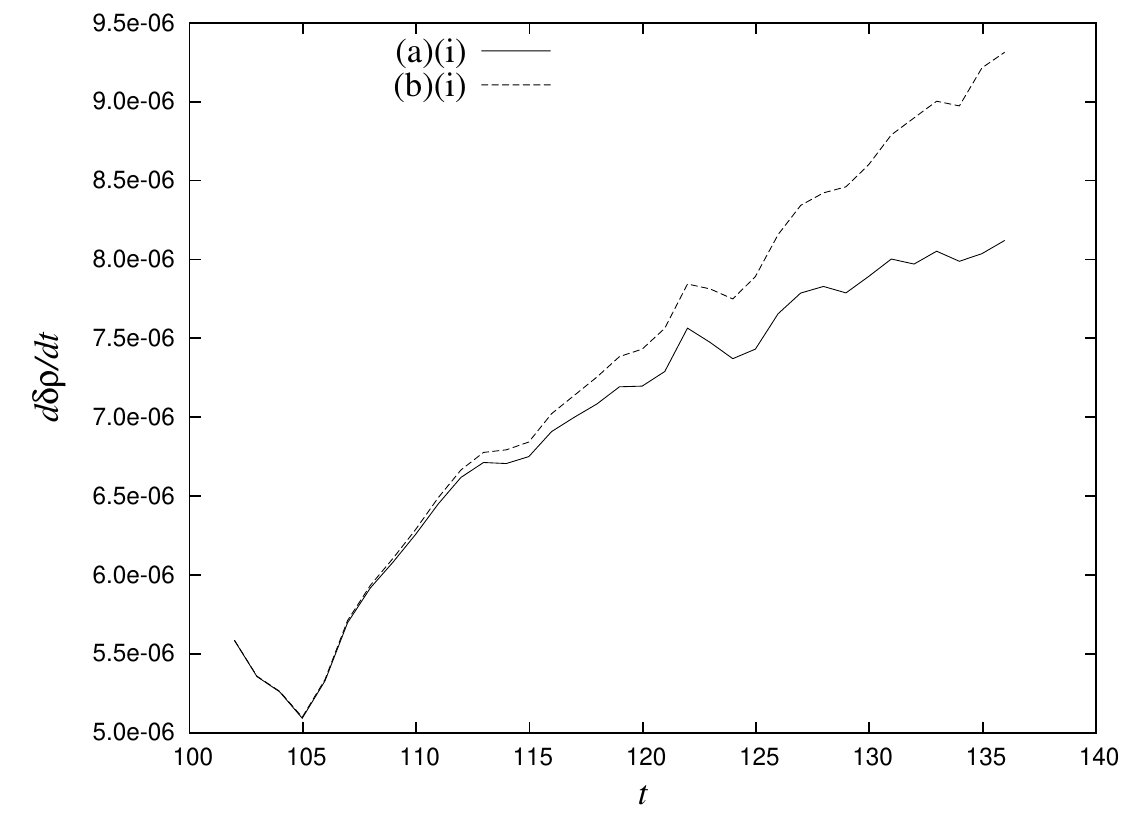}}
     \resizebox{75mm}{!}{\includegraphics{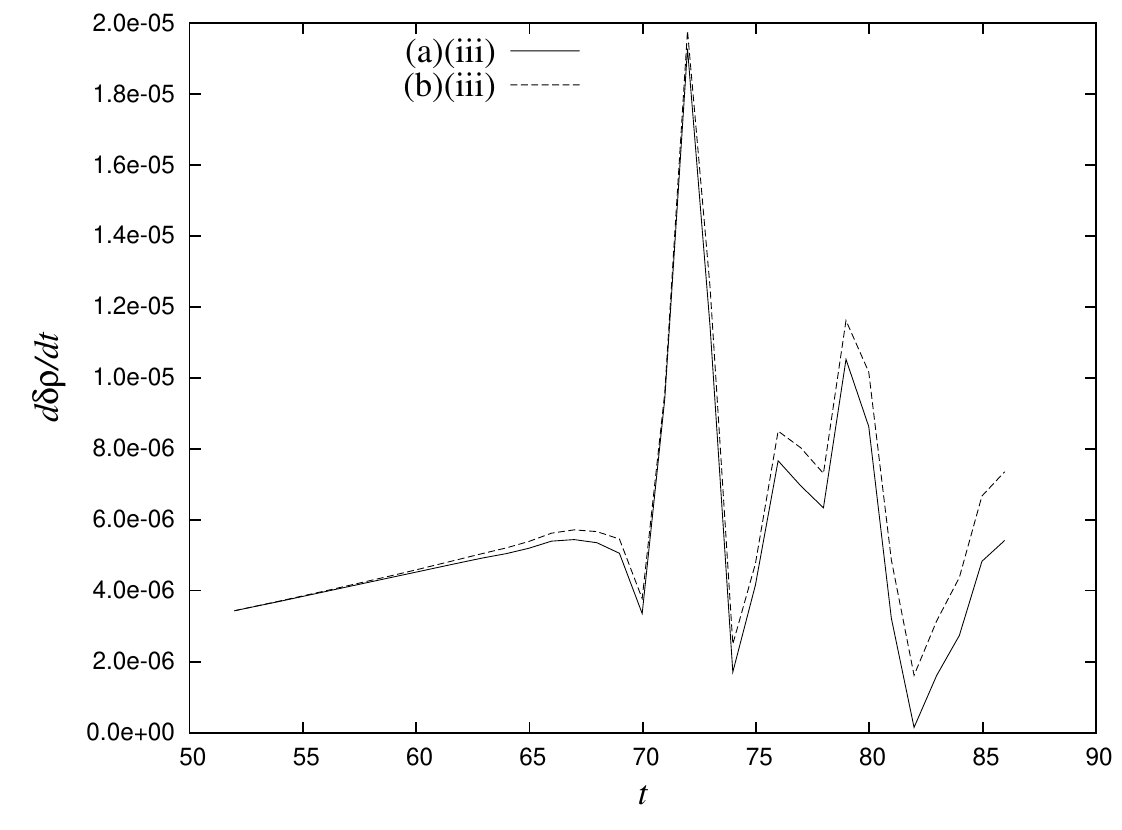}}
  \end{tabular}
\caption[Change of $\frac{d\delta\rho}{dt}$ with respect to $t$ for two transition scenarios.]{}
\label{fig:4d}
\end{figure}
\begin{figure}
  \begin{tabular}{cc}
     \resizebox{75mm}{!}{\includegraphics{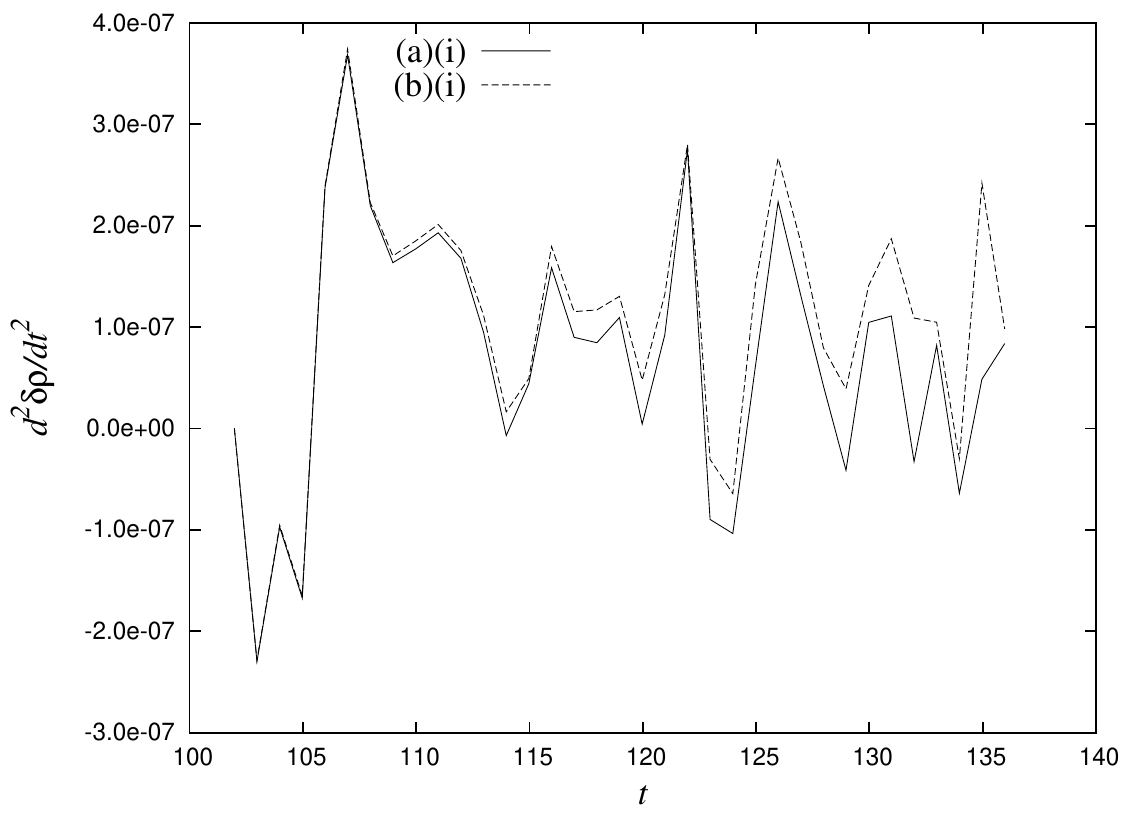}}
     \resizebox{75mm}{!}{\includegraphics{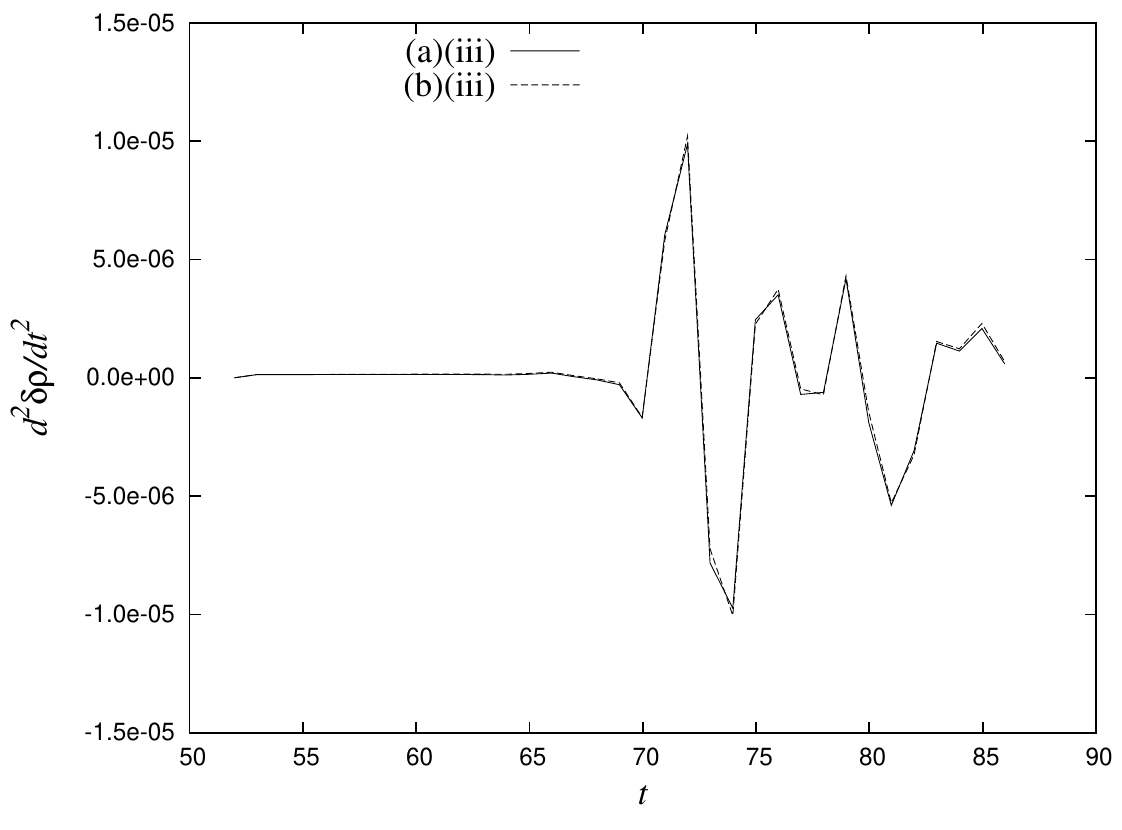}}
  \end{tabular}
\caption[Change of $\frac{d^2\delta\rho}{dt^2}$ with respect to $t$ for two transition scenarios.]{}
\label{fig:4e}
\end{figure}

Fig.~\ref{fig:4b} shows the change of $\delta\rho$, $d\delta\rho/dt$ and $d^2\delta\rho/dt^2$ with time for each of the transition speeds 
(i),(ii) and (iii). From this figure, we observe that $\delta\rho$ changes
at a constant speed for each of the transition speeds, with $d^2\delta\rho/dt^2\approx 0$. 

Fig.s~\ref{fig:4c} to ~\ref{fig:4e} show the comparison of the $\delta\rho$, $d\delta\rho/dt$ and $d^2\delta\rho/dt^2$ changes with time, between
scenario (a) and (b). From these figures, we observe that the changes of $\delta\rho$ and their respective rates of change 
do not differ much between a neutron star that undergoes a phase transition up till the threshold point, and a neutron star that undergoes a phase transition 
\textit{past} the threshold point reaching a state which cannot be described by an equilibrium configuration. Whilst a small difference is seen only toward the late 
stage of the collapse, the difference in the $d^2\delta\rho/dt^2$ is negligible when the transition speed reaches the order of magnitude of $10^{-4}$.  
 
   \chapter{Critical gravitational collapse}

\section{Critical phenomena in general relativity}

Critical phenomena in general relativity were discovered by Choptuik \cite{Choptuik93} in 1993 in numerical simulations of spherical scalar field collapses. 
The phenomena of universality, mass scaling and self-similarity observed in these gravitational collapses garnered the name of critical phenomena, 
analogous to phase transition phenomena observed in condensed matter systems. 

In particular, as a parameter of a set of general relativistic initial data is varied, the evolution of this initial data 
passes through a threshold between black hole formation and dispersion to infinity. They evolve toward a spacetime which
can be stationary or scale-free \cite{Gundlach07}. We shall call this spacetime the \textit{critical~set} henceforth.
These evolutions leave the threshold via the one unstable mode of the critical set, and the time scale of this departure is denoted as the critical index.
In the present context, universality means that the critical index is independent of the initial data parameter that is varied.

Mass scaling occurs when the black holes that form at this brink of 
collapse begin with infinitesimal masses that scale with respect to the distance of the initial data from the threshold, as follows:
\begin{equation}
\label{eq:crit1}
M\propto(p-p*)^{\gamma},
\end{equation}
where $M$ denotes the mass of the black hole, $p$ denotes the parameter of the initial data that is varied, $p*$ the value of this parameter when the 
initial data evolution stays on the threshold as time goes to infinity, and $\gamma$ the critical index. 
According to \cite{Gundlach07}, an initial data set in general relativity can be denoted
as a function of the spatial coordinates, eg. $S(x)$ where we express it in a 1-dimensional spatial coordinate system, and its evolution in the coordinate time can be denoted as 
$S(x,t)$. $S(x)$ can be the density distribution of the matter configuration or the 3-metric of the spacetime, as mentioned in Section 3.4.
Denoting the critical set itself as $S^{*}(x)$, a solution of the initial data $S(x)$ that produces an evolution very near the critical set can be linearized in a 
small neighborhood of the critical set as follows:
\begin{equation}
\label{eq:crit2}
S(x,t)\approx S^{*}(x)+\sum^{\infty}_{k=0}C_k(p)e^{\lambda_k t}S_k(x),
\end{equation}
where the $\lambda_k$'s are eigenvalues of the critical set which can be purely real, complex or purely imaginary.
As the critical set by definition possesses one unstable mode, as $t\rightarrow\infty$, all perturbations about the critical set vanish leaving one with 
positive real $\lambda_k$, for which we denote the amplitude of the constant as $C_0(p*)$. 
We can then further linearize in a small neighborhood about the critical set via a Taylor's expansion as follows:
\begin{equation}
\label{eq:crit3}
\lim_{t\rightarrow\infty}S(x,t)\approx S^{*}(x)+\frac{dC_0}{dp}(p-p*)e^{\lambda_0 t}S_0(x).
\end{equation}
We can choose a $t=t*$ where Eq.~\eqref{eq:crit3} still holds and denote:
\begin{equation}
\label{eq:crit4}
S(x,t*)\approx S_{*}(x)+\epsilon S_0(x),
\end{equation}
where:
\begin{equation}
\label{eq:crit5}
\epsilon\equiv\frac{dC_0}{dp}(p-p*)e^{\lambda_0 t*}. 
\end{equation}
A black hole that forms in the solution $S(x,t*)$ starts with a mass that grows proportionately to $e^{-t*}$, 
which, in line with Eq.~\eqref{eq:crit5}, is in turn proportional to $(p-p*)^{1/\lambda_0}=(p-p*)^{\gamma}$.
From Eq.~\eqref{eq:crit5} too, we can calculate the critical index as:
\begin{equation}
\label{eq:crit5a}
\gamma=\frac{1}{\lambda_0}=\frac{|t*-C|}{\ln|p-p*|},
\end{equation}
where we have taken $C=\ln(\epsilon/\frac{dC_0}{dp})/\lambda_0$. Analogous to critical phase transitions in materials, ie. first and second phase 
transitions, with discontinuous and continuous order parameters, we can categorize critical gravitational collapse scenarios between those that exhibit mass scaling and 
those that do not. Scenarios that do exhibit mass scaling are called Type II scenarios and those that do not are called Type I. Type II scenarios occur when the system in 
question does not have a preferred scale, eg. in scalar field systems, 
whereas in Type I scenarios, there exists a scale that typifies the Einstein field equations for the system in question, and which cannot be neglected. Therefore, in Type I 
scenarios, the mass of the black holes formed start with finite value. The critical index in Type I collapse scenarios is similarly defined as in Eq.~\eqref{eq:crit5a}.

\begin{figure}  
\begin{center}
\includegraphics[scale=0.8]{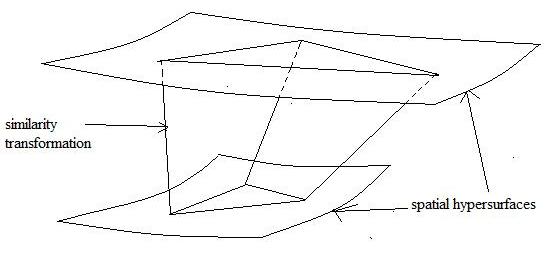}
\caption[Schematic diagram of a similarity transformation.]{}
\label{fig:similarity}
\end{center}
\end{figure}

The next feature seen in critical gravitational collapses reminiscent of phenomena seen in condensed matter systems is the feature of self-similarity.
Self-similarity describes the symmetries of the critical set of the collapse scenario, where the system exhibits scale-free behavior. There are two types of self-similarity
observed in collapse scenarios, namely discrete self-similarity (DSS) and continuous self-similarity (CSS). Cahill and Taub \cite{Cahill71} first computed the equations for CSS of a spherically symmetric configuration of perfect fluid, as follows:
\begin{equation}
\label{eq:crit6}
\mathcal{L}_{\xi}g_{\alpha\beta}=2g_{\alpha\beta},
\end{equation}
where $\xi$ is a homothetic Killing vector producing this self-similarity transformation, which 
in terms of the coordinate system $x^{\mu}=(t,x^i)$, can be denoted as:
\begin{equation}
\label{eq:crit7}
\xi=-\frac{d}{dt},
\end{equation}
and $g_{\alpha\beta}$ is the 4-metric of the spacetime exhibiting CSS.
Eq.~\eqref{eq:crit6} reduces Einstein's field equations to a set of ordinary differential equations.
The resulting equations for the spacetime in such a collapse scenario are:
\begin{equation}
\label{eq:crit9}
\mathcal{L}_{\xi}R^{\alpha}_{\mu\beta\gamma}=0; \mathcal{L}_{\xi}R_{\alpha\beta}=0; \mathcal{L}_{\xi}G_{\alpha\beta}=0.
\end{equation}
This leads to the following equation for the energy-momentum tensor:
\begin{equation}
\label{eq:crit10}
\mathcal{L}_{\xi}T_{\alpha\beta}=0,
\end{equation}
which, coupled with Eq.~\eqref{eq:crit6}, produces, in the case of a perfect fluid, the following for the matter variables:
\begin{equation}
\label{eq:crit11}
\mathcal{L}_{\xi}p=-2p; \mathcal{L}_{\xi}u_{\alpha}=-u_{\alpha}; \mathcal{L}_{\xi}\rho_e=-2\rho_e.
\end{equation}
Eq.~\eqref{eq:crit9} describe geometric self-similarity whereas Eq.~\eqref{eq:crit11} describe physical self-similarity, which is equivalent to geometric self-similarity only in the case of perfect fluids. 
Eq.s~\eqref{eq:crit6} and ~\eqref{eq:crit11} can be viewed as a similarity transformation in general relativity whereby a spatial hypersurface is pushed
forward into itself as depicted in Fig.~\ref{fig:similarity} (refer to Appendix B).
DSS is a more general form of self-similarity \cite{Coley97} and the 4-metric $g_{\alpha\beta}$ for the spacetime undergoing such a collapse scenario can be written in the form:
\begin{equation}
\label{eq:crit8}
g_{\alpha\beta}(t,x^i)=e^{-2t}\tilde{g}_{\alpha\beta}(t,x^i).
\end{equation}
Due to the scale-invariance of Type II critical gravitational collapses, the self-similarity feature can only be observed in Type II scenarios.

Since the first discovery of critical gravitational collapse by Choptuik in spherical scalar fields, universality, mass scaling and self-similarity have since been 
observed in a variety of other systems such as perfect fluids \cite{Evans94},\cite{Neilsen00}, 2D sigma models \cite{Hara96},\cite{Liebling96},\cite{Liebling98}, SU(2) models \cite{Choptuik96},\cite{Bizon98},\cite{Bizon99},\cite{Choptuik99},\cite{Millward03} and primordial density fluctuations in cosmological models \cite{Niemeyer97},\cite{Niemeyer99},\cite{Yokoyama98},\cite{Jedamzik99},\cite{Green99}. However, these systems are largely restricted to spherical symmetry and 
require fine tuning of initial data and are therefore less prevalent in 
realistic astrophysical systems. Later on in 2003, Choptuik and his collaborators ventured into the study of scalar fields in axisymmetry \cite{Choptuik03}. Abrahams and Evans studied 
axisymmetric collapse in vacuum gravity \cite{Abrahams93} and Lai \cite{Lai04} found Type I critical phenomena in boson star systems. Noble and Choptuik then reported Type II critical 
phenomena in spherically symmetric static neutron stars \cite{Noble08}. In 2007, Jin and Suen \cite{Jin07} produced evidence that Type I critical collapse occurs 
in axisymmetric neutron star systems. A crucial finding in the latter that differs from the previous ones is that when the adiabatic index of the equation of state is 
varied, the same Type I critical collapse is seen to occur in the neutron star system. This presents evidence that critical phenomena not only occur via fine-tuning of 
initial data, but can occur as the equation of state of realistic star systems softens and when this time scale of equation of state change is longer than the time scales involved in the critical collapse.

\section{Universality and the critical index}

In the previous section, we reviewed the basic concepts of critical gravitational collapse and the gravitational systems in which the phenomena has been observed. In this 
section, we focus on several aspects of the concept of universality, which drive the main analyses work in this thesis. 

As mentioned in the previous section, critical collapses are said to be universal as they produce the same critical index even when different parameters of the initial data 
are varied. This is seen to be the case due to Eq.~\eqref{eq:crit3} where only one growing mode survives in the critical solution carrying the system away from the critical 
set - the main criterion for a solution to be critical. Let us take a certain set of initial data that is tuned via variation of a certain parameter to form a 
critical solution. We then fix this parameter and vary another parameter to tune the initial data until it produces a critical solution. We note that these two initial data 
sets that produce the critical solution lie very close to each other. Therefore, even when the critical index extracted from these two sets of 1-parameter 
variations are the same, we cannot conclude that we observe universality in the strictest sense. 

Another aspect of universality is the characteristic of the critical solution whereby all details of the initial data, 
except for the distance of the initial data from the threshold, and its conserved geometric quantities, are washed away in the dynamical evolution, thus 
arriving at a universal configuration. In scale-invariant spacetimes where self-similarity is observed, 
universality in this aspect indicates that the system has entered a region of \textit{intermediate~asymptotics} which simultaneously ceases to depend on details of the 
initial data as well as being far away from equilibrium. In Type I collapses of neutron star systems where there exists a constant entropy in the evolution, it is 
especially unclear how dynamical processes drive the system toward a universal configuration \cite{Gundlach07}. In light of this, observing evolutions of initial data sets that 
are very close to each other does not enable us to strictly conclude universality. 

\section{The dynamical systems picture}

Another approach in understanding critical gravitational collapse which proved to be very useful, is the dynamical systems picture. 
The theory of dynamical systems is generally employed in studying physical systems that evolve in time, where the state of the system at an instant in time is described by 
an element $\underline{x}$ in a phase space $X$, which can be finite or infinite-dimensional. The evolution of the system is described by an autonomous system of 
differential equations on $X$ as follows:
\begin{equation}
\label{eq:dyn1}
\frac{d\underline{x}}{dt}=f(\underline{x}), \underline{x}\in X, 
\end{equation}
and: 
\begin{equation}
\label{eq:dyn2}
f:X\rightarrow X,
\end{equation}
where we see that the right hand sides do not have an explicit time dependence. When $X$ is finite-dimensional, Eq.~\eqref{eq:dyn1} becomes an autonomous system of 
ordinary differential equations. When $X$ is infinite-dimensional, Eq.~\eqref{eq:dyn1} becomes an autonomous system of partial differential equations. Eq.~\eqref{eq:dyn2}
is a smooth function which generates a \textit{flow} $\phi(t,x)$. Given an initial condition $x(0)=x_0$, a solution $\phi(t,x_0)$ is obtained and is denoted as the 
trajectory of Eq.~\eqref{eq:dyn1} based at $x_0$. An important feature of this equation is the invariance of $f$ up to translations in time, which therefore enables the 
translation of solutions based at $t_0\neq 0$ to that based at $t_0=0$. Among the solutions to this system of differential equations, there exists an 
important class called \textit{fixed~points}, where the $f(x)=0$. A fixed point $\bar{x}$ is said to be \textit{stable} when every solution $\phi(t,x_0)$ with $x_0$ lying 
in a neighborhood of $\bar{x}$ approaches and remains close to $\bar{x}$ as $t\rightarrow\infty$. The fixed point is further said to be \textit{asymptotically~stable} when 
$\phi(t,x_0)\rightarrow\bar{x}$ as $t\rightarrow\infty$. Conversely, the fixed point is said to be \textit{unstable} when every solution $\phi(t,x_0)$ leaves $\bar{x}$ as 
$t\rightarrow\infty$. Linearization of Eq.~\eqref{eq:dyn1} in a small neighborhood of $\bar{x}$ enables one to cast the system of equations into a \textit{characteristic} 
form and obtain its eigenvalues and their corresponding eigenvectors, as follows:
\begin{eqnarray}
\label{eq:dyn3}
\frac{dx}{dt}\equiv\dot{x} & = & Ax \nonumber \\
x^j(t) & = & e^{\lambda_j t}v^j,
\end{eqnarray}
where $\lambda_j$ is the eigenvalue associated with the eigenvector $v^j$. Solutions that lie in the linear subspaces spanned by the eigenvectors form what is called 
\textit{invariant~sets}. These invariant sets can be further and correspondingly categorized as stable and unstable depending on the directions of the trajectories.
The eigenvalues are crucial in determining the nature of the fixed point as well as the trajectories in its neighborhood. If there exists more than one eigenvalue, and if 
all of them possess negative real parts, the fixed point is determined as asymptotically stable. If one of the eigenvalues have a positive real part, the fixed point is 
unstable. If all the eigenvalues have non-zero real parts, the fixed point is called \textit{hyperbolic}. Eigenvalues that are all purely imaginary indicate that the fixed 
point is stable but not asymptotically stable. Different combinations of the eigenvalues also produce different trajectorial behaviors near the fixed point. We present 
here a 3-dimensional example where the eigenvalues consist of a conjugate pair of complex numbers, and a real number. Let us take the system of ordinary 
differential equations as follows:
\begin{equation}
\label{eq:dyn4}
\begin{bmatrix}
\dot{x} \\
\dot{y} \\
\dot{z}
\end{bmatrix}=\begin{bmatrix}
\alpha & -\beta & 0 \\
\beta & \alpha & 0 \\
0 & 0 & \lambda 
\end{bmatrix}
\begin{bmatrix}
x \\
y \\ 
z
\end{bmatrix}; \omega=\alpha+i\beta,
\end{equation}
which has the following solution:
\begin{equation}
\label{eq:dyn5}
\begin{bmatrix}        
x \\
y \\ 
z
\end{bmatrix}(t)=\begin{bmatrix}
e^{\alpha t}((\cos\beta t)x(0)-(\sin\beta t)y(0)) \\
e^{\alpha t}((\sin\beta t)x(0)+(\cos\beta t)y(0)) \\           
e^{\lambda t}z(0)                 
\end{bmatrix}. 
\end{equation}
The fixed point for Eq.~\eqref{eq:dyn4} is the point $(0,0,0)$. In this case, the trajectories of solutions in the small neighborhood about this fixed point form a spiral 
structure. As the radius $\sqrt{x^2+y^2}$ of the trajectory shrinks, the $z$ component of it increases. 

\begin{figure}
\begin{center}
\includegraphics[scale=0.7]{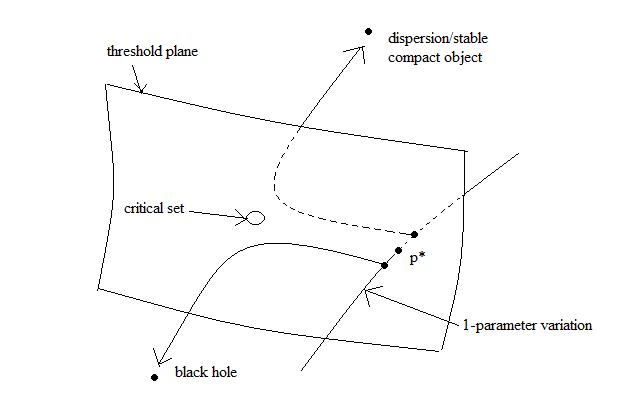}
\caption[Schematic diagram of phase space dynamics.]{}
\label{fig:phasespace1}
\end{center}
\end{figure}

\begin{figure}
\begin{center}
\includegraphics[scale=0.7]{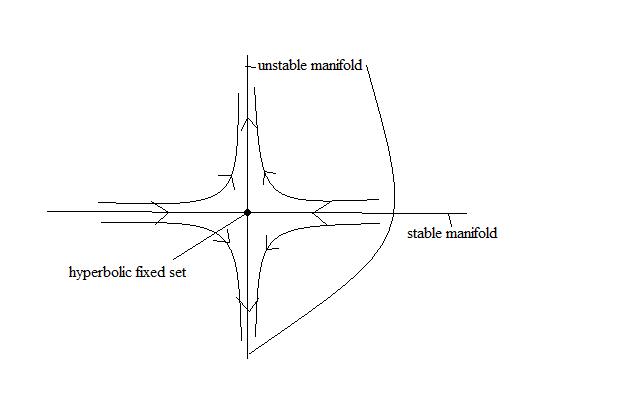}
\caption[Schematic diagram of phase space dynamics with one dimension suppressed.]{}
\label{fig:phasespace2}
\end{center}
\end{figure}

Using the dynamical systems approach, general relativistic initial data sets that evolve and form a \textit{critical~solution} can be seen as points in an 
infinite-dimensional phase space. As shown in Section 5.1, linearization can be performed in a small neighborhood of the critical set, and its eigenvalues determine the 
nature of the trajectories of these initial data sets in the neighborhood of the critical set. The critical set is a \textit{semi-attractor} that serves as a 
\textit{hyperbolic} limit set possessing infinite decaying modes but one growing unstable mode in its neighborhood. Thus, trajectories are attracted to the critical                  
set via decaying modes that are tangent to the trajectories in the phase space, but as they reach the critical set, they are repelled from the threshold 
via the one growing unstable mode (Fig.~\ref{fig:phasespace1}). 
The subspaces that are inhabited by these trajectories are manifolds. Manifolds that contain trajectories leaving the critical 
set are unstable while those that contain trajectories attracted to the critical set are stable. These manifolds are tangent to their corresponding linear subspaces which 
are obtained in the former linearization in the small neighborhood of the critical set, and are of the same dimension as the latter \cite{Guckenheimer83}. The nonlinearity of 
the system is what allows for the possiblity of the critical set being a \textit{closed~or~periodic~orbit} rather than just a point. In the case of a periodic orbit, 
the solution takes the form $\phi(t,x_0)=\phi(t+T,x_0)$ as $t\rightarrow\infty$. As mentioned in Section 5.1, the critical set in gravitational collapse scenarios can thus 
be either a limit cycle or a fixed point. In Type II collapses, the critical set is a limit cycle when the the system exhibits CSS, and a
fixed point when it exhibits DSS. The stationarity of the critical set thus precludes any possibility of its emission of gravitational radiation by definition.

Recalling Section 5.1, in critical gravitational collapses, 1-parameter variations of the initial data produce groups of initial data that arrive at different end states. 
These end states are separated by a threhold into two basins, ie. that of a black hole and that of dispersion into infinity or a stable compact object. 
The closer the initial data point is to the threshold, the longer its 
trajectory will stay parallel to the threshold and the easier it gets attracted to the critical set. As these initial data points are only very near but not directly on the 
threshold, their trajectories are repelled by the one unstable mode of the critical set and leave the threshold in a path perpendicular to the threshold even 
\textit{before} it reaches the critical set (Fig.~\ref{fig:phasespace1}). The manifold that contains only trajectories leaving the critical set is called the unstable manifold of 
the critical set. Points that lie directly on the threshold by definition produce trajectories that never leave the threshold (Fig.~\ref{fig:phasespace2}).
This is due to the fact that these trajectories will all lie within the stable manifold of the critical set. We thus note that the critical set 
defines an \textit{attraction~basin}, a region of the phase space near the threshold inhabited by initial data 
points that produce trajectories that are attracted to the critical set. The size of this basin can be described in terms of its width along the direction perpendicular 
to the threshold, and its breadth along the tangential direction. The threshold, on the other hand, is a plane of codimension one in the phase 
space separating between the black hole and dispersion/stable compact object attraction basins. We denote the region in the phase space situated between the stable and 
unstable manifolds as a hyperbolic region where trajectories are both moving toward the critical set and also moving away from the threshold. Strictly speaking, in the theory of dynamical systems, the trajectories in this region are sequences of points rather than a curve in the phase space.

The critical set itself however is denoted as the \textit{center} manifold, as all its eigenvalues have zero real parts, in contrast with the unstable manifold in its 
neighborhood whose eigenvalues have positive real parts and the stable manifold where the eigenvalues have negative real parts. 
In dynamical systems theory, the center manifold is where \textit{bifurcation} occurs. Bifurcation describes the \textit{change} in the qualitative structure of the 
solutions when parameters in the defining differential equations are varied and when they enter a range of values, called \textit{bifurcation~values}. 
The \textit{bifurcation~point} is 
the point where the qualitative behavior switches. The center manifold thus contains the bifurcation point, and is tangent to the center linear subspace. Codimension 
one bifurcations depend only on one parameter. We present here an example where a particular type of bifurcation occurs, called the Hopf bifurcation. Let us take the system of 
ordinary differential equations as follows:
\begin{eqnarray}
\label{eq:dyn6}
\dot{x} & = & y+x(1-x^2-y^2)(4-x^2-y^2) \nonumber \\
\dot{y} & = & x+y(1-x^2-y^2)(4-x^2-y^2),
\end{eqnarray}
which can be rewritten as follows:
\begin{eqnarray}
\label{eq:dyn7}
r^2 & = & x^2+y^2 \nonumber \\
r\dot{r} & = & x\dot{x}+y\dot{y} \nonumber \\
\dot{r} & = & r(1-r^2)(4-r^2).
\end{eqnarray}
The trajectories of solutions to Eq.~\eqref{eq:dyn6} will be governed by the signs of $\dot{r}$ in Eq.~\eqref{eq:dyn7} for different ranges or $r$. If the initial condition 
$r_0$ is such that $0<r_0<2$, the solution $r(t,r_0)$ approaches the value of 1 as $t\rightarrow\infty$. However, when $r_0>2$, the solution goes to infinity as 
$t\rightarrow\infty$. Thus, the limit cycle where $\dot{r}=0, r=2$ is a repelling set, with the direction of the trajectories changing with respect to the variation of the 
parameter $r_0$ in the initial condition. We note however, in numerical simulations of nonlinear infinite-dimensional systems where no exact solutions exist, 
bifurcation is never actually seen, due to the fact that the evolutions cannot in principle be carried out for $t\rightarrow\infty$.
In effect, what we see is only a jumping back and forth of initial data points in a small neighborhood about the threshold. This is analogous to the introduction of a small 
number $|\epsilon|$ in Eq.~\eqref{eq:dyn6}. The trajectories are therefore seen to leave the threshold in either direction, never actually dwelling on the critical solution as $t\rightarrow\infty$.

   \chapter{Critical gravitational collapse of neutron stars}

\section{Time scale comparisons with non-radiative pulsation modes of neutron stars}

\begin{figure}
\begin{center}
\includegraphics[scale=1.0]{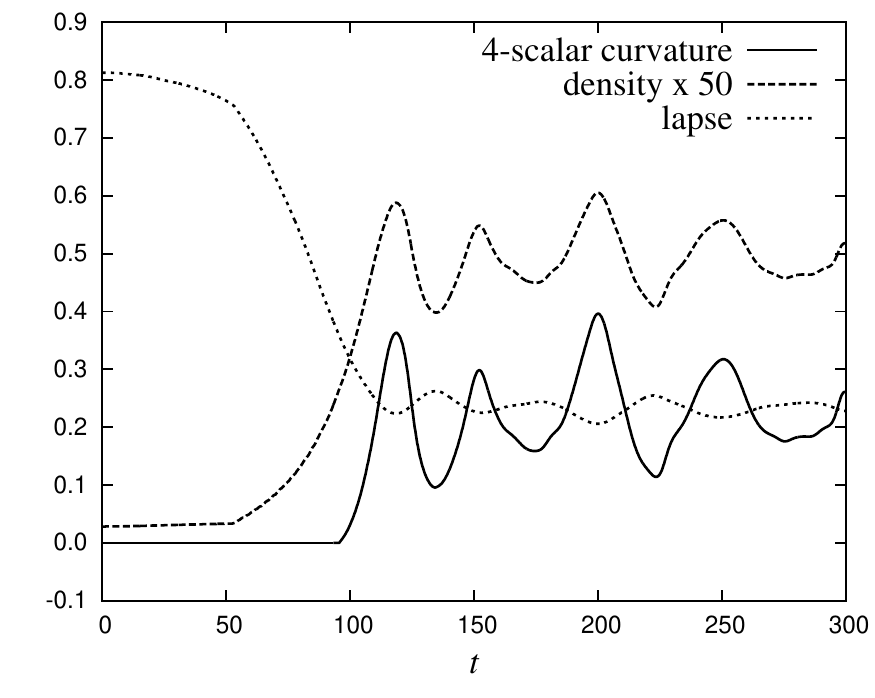}
\caption[Evolution of central matter density, lapse and 4-scalar curvature of the neutron star critical solution.]{}
\label{fig:6a}
\end{center}
\end{figure}
\begin{figure}
\begin{center}  
\includegraphics[scale=1.0]{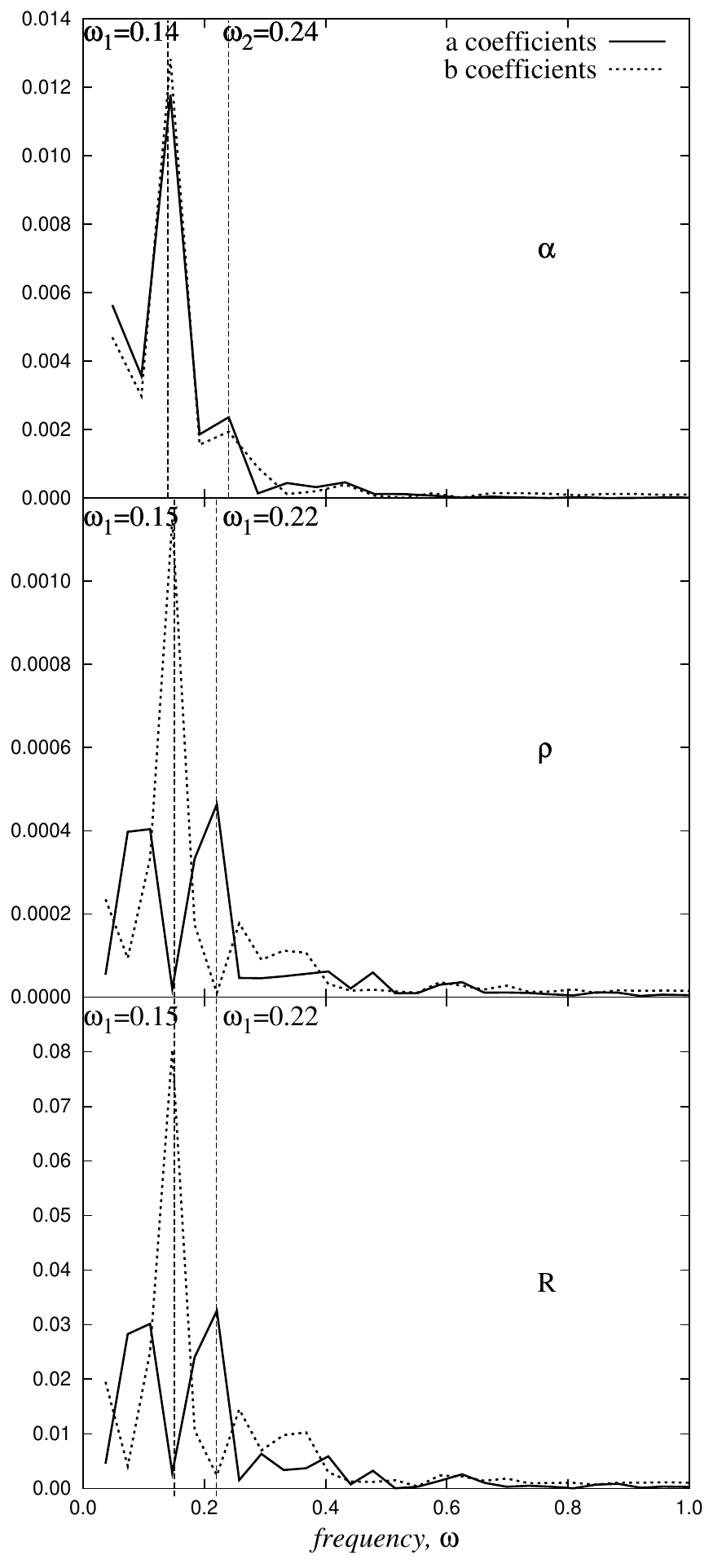}
\caption[Fourier transforms of the central matter density, lapse and 4-scalar curvature oscillations.]{}
\label{fig:6b}  
\end{center}
\end{figure}
\begin{figure}
  \begin{center}
    \begin{tabular}{c}
      \scalebox{0.8}{\includegraphics{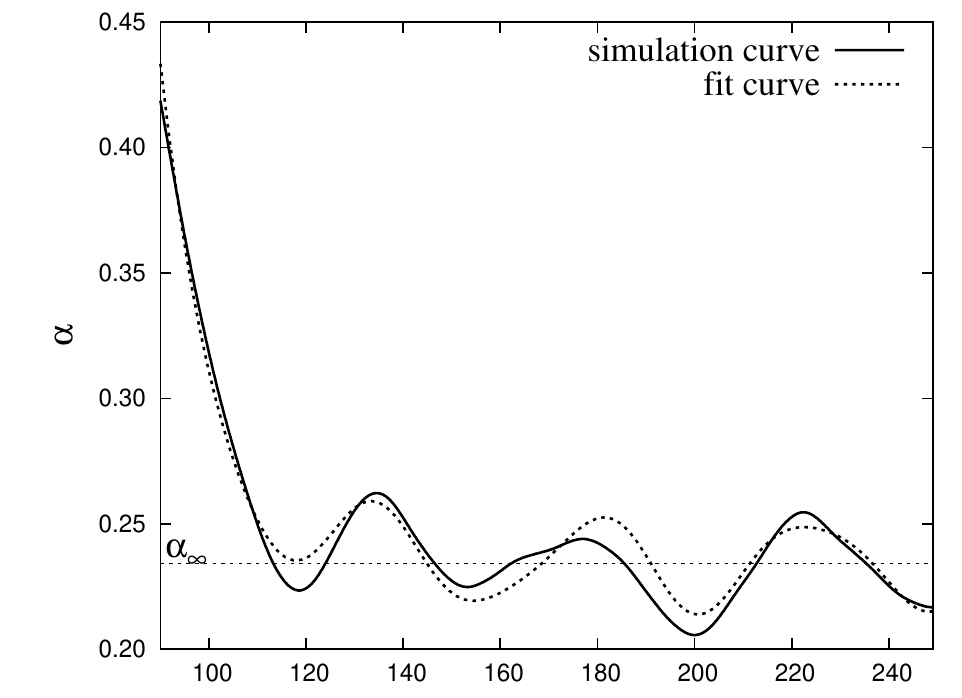}} \\
      \scalebox{0.8}{\includegraphics{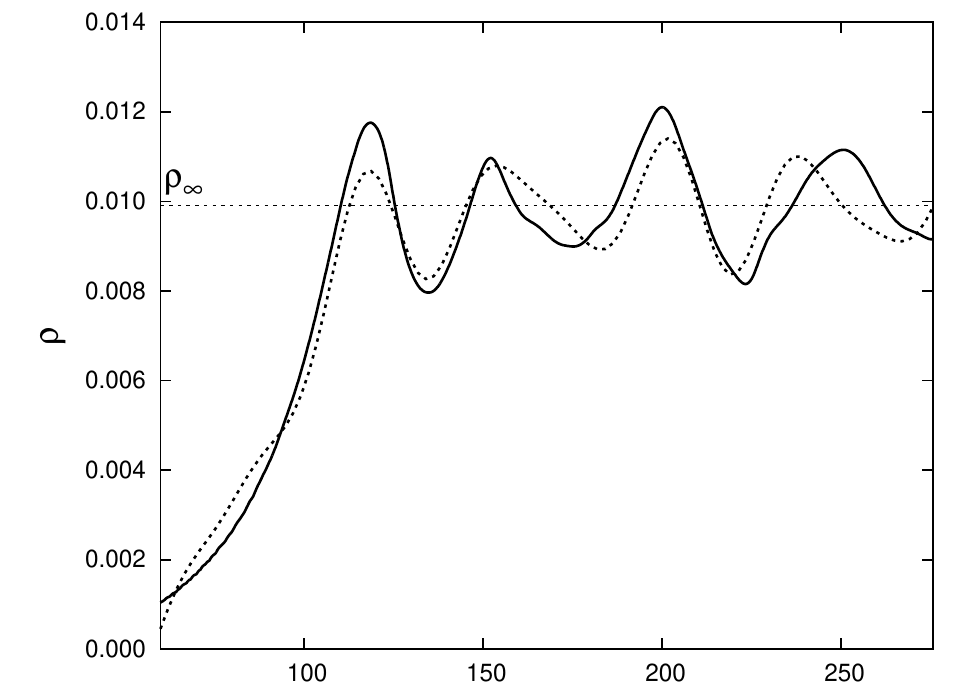}} \\
      \scalebox{0.8}{\includegraphics{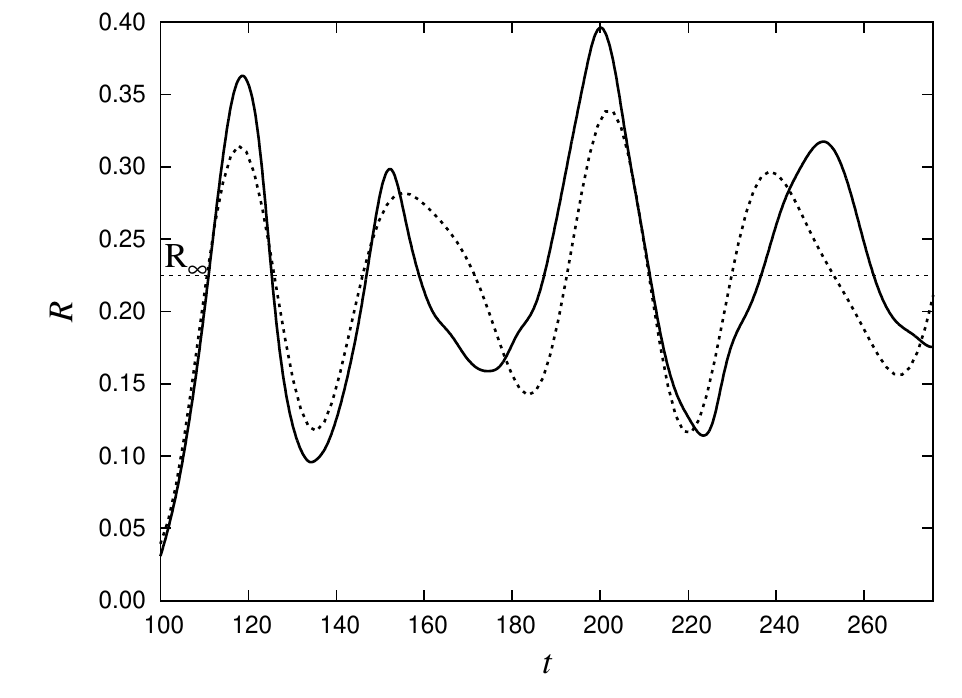}}
    \end{tabular}
    \caption[Fits of the central matter density, lapse and 4-scalar curvature oscillations.]{}
    \label{fig:6c}
  \end{center}
\end{figure}
In this section, we shall seek to clarify the different time scales involved in axisymmetric neutron star critical collapse.
Considering the results obtained by Jin and Suen \cite{Jin07}, we obtain the time scale of attraction of the system toward the critical solution and the dominant frequencies of 
the oscillations of the critical solution. Fig.~\ref{fig:6a} shows the evolution of the lapse, density and 4-scalar curvature at the center of collision. 
Via a Fourier analysis, the dominant frequencies of the oscillations of these three evolution variables are obtained. Fig.~\ref{fig:6b} indicates two dominant 
frequencies, namely, $\omega_1=0.15$ and $\omega_2=0.23$, corresponding to the oscillation periods of $T_1=0.21$ and $T_2=0.13$ 
respectively. These values are in units of solar masses and all values henceforth in this chapter will be designated in these units. 
These frequencies are found to be similar across all three variables. Using the dominant frequencies, we then fit the critical solution with the following 
equations:\\
lapse:
\begin{equation}
\label{eq:resu1}
\alpha=\alpha_{\infty}+ae^{-\lambda(t-90)}+b(\cos(\omega_{1}(t-90)+c)+ratio_{\alpha}\cos(\omega_{2}(t-90)+d))
\end{equation}
density:
\begin{equation}
\label{eq:resu2}
\rho=(\frac{1}{\frac{1}{\rho_{\infty}}+ae^{-\lambda(t-60)}})-b(\cos(\omega_{1}(t-60)+c)+ratio_{\rho}\cos(\omega_{2}(t-60)+d))
\end{equation}
4-scalar curvature:
\begin{equation}
\label{eq:resu3}
R=(\frac{1}{\frac{1}{R_{\infty}}+ae^{-\lambda(t-100)}})-b(\cos(\omega_{1}(t-100)+c)+ratio_{R}\cos(\omega_{2}(t-100)+d)),
\end{equation}
where $a$, $b$, $c$ and $d$ are fit constants, $\alpha_{\infty}$, $\rho_{\infty}$ and $R_{\infty}$ are respectively the averages of the
lapse, density and 4-scalar curvature throughout the oscillation phase of the critical solution, $ratio_{\alpha}=0.2$,
$ratio_{\rho}=ratio_R=0.4$ are the ratio of the amplitudes of $\omega_2$ over $\omega_1$ obtained from the Fourier transforms of the
respective evolution variables. The fit results are shown in Fig.~\ref{fig:6c}. From the fit, we find that $\lambda=0.09$, which corresponds to the
attraction time, $t=0.05ms$, of the system toward the critical solution, is similar across the three evolution variables. We further observe
that the attraction time is similar to the departure time found in \cite{Jin07} which indicates that the eigenvalues of the decaying mode and the unstable mode have degenerate real parts. We note that these time
scales are of an order of magnitude smaller than the cooling time scales of neutron stars reported in the astronomy and astrophysics
literature \cite{Page04},\cite{Page06},\cite{Pavlov03}. Therefore, there is a high possibility that real astrophysical systems of cooling neutron stars pass through the threshold of
critical collapse. Further, systems that undergo slow accretion and angular momentum loss with time scales larger than that reported
here for the neutron star critical solution may also pass through the threshold and exhibit critical behavior.

In order to understand the time scales of the oscillation phase of this critical solution, we now seek to compare the frequencies to that obtained via a perturbative 
approach. We observe that the rest mass of the object undergoing critical collapse is higher than the maximum allowed mass for a TOV configuration with the same 
equation of state. Therefore, we consider a static non-rotating equilibrium TOV configuration having the same rest mass as the rest mass of the two TOV configurations 
producing the critical solution added together, namely at rest mass $M_b=1.5$.  
We recall from Chapter 5 that the critical set is a stationary or self-similar spacetime. In the case of axisymmetric neutron star systems considered in \cite{Jin07}, 
black holes form with a finite mass, and scale-invariance is not seen, therefore the critical collapse is of Type I, and the critical set is thus a stationary spacetime. 
In light of this, the critical set of this solution is non-radiative in principle, and we can then employ the tools presented in Chapter 4, ie. the l=0 and even parity l=1 mode 
perturbations, to determine the non-radiative pulsations of a corresponding non-rotating TOV configuration. We obtain the finite-differenced TOV solution, namely the finite-differenced 
matter-energy density, the specific heat enthalpy, the pressure and the ADM mass function of the solution with respect to the isotropic radius of the configuration. We then 
manipulate these functions accordingly and insert them into Eq.s (4.70) to (4.72), and follow the rest of the procedure as mentioned in Section 4.3. Via this method, the 
fundamental $l=0$ frequency for a static equilibrium non-rotating TOV configuration is found to be $\omega_{1p}=6.858\times 10^{-3}$. Similarly, for the 
even-parity $l=1$ mode, we insert the manipulated functions of the TOV solution into Eq. (4.75). The estimated fundamental frequency for this mode is found to be 
$\omega_{2p}=1.813\times 10^{-2}$. We note that $\omega_{1p}$ and $\omega_{2p}$ are both one and two orders of magnitude smaller than $\omega_1$ and $\omega_2$ 
respectively. This is a confirmation that the oscillations of the critical set of the solution are \textit{not} caused by non-radiative pulsations of an equilibrium TOV configuration. There therefore exists the need of performing perturbative analysis on \textit{non}-equilibrium rather than equilibrium background spacetimes in order to ascertain the nature of the oscillations exhibited by the neutron star critical solution. 

\section{The neutron star critical gravitational collapse solution as a semi-attractor}

\begin{table}[ht]
\centering
\begin{tabular}{ccccc}
\hline\hline
Configuration & $M_b$ & $\rho_c$ & $d$ & $v_z$ \\ [0.5ex]
\hline
NS & 1.6 & 0.00056595 & 27.6 & 0.15 \\
1 & 1.5 & 0.00038698 & 29.6 & 0.1 \\
2 & 1.5 & 0.00039192 & 29.6 & 0.1 \\
3 & 1.6 & 0.00055501 & 27.6 & 0.12 \\
4 & 1.6 & 0.00064912 & 27.6 & 0.12 \\ [1ex]
\hline
\end{tabular}
\label{table:t1}
\caption[Initial data configurations for neutron star and Gaussian packet systems.]{$M_b$ is the rest mass, $\rho_c$ is the neutron star central matter density/height of the Gaussian matter density distribution at $t=0$, $d$ is
the center-to-center separation between the neutron stars/Gaussian packets, and $v_z$ is the boost velocity of the neutron stars/Gaussian packets along the
z-direction of the grid.}
\end{table} 
\begin{figure}
\begin{center}
\includegraphics[scale=1.0]{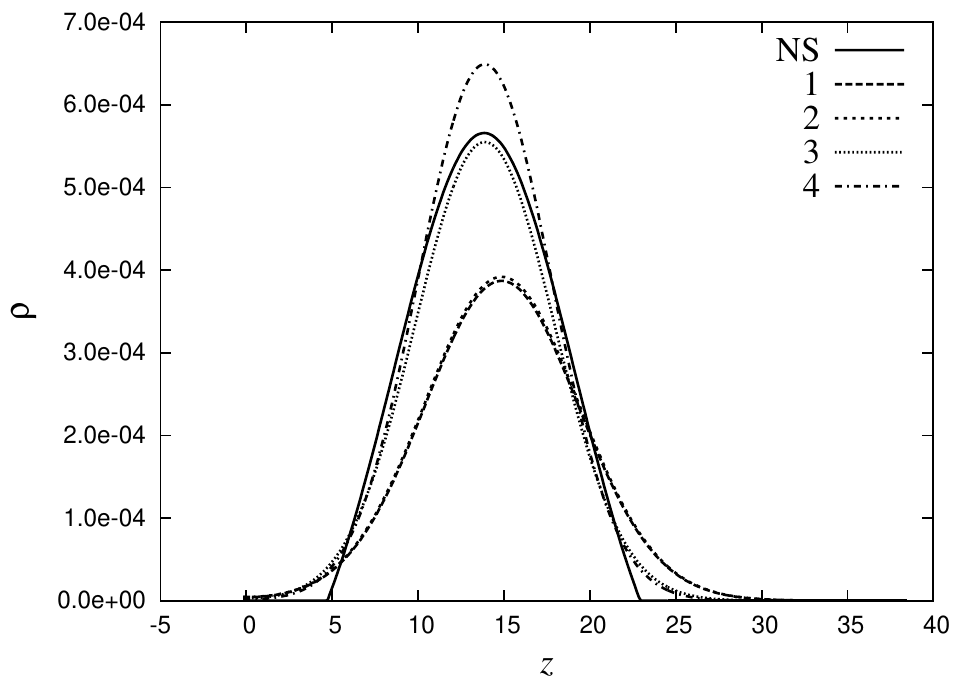}
\caption[Density distributions for neutron star and Gaussian packet systems.]{The Gaussian matter density distributions are constructed so as to maintain the rest mass of the system as a constant. Therefore, when
the Gaussian heights are increased, their widths are correspondingly decreased. Configurations 1 and 2 are less compact distributions
whereas that of 3 and 4 are more compact.}
\label{fig:6d}
\end{center}
\end{figure}

In \cite{Jin07}, the critical index is seen to be similar across variations of different parameters of the initial data ie. the boost velocity, initial separation and 
equation of state. In line with the reasoning presented in Section 5.2, we acknowledge that via this method, universality cannot be strictly claimed for the neutron star 
critical solution. In this section, we present another method to prove universality. In this approach, we construct a neutron-star like 
initial data by solving the initial value problem where the matter field consists of two packets of matter whose densities are characterized by Gaussian distributions 
rotated about the axis of axisymmetry and follow a polytropic equation of state, $p=\kappa\rho_e^{\Gamma}$ with $\kappa=80$ and $\Gamma=2$. The Gaussian density distribution is as follows:
\begin{equation}
\label{eq:gausseq}
\rho(r_{iso})=A+Be^{r_{iso}^2/2C},
\end{equation}
where $A$, $B$ and $C$ are constants that can be adjusted to yield different spacetimes. The normal coordinate condition 
$\mathbf{\beta}=0$ and the geodesic slicing condition $N=1$ are chosen for the initial data. The 3-metric $\mathbf{\gamma}$ for the spacetime of this initial data is 
approximately asymptotically flat. We shall comment further on how an approximately asymptotically flat spacetime differs from one that is exactly asymptotically flat in 
our analysis in Section 6.5. The tuning of the height and width of the Gaussian density distributions produce a new 1-parameter family of initial data. The evolution of 
these initial data sets are carried out using the same numerical setup as used in \cite{Jin07}, namely, using the $\Gamma$-freezing shift and $1+\log$ slicing condition. The 
BSSN scheme is also used in all our simulations as in \cite{Jin07}. 

We first set up this initial data set on the same numerical finite-differenced grid as that used in \cite{Jin07}, namely, on a thin slab of $323$ points in the $x$ and 
$z$-directions and $5$ points along the $y$ direction. The Gaussian packets are then boosted to a head-on collision. The boost is performed via a coordinate transformation 
of the spacetime as follows:
\begin{equation}
\label{eq:boost}
\begin{bmatrix}
t' \\
x' \\
y' \\
z' 
\end{bmatrix}=\begin{bmatrix}
w & v_x w & v_y w & v_z w \\
v_x w & (1+\frac{(w-1)v_x^2}{v^2}) & (\frac{(w-1)v_x v_y}{v^2}) & (\frac{(w-1)v_x v_z}{v^2}) \\
v_y w & (\frac{(w-1)v_x v_y}{v^2}) & (1+\frac{(w-1)v_y^2}{v^2}) & (\frac{(w-1)v_y v_z}{v^2}) \\
v_z w & (\frac{(w-1)v_x v_z}{v^2}) & (\frac{(w-1)v_y v_z}{v^2}) & (1+\frac{(w-1)v_z^2}{v^2}) 
\end{bmatrix}\begin{bmatrix}
t \\
x \\
y \\
z
\end{bmatrix},
\end{equation} 
where $\mathbf{v}=(v_x,v_y,v_z)$ is the boost velocity vector, $v^2=v_x^2+v_y^2+v_z^2$ and $w=(1-v^2)^{-1/2}$. Therefore, the values of the boost velocity vector 
components do not possess intrinsic physical meaning. However, the variation of the boost velocity is non-degenerate and well-behaved, ie. we can say that 
the increase or the decrease of the value of the boost velocity does indicate the isomorphic increase or decrease in the physical boost velocity of the objects in the system.
The behavior of the axisymmetric object formed is observed. We note that this initial data is in a non-equilibrium state at the initial time, although it satisfies the constraint equations. The constraint equations with the two Gaussian distributions of matter and their boost velocities are solved using York's formulation (Section 3.4). 
Table~\ref{table:t1} and Fig.~\ref{fig:6d} show the configurations that are constructed and their density distributions along the direction of collision respectively. These 
initial data sets are distinct from the neutron star configuration in terms of their spacetime and matter properties, as they are clearly not obtained 
by varying different parameters of the neutron star configuration, nor are bound by the TOV spacetime and matter configuration. They are also different from each other in terms of their matter configurations. We examine the evolutions of these initial data sets to see whether they are attracted to the same 
oscillatory critical set of the neutron star solution. 

\begin{figure}
\includegraphics[scale=1.0]{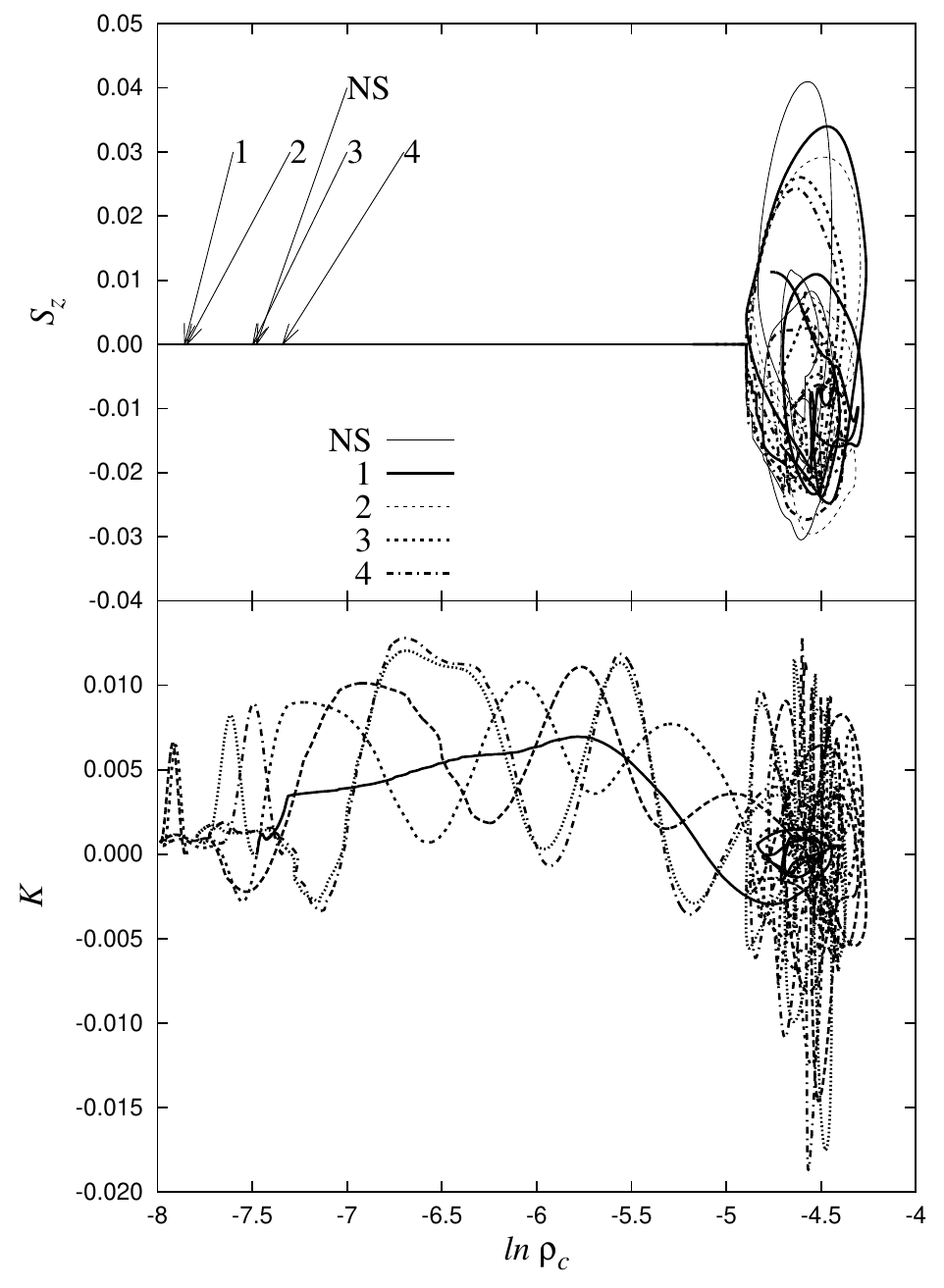}
\caption[Time-based phase spaces.]{The trajectories of all the configurations start on the left and are attracted to the critical set on the right. The arrows point to the differing starting points of the trajectories. $K$ denotes the trace of the
extrinsic curvature at the coordinate (2.1,0.06,0.06) and $S_z$ denotes the z-component of the 3-momentum, $S_i$, of the fluid element
on $7.5\times 10^{-3}$ density contour, respectively. The natural logarithm is taken of the density at the center of the grid, namely
$ln \rho_c$, so as to greater distinguish the differing starting points of the trajectories in the phase space.}
\end{figure}
In order to clearly present this behavior of the evolution of these initial data sets, we choose several variables, namely the trace of the extrinsic curvature $K$ at the 
coordinate $(2.1,0.06,0.06)$, the $z$-component of the momentum density at the matter density contour of $7.5\times 10^{-3}$, and the maximum matter density of the system. 
In Fig. 6.5, we draw the trajectories of these configurations using these variables. From this figure, we see that these variables correlate with each other such that 
they form bounded periodic orbits. In particular, we note that the correlation between the momentum density of a fluid element away from the center of collision and the 
matter density at the center of collision is analogous to that between the position and momentum of a simple pendulum. The oscillations of $K$ before reaching the 
critical set is caused by the fact that the geodesic slicing is used at initial time. As the slicing switches from geodesic to $1+log$ after the first time step, the 
value of the lapse at the grid boundary jumps discontinously from one value to another. 
We observe that these configurations are all attracted to the same critical set of the neutron star solution. This provides evidence that the neutron star 
critical solution is a semi-attractor. 

\section{Phase space analysis on the neutron star critical gravitational collapse threshold}

\begin{figure}
\begin{center}
\includegraphics[scale=1.0]{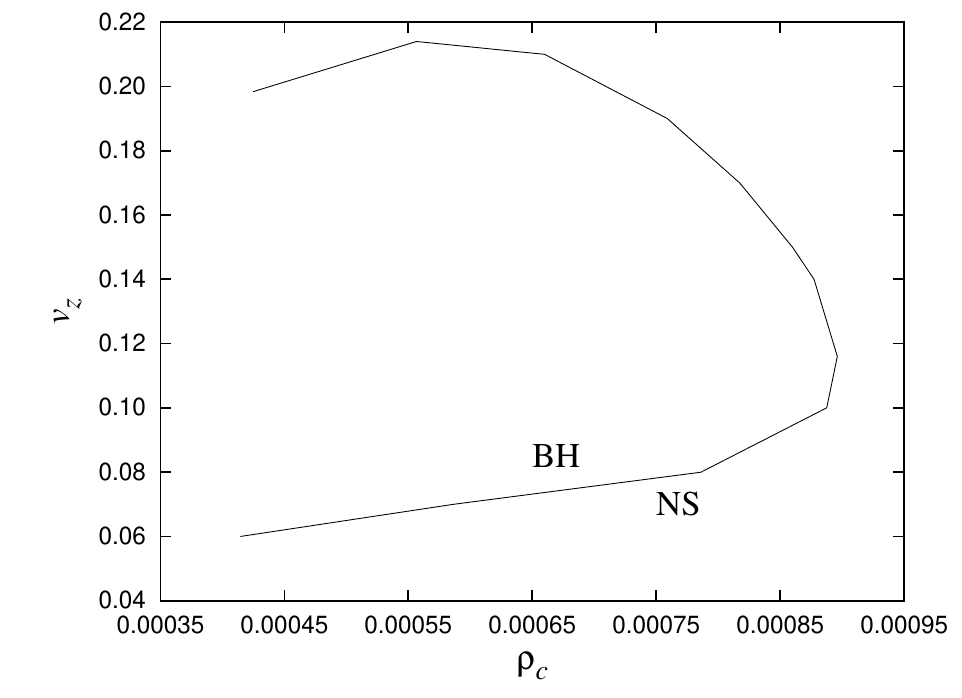}
\caption[Boost velocity-central matter density phase space.]{}
\label{fig:6f}
\end{center}
\end{figure}

\begin{figure}
\begin{center}
\includegraphics[scale=1.0]{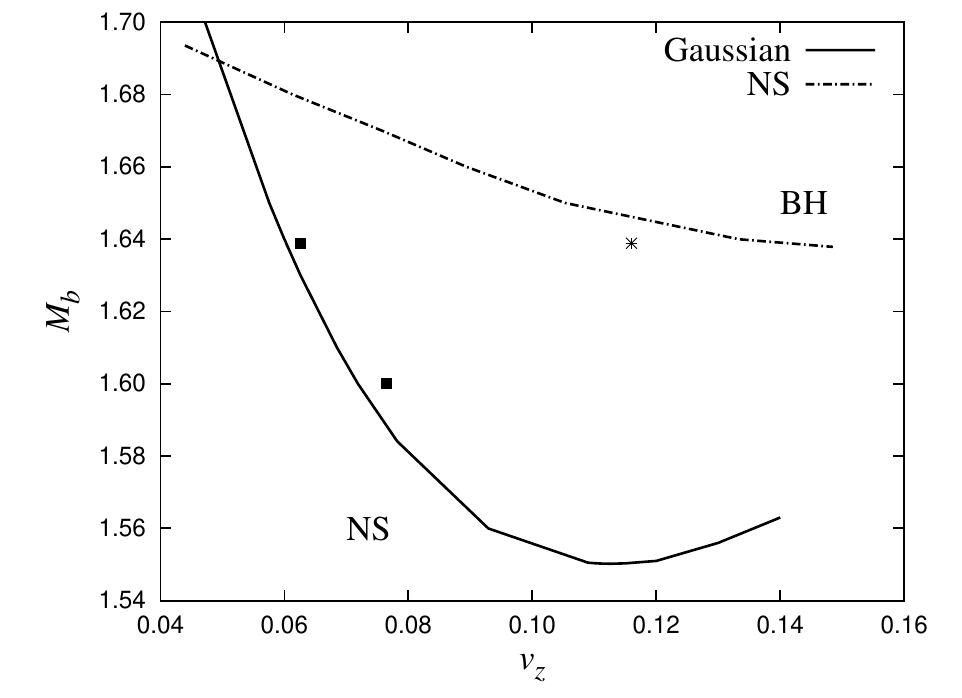}
\caption[Rest mass-boost velocity phase space.]{Comparison between the neutron star and Gaussian packet critical surfaces in the rest mass-boost velocity phase space.}
\label{fig:6g}
\end{center}
\end{figure}
\begin{figure}
\begin{center}
\includegraphics[scale=1.0]{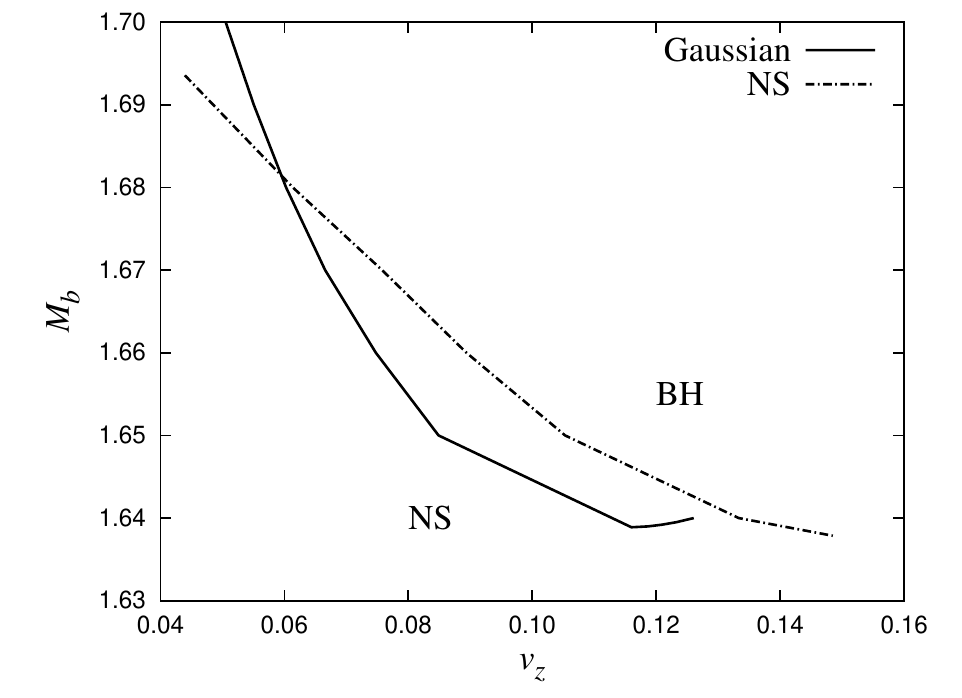}
\caption["Corrected" rest mass-boost velocity phase space.]{Comparison between the neutron star and the "corrected" Gaussian packet critical surfaces in the rest mass-boost velocity phase space.}
\label{fig:6ga}
\end{center}
\end{figure}

\begin{figure}
\begin{center}
\includegraphics[scale=1.0]{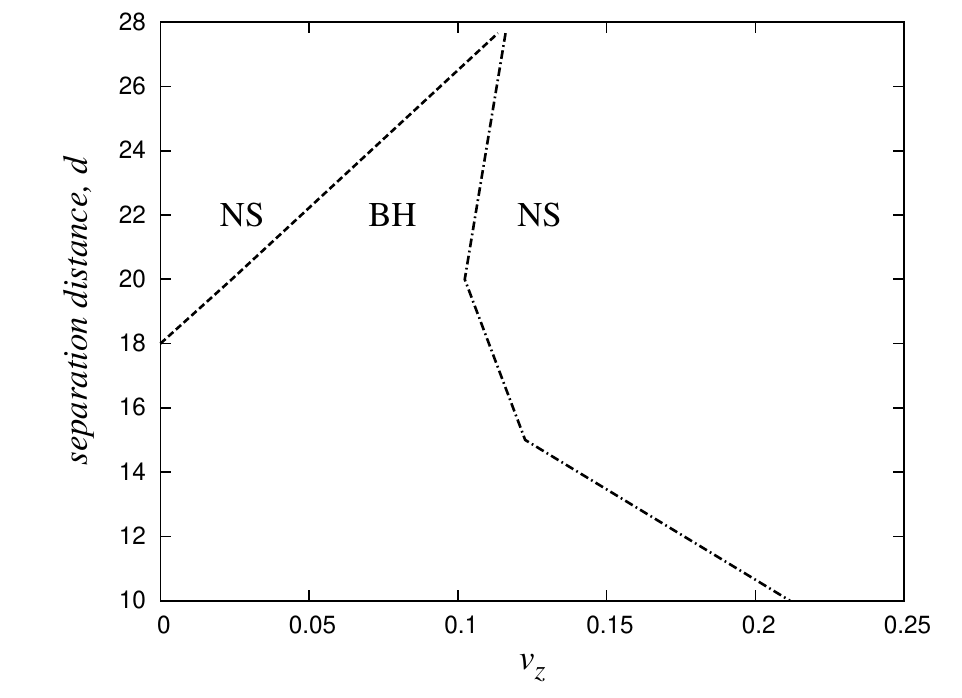}
\caption[Separation distance-boost velocity phase space.]{}
\label{fig:DvV}
\end{center}
\end{figure}

In this section, we probe the structure of the critical solution in the framework of phase spaces. We note that Gaussian initial data sets possess 
additional degrees of freedom in the matter and in the spacetime as compared to the neutron star initial data. Using the additional 
degree of freedom in the matter configuration, we 
vary the boost velocity of the Gaussian packets together with the Gaussian heights and widths while maintaining the rest mass of the system at $M_b=1.6389$. In 
Fig. 6.6, we plot boost velocity with respect to Gaussian height. In this phase space, we use the maximal slicing for the Gaussian packet initial data. We 
observe negligible deviation of the evolutions of these initial data sets with those of the Gaussian packet initial data sets in the previous section that were set with 
geodesic slicing. A main feature of the phase space depicted in Fig. 6.6 is that there is a turning point at approximately 
$(\rho_c,v)=(0.0009,0.12)$. This is an indication that for a certain compactness, the Gaussian packet system crosses two phase thresholds as the boost velocity is increased, 
ie. from a neutron star-like phase to a black hole phase and back to a neutron star-like phase. The cross from the neutron star-like phase to the black hole phase is 
expected as the boost velocity increases entails the increase of gravitational energy. However, we note that the cross from the black hole phase back to the neutron 
star-like phase when the boost velocity is increased further, is a surprising physical behavior.  
 
Given the limitation imposed by the grid size, we are unable to extend the curve further to the left of the phase space. 
An initial data configuration to the left of the \textit{critical~surface} collapses to a black hole and one to the right results in a 
neutron star-like object. 
The extent of this critical surface is indicative of the size of the attraction basin of the 
neutron star critical set with rest mass 1.6389. 
Beyond the right edge of the critical surface at this rest mass, no critical collapse behavior is observed, ie. configurations with Gaussian packets with 
compactness $\rho_c\gtrsim 0.0009$ are not attracted to the critical surface at all. 
To further explore the extent of the attraction basin, we construct another phase space using the boost velocity and the separation distance between the Gaussian packets. 
Fig.~\ref{fig:DvV} shows two critical surfaces of which the left is where the system at a fixed separation distance passes the threshold from the neutron star-like phase into 
the black hole phase and passes back into the neutron star-like phase through the right critical surface.
Both the critical surfaces end at $d/2\sim 14$ at the top of the phase space. 
The bottom extent of the right critical surface is limited when the separation distance between the two Gaussian packets
decreases to the point that the packets merge to become one single packet.    
As the total rest mass of this single packet, ie. $1.6389$, is more than the maximum rest mass of an equilibrium neutron star configuration, ie. $1.6087$ for configurations with $\Gamma=2.0$, the system collapses into a black hole even when there is no implosion velocity.
We thus set the rest mass of the system to be $1.6$, which is below the maximum rest mass. As the maximum point of the rest mass-central density relation for equilibrium star configurations  (Fig. 4.1) is not attracted to the critical set, 
we study the transition of the neutron star critical set attraction basin by increasing the central density of the Gaussian packets so that it approaches the maximum point, ie. $\rho_c\sim 4.0\times 10^{-3}$.
We find that the oscillations of the critical set decrease in amplitude. The average of the oscillations in the central lapse function $N$ shifts up whilst that of the central density shifts down. At $\rho_c\sim 1.833\times 10^{-3}$, the threshold between the black hole and neutron star-like phases disappear. At central densities beyond this value, 
the system oscillates at the normal mode frequencies of its corresponding equilibrium star configuration not only when there is no implosion velocity but at all other values of the implosion velocity up till $0.8$.     
However, we observe that the transition point across $\rho_c\sim 1.833\times 10^{-3}$ itself is characterized by a critical set between the black hole phase and the neutron star-like phase, namely, when the central density is more than this threshold value, the single Gaussian packet settles into an oscillating state about a corresponding equilibrium star configuration, while it collapses into a black hole when the central density is less.   

We also construct a phase space of the boost velocity and the rest mass to compare with 
the critical surface found with that for the neutron star initial data (Fig.~\ref{fig:6g}). 
We note that there is also a turning point behavior in both critical surfaces, similar to that 
found in the boost velocity-Gaussian height phase space above. 
However, another striking feature of Fig.~\ref{fig:6g} is that the Gaussian packet critical surface is shifted to the bottom of the neutron star critical surface 
due to the fact that the spacetime for the Gaussian packet initial data is only approximately asymptotically flat, ie. the asymptotic value of the 3-metric is slightly less 
than 1. This difference in the spacetime causes the rest mass summed up over each hypervolume element of the Gaussian packet initial data, given by:
\begin{equation}
\label{eq:barymass}
M_b=\int\sqrt{\gamma}\rho W d^3 x,
\end{equation}
to be less by a scale with respect to that for the neutron star initial data. 
However, based on the similarity of their evolution trajectories, we emphasize that the critical solution 
represented by the Gaussian packet initial data point with $M_b=1.5505$ is the same as that represented by the neutron star initial data point with $M_b=1.6379$, 
Gaussian $M_b=1.5600$ the same as neutron star $M_b=1.6382$, and so on. This represents a scaling effect in the critical surface of the rest mass-boost velocity phase space 
caused by the difference in spacetime as mentioned earlier in this paragraph. 
We also vary the spacetime by changing the height and width of the metric function in the region occupied by the Gaussian packets.
The black boxes in Fig.~\ref{fig:6g} represent critical solutions found when the metric function is varied in such a manner.

To test universality, we set the spacetime of the Gaussian packet initial data back to asymptotic flatness. 
The Gaussian critical surface in this phase space shifts back up approaching that of the neutron star
as the asymptotic behavior of the Gaussian packet initial data spacetime approaches that of the neutron star initial data (Fig.~\ref{fig:6ga}).
We see a common rest mass range but not a common boost velocity range. The discrepancy in the boost velocity ranges is due to the fact that the Gaussian packet initial data possess different coordinate systems than the neutron star initial data.  
We observe that a critical surface with a common boost velocity range can also be obtained, indicating that the coordinate system of the initial data can be freely adjusted using the additional degrees of freedom in the Gaussian packet configuration. 
In fact, we are able to obtain a family of critical surfaces in the rest mass-boost velocity phase space with a common rest mass range and within the boost velocity range indicated in Fig.~\ref{fig:6f} for the rest mass of $1.6389$, ie. between $0.06$ and $0.21$.  
As rest mass is conserved, 
the overlap in the rest mass range of both critical surfaces confirms that neutron star-like systems evolve toward the same critical sets as the neutron star system.

\begin{figure}
\begin{center}
\includegraphics[scale=1.0]{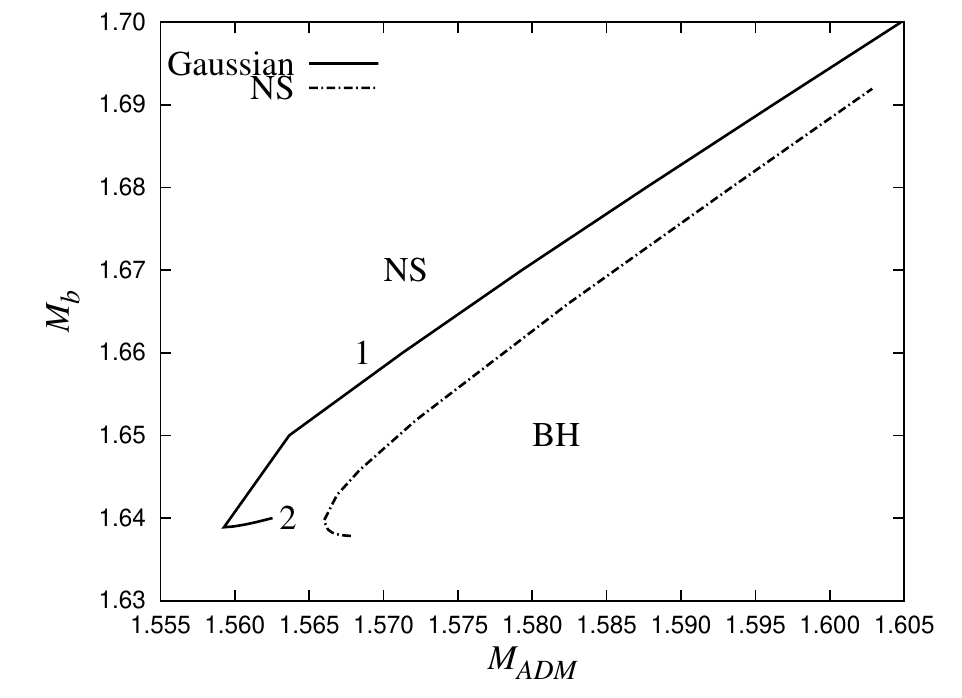}
\caption[Rest mass-ADM mass phase space.]{$1$ and $2$ represent two branches of the critical surface.}
\label{fig:6gT}
\end{center}
\end{figure}

\begin{figure}
\begin{center}
\includegraphics[scale=1.0]{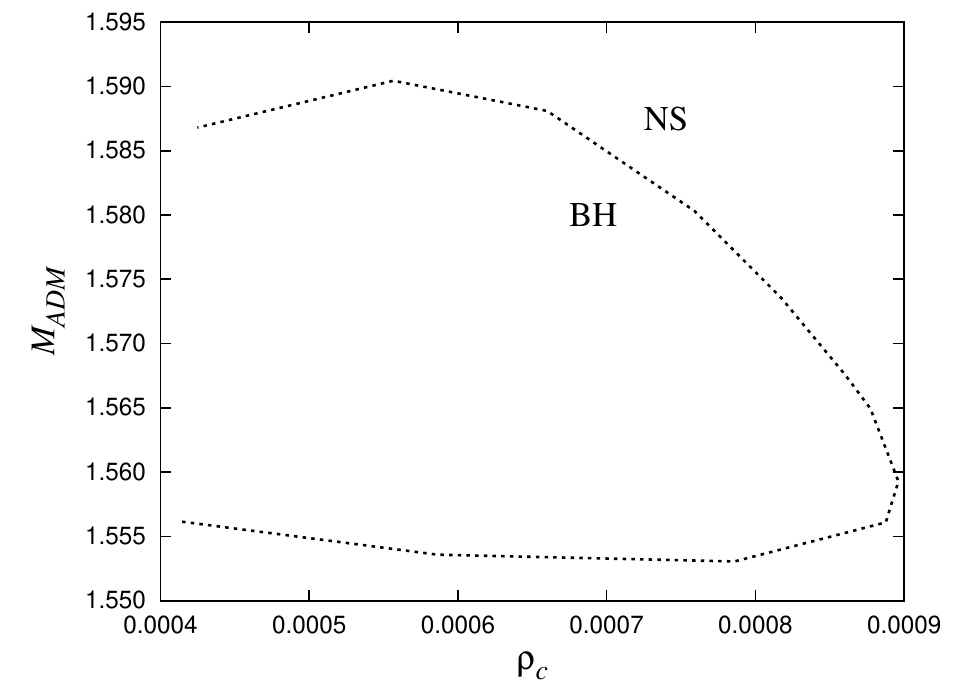}
\caption[ADM mass-central matter density phase space.]{}
\label{fig:6fT}
\end{center}
\end{figure}

Using the ADM mass in replacement of the boost velocity, we find similar critical surfaces with turning points (Fig.~\ref{fig:6gT} and Fig.~\ref{fig:6fT}).
The ADM mass is calculated in our numerical simulations using a volume integral as follows:
\begin{equation}
\label{eq:ADMmass}
M_{ADM}=\int\Psi^5[\rho_e+\frac{1}{16\pi}(\tilde{A}_{ij}\tilde{A}^{ij}-\frac{2}{3}K^2-\tilde{R}\Psi^{-4})]\sqrt{\tilde{\gamma}}d^3 x,
\end{equation}
using the conformal variables that have been defined in Section 3.2 and the matter-energy density as mentioned in Section 4.1.
In line with the definition that is also given in Section 4.1, this mass is the total gravitational energy of a system that is isolated at spatial infinity, 
and requires a quasi-isotropic gauge choice in which $\tilde{\gamma}_{ij,j}\sim \mathcal{O}(r^{-3})$ as $r\rightarrow\infty$ \cite{York79}, a condition that is instantaneously realized by the initial data before it gravitationally radiates, even though in non-equlibrium states. 

In Fig.~\ref{fig:6gT}, we also compare the rest mass-ADM mass critical surface between that of the Gaussian system and that of the neutron star system.
Branch $1$ and branch $2$ of the critical surface in the rest mass-ADM mass phase space represent two different phase thresholds.  
In Section 6.5, we comment on the indices of these critical solutions and their comparisons between the Gaussian system and the neutron star system.

Results of this section thus indicate that the neutron star critical solution is universal with respect to spacetime and matter configuration and reinforces 
the claim in the previous section that the neutron star critical solution is a semi-attractor. We will further support this evidence by analyzing the properties of their 
critical indices in Section 6.5. 

\section{Spacetime and matter properties of the critical solution}

\begin{figure}
\begin{center}
\includegraphics[scale=1.0]{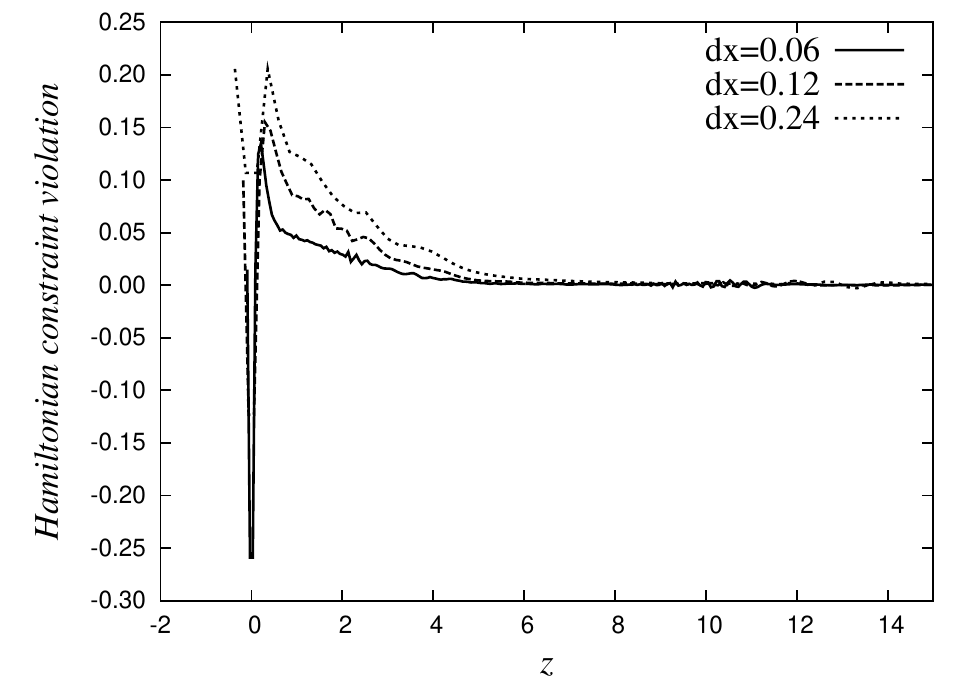}
\caption[Convergence of the Hamiltonian constraint violation for GRAstro-2D.]{}
\label{fig:conv}
\end{center}
\end{figure}  
\begin{figure}
\begin{center}
\includegraphics[scale=0.8]{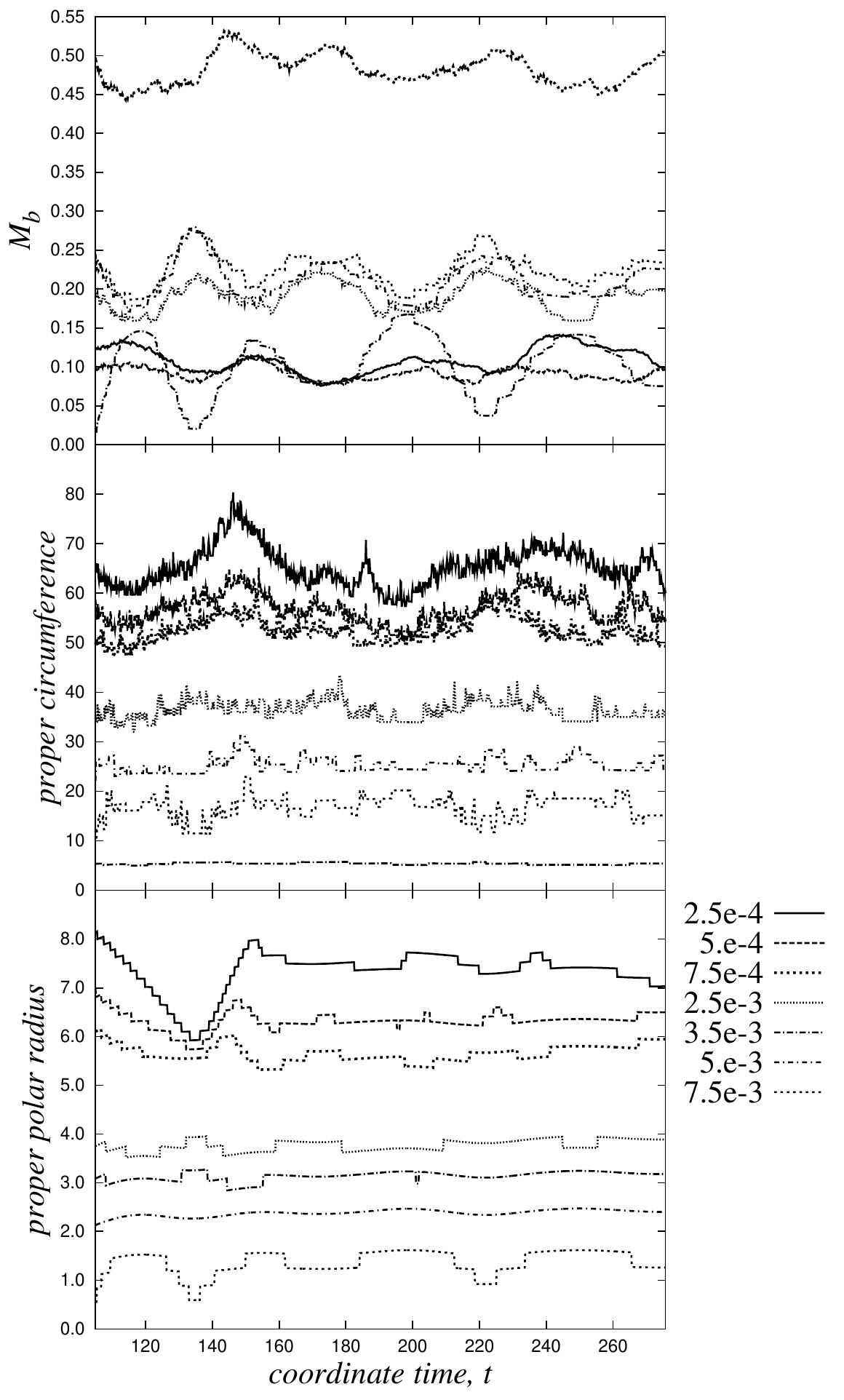}
\caption[Spacetime and matter measures of the neutron star critical solution.]{Right legend shows contour densities at which measures are performed.
Polar and equatorial proper radii shown on the right are measured at the $5.0 \times 10^{-4}$ density contour. The polar
direction is taken to be the x-axis on the xz-plane of the grid and the equatorial direction is taken to be the z-axis on the same
plane. The equatorial direction is thus along the direction of collision of the neutron stars.}
\label{fig:6h}
\end{center}
\end{figure}
\begin{figure}
\begin{center}
\includegraphics[scale=1.0]{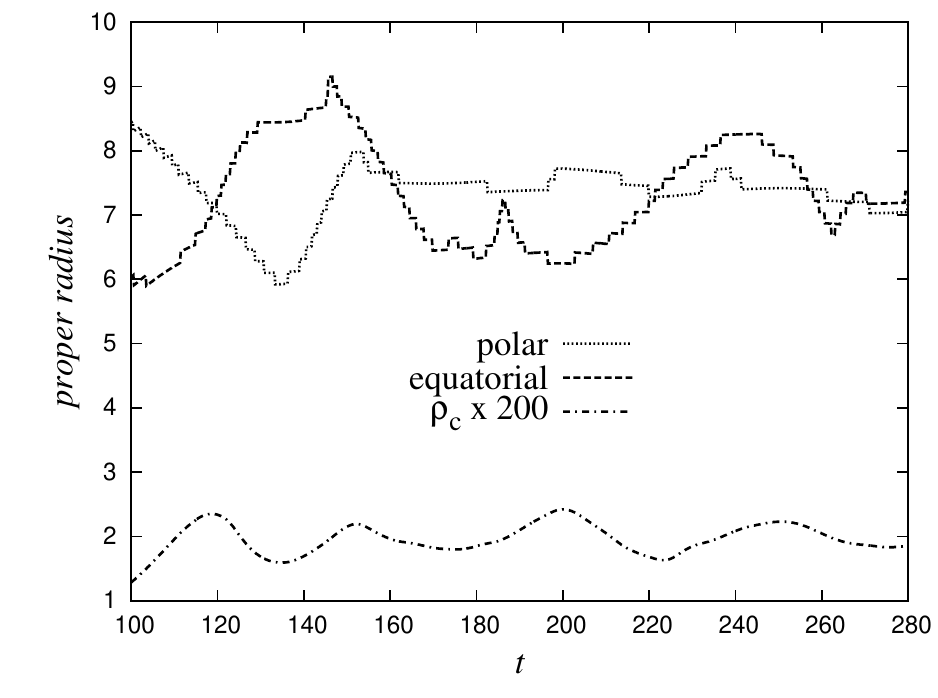}
\caption[Phase difference between the proper radii oscillations for the neutron star critical solution.]{}
\label{fig:6i}
\end{center}
\end{figure}
In this section, we explore the various spacetime and matter properties of the neutron star critical solution. For this purpose, we first ensure that our 
critical collapse simulations are converging with respect to the finite-differencing resolution up till the duration of the critical solution (refer to Appendix A).
Fig.~\ref{fig:conv} shows the first-order convergence in the Hamiltonian constraint violation of our simulations using the GRAstro-2D code at $t=243M_{\odot}$, 
which is well into the fourth oscillation of the matter density and 4-scalar curvature of the critical solution at the center of collision.
\begin{figure}
\begin{center}
\includegraphics[scale=1.0]{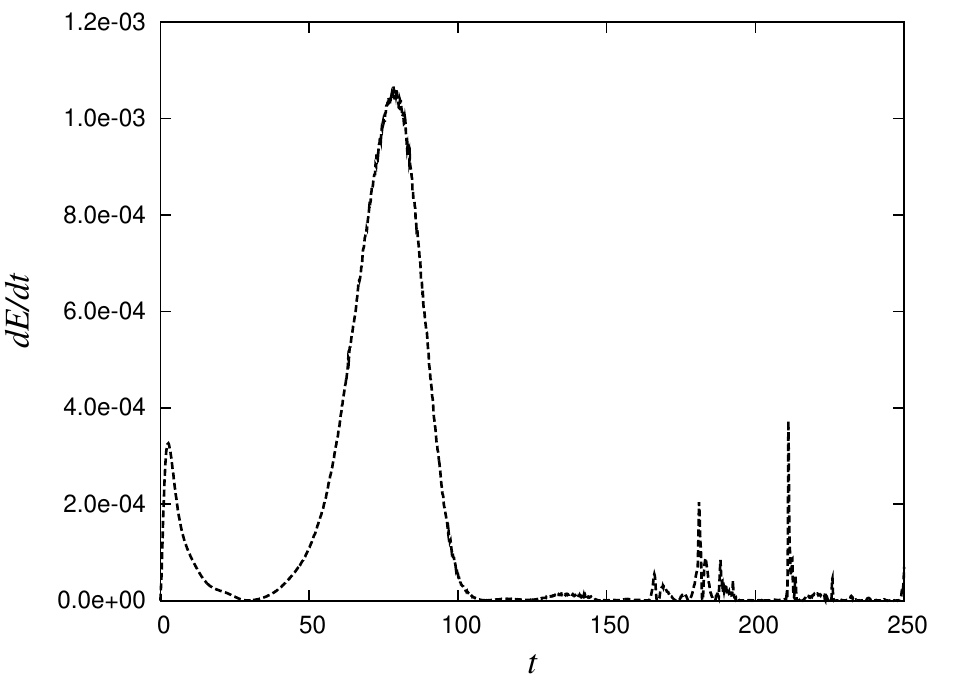}
\caption[Gravitational radiation emission power via the quadrupole formula.]{}
\label{fig:quad}
\end{center}
\end{figure}
The anomaly at $z=0$ is an expected effect caused by the truncation in the finite-differencing at the center of symmetry which has been shifted (to $z=0.03$ for $dx=0.06$, 
$z=0.06$ for $dx=0.12$, and so on) by a staggered grid.
We then measure the baryonic mass enclosed within different density contours throughout
the merged object, the proper polar circumferences of these density contours, and their equatorial and proper radii, and observe their evolution 
throughout the critical set oscillation phase. The measure of the baryonic mass enclosed between different density contours
throughout the merged object has the advantage of being a fully-geometric measure. In Fig.~\ref{fig:6h}, we observe that this measure exhibits 
oscillations that are correlated between different regions of the merged object, whereas the corresponding proper polar circumferences 
do not exhibit such well-defined oscillations throughout the critical set dynamics, ie. on the time scale of $t\sim 40$. 
The well-defined oscillations in the baryonic masses summed up between the density contours thus indicate the existence of 
matter fluxes or a slushing back and forth of the matter through the different density contours inside the object 
and are not due to any correlated changes in contour circumferences.
The proper radii however exhibit prolonged oscillations only within the region bounded by a
specific density contour. Fig.~\ref{fig:6i} shows the oscillations of these radii outside this region exhibiting damping into a stationary state. These
oscillations are correlated between the polar and equatorial directions ie. the polar proper radius is in
phase with the oscillation of the central density while the equatorial radius is out of phase. However, we note that the proper radii oscillations presented here have not 
eliminated the effect of the lapse at the outer limit of the proper radii integration. A good method to isolate this effect is to set a diagnostic spacetime with the normal 
coordinate choice and where the 11-component of the 3-metric is set to 1. With such a spacetime, the hypersurface will always be perpendicular to the timelike normal unit 
vector $\mathbf{n}$ and the proper radius oscillation will reflect the real oscillation of the density contour of the neutron star \cite{Gundlachpr}.
In order to gauge the stationarity of the merged object, we also measure the power of gravitational radiation emission throughout the dynamical process leading to and that of the critical set itself. Fig.~\ref{fig:quad} shows the power that is calculated using the quadrupole formula (\cite{Shibata03},\cite{Lin06}), 
namely:
\begin{equation}
\label{eq:quadeq}
\frac{dE}{dt}=\frac{1}{5}Q^{(3)}_{ij}Q^{(3)}_{ij},
\end{equation}
where $Q_{ij}=\int\rho(x^i x^j-\frac{1}{3}\delta^{ij}r^2)d^3 x$.  
From this figure we see that the energy radiated gradually decreases as the system approaches the critical set at $t\sim 110$,
which is the approximate time of merge of the two neutron stars.
The power remains at values very close to zero until $t\sim 165$ when it jumps up to positive values that increase with time.
We note that these sharp jumps may be due to noise created by finite-differencing the time derivatives of $Q_{ij}$ \cite{Lin06}.
The average of the power throughout the critical set dynamics remains very close to zero.
This suggests that the oscillations of the critical set may not damp out with time thus pointing to the possibility that the critical set is a limit cycle 
rather than a limit point. Also, it shows that at $t\sim 110$, when the oscillation phase of the critical solution starts, the system has already settled down to a configuration that is stationary enough and therefore very near to an equilibrium state. 

\section{Properties of the critical index}

As mentioned in Section 6.3, this section will contain further elaboration of the properties of the critical indices of the different Gaussian packet initial data 
configurations as compared to that of the neutron star. 

\begin{figure}
\begin{center}
\includegraphics[scale=1.0]{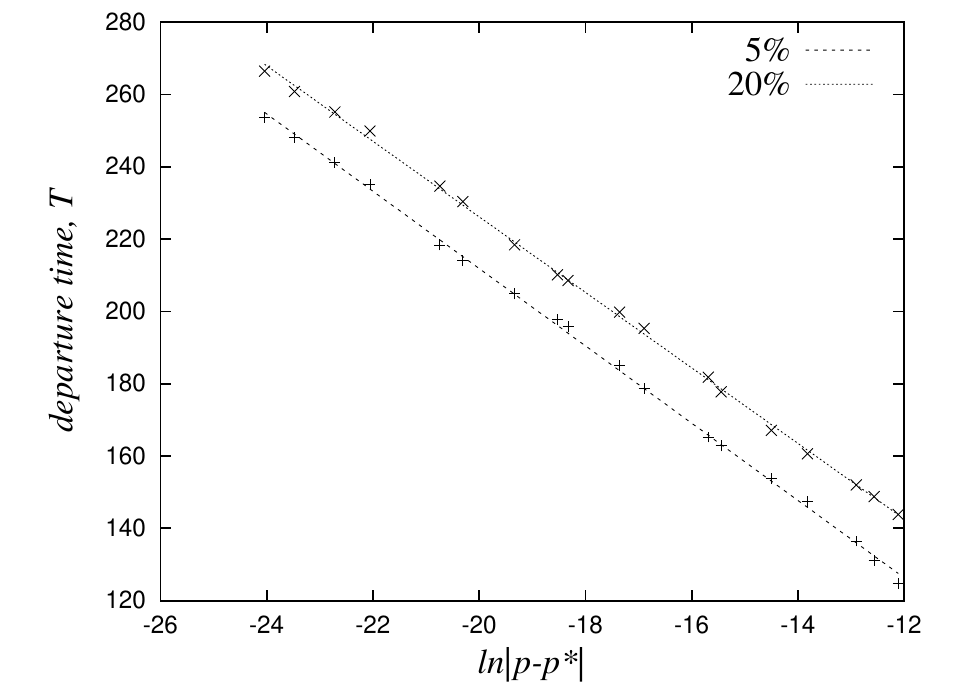}
\caption[Sample critical index extraction.]{}
\label{fig:crez5h}
\end{center}
\end{figure}

\begin{figure}
\begin{center}
\includegraphics[scale=1.0]{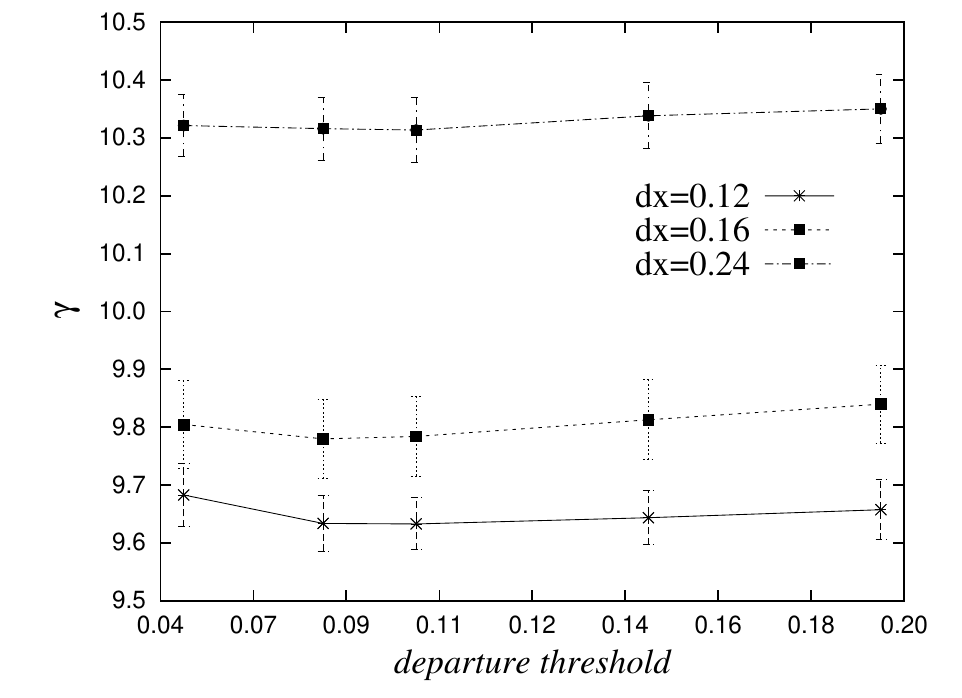}
\caption[Convergence of the critical index with respect to departure threshold.]{}
\label{fig:6k}
\end{center}
\end{figure}

\begin{figure}
\begin{center}
\includegraphics[scale=1.0]{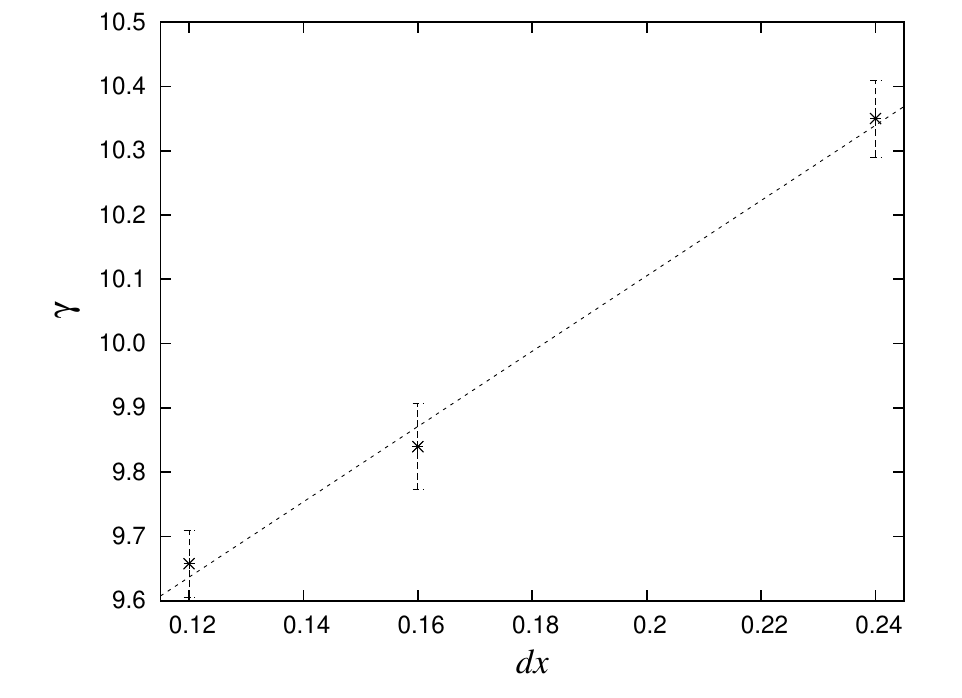}
\caption[Convergence of the critical index with respect to grid resolution at departure threshold 0.2.]{}
\label{fig:6m}
\end{center}
\end{figure}

\begin{figure}
\begin{center}
\includegraphics[scale=1.0]{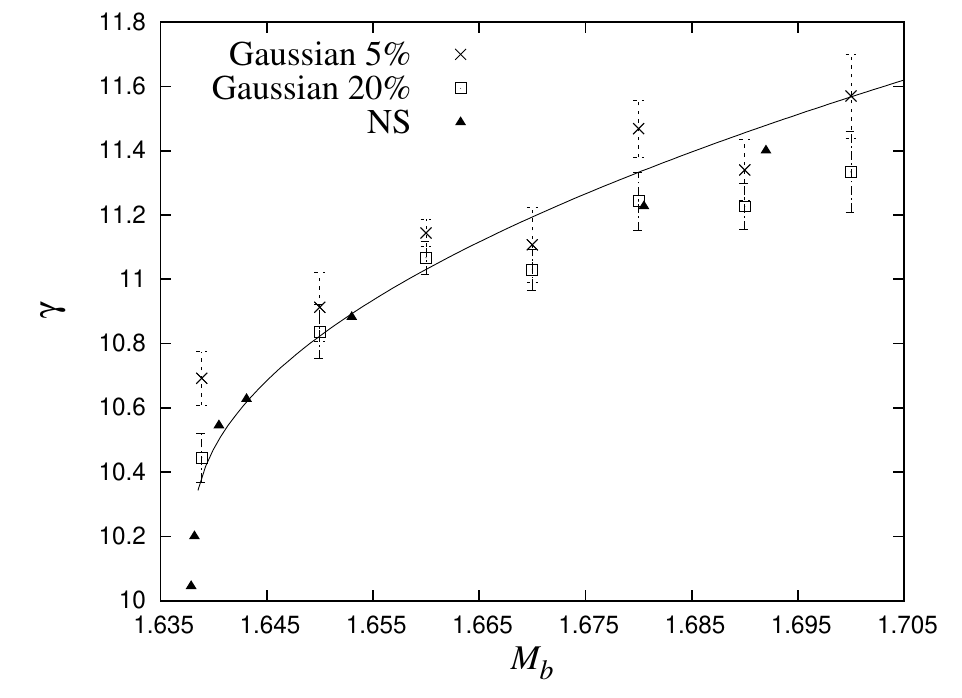}
\caption[Comparison of critical indices for Gaussian packet and neutron star initial data.]{}
\label{fig:6jT}
\end{center}
\end{figure}

\begin{figure}
\begin{center}
\includegraphics[scale=1.0]{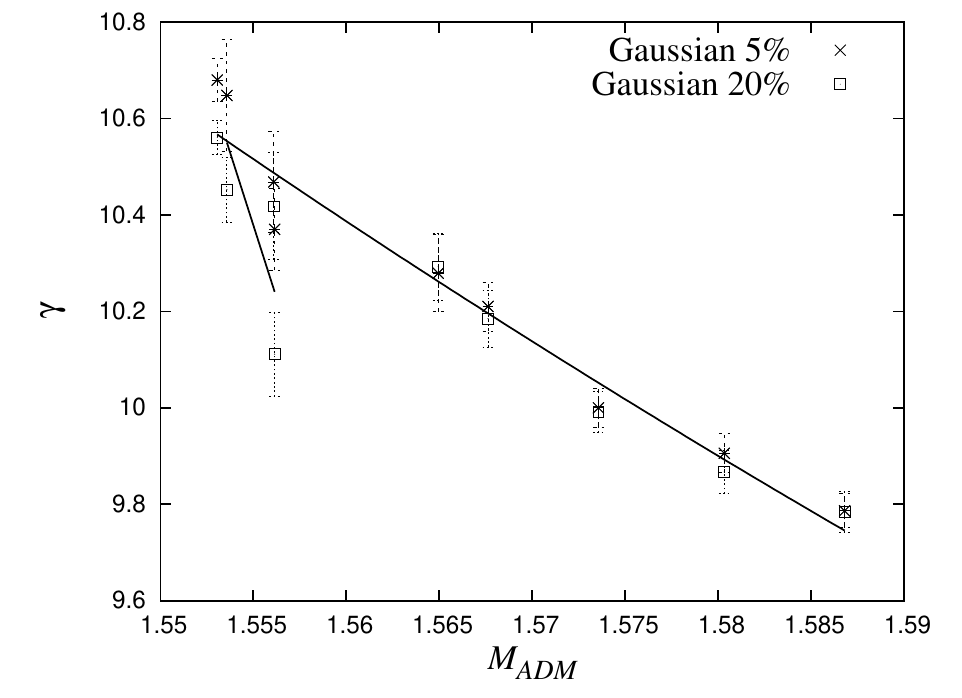}
\caption[Variation of critical index with respect to ADM mass for rest mass 1.6389.]{}
\label{fig:GAvV}
\end{center}
\end{figure}

Analogous to Eq. (5.6), the critical index is calculated using two departure thresholds $5\%,20\%$,
where the departure threshold is taken as the time when the percent difference between the evolution variable of a solution leaving the
critical set and that of the critical solution reaches a certain value \cite{Jin07}. In this case, we take the evolution of the lapse at the center of collision.
We denote the departure thresholds as $T_{0.05},T_{0.2}$ and so on according to
the percent difference used. For each critical point, we find that $T_p$ is linearly proportionate to $\ln|p-p*|$, where $p$ is any initial data parameter that we choose to vary 
while fixing all others. In the case of Fig.~\ref{fig:crez5h}, the initial data parameter varied is the boost velocity of the Gaussian packets.  
Straight best fit lines are therefore obtained and the standard deviations
of their slopes are plotted as errorbars in this figure. We note that this errorbar only represents one type of error,
ie. the error affecting how the evolution trajectories follow the power-law behavior characteristic of critical and near-critical solutions.
Other errors that affect the numerical results include the error caused by the finite differencing scheme and
the error affecting how the evolution trajectories leave the critical set in an exponential manner.

We check the convergence of the critical index with respect to the finite-differencing resolution $dx$. Fig.~\ref{fig:6k} shows the critical indices for different departure thresholds at three different resolutions for a sample Gaussian system. At the departure threshold of 0.2, Fig.~\ref{fig:6m} shows an approximate first order convergence with respect to the resolution.
The grid boundary used for the simulations with resolution $dx=0.16$, namely, $41.0$, is slightly larger than that with $dx=0.12$ and $dx=0.24$, namely $38.5$, due to the restrictions on the number of grid points imposed by the multigrid initial value problem solver, which accounts for the slightly larger than first order convergence demonstrated by the $dx=0.16$ point in this figure.

Fig.~\ref{fig:6jT} shows the rest mass dependence of the Gaussian packet critical index compared with that of the neutron star, 
where we observe that the critical index for both the Gaussian packet and neutron star initial data increases with rest mass of the system. 
This indicates that 
the time scale of departure of a near-critical solution from the critical set decreases as the rest mass of the system increases.
The overlapping of the critical indices of both the Gaussian and the neutron star systems shows that the neutron star critical set is universal with respect to different initial data, where for the same rest mass and within a small interval of ADM masses, the Gaussian initial data produces critical indices that are within the errorbars of those produced by the neutron star initial data. 
The universality of the neutron star critical index with respect to these different initial data configurations is a crucial point in establishing the quality of the neutron star critical set as a semi-attractor.

To answer the question of the parameter dependence of the critical index, we consider the critical solutions represented by each point 
in the ADM mass-Gaussian height phase space and study how the critical indices change along this critical surface. 
Fig.~\ref{fig:GAvV} shows the variation of critical index along this critical surface 
with respect to the ADM mass fixing the rest mass at $1.6389$. The solid line is a fit of the average of the indices obtained at departure thresholds $0.05$ and $0.2$. 
We note that the first order convergence of these critical indices with respect to resolution holds for different departure threshold regimes for configurations with 
different ADM masses. This is due to the fact that for configurations with higher compactness, and hence lower ADM mass, the grid boundary is further away from the physical system than for configurations with lower compactness, or higher ADM mass. Convergence is better achieved when the grid boundary is further away from the physical system as there is 
less interference in the spacetime due to gravitational radiation.  

In Fig.~\ref{fig:6gT}, we take the rest mass of $1.65$ and extract the critical index for the critical points on both branches $1$ and $2$ of the critical surface.
In accordance with Fig.~\ref{fig:GAvV}, we also find a discrepancy between these critical indices that goes beyond their errorbars, indicating that
critical points on branches $1$ and $2$ of the critical surface represent two different phase thresholds respectively.
On branch $1$ of the critical surface with this rest mass, the critical index obtained is $10.9\pm 0.1$ at departure threshold $0.05$ and $10.84\pm 0.08$ 
at departure threshold $0.2$, whilst on branch $2$,
the critical index is found to be $10.26\pm 0.06$ at departure threshold $0.05$ and $10.24\pm 0.07$ at departure threshold $0.2$.

As mentioned in Section 6.3, the ADM mass of the initial data is evaluated using Eq.~\eqref{eq:ADMmass}. 
The constancy of the ADM mass guarantees that Einstein's field equations for the system are satisfied for all time \cite{Gourgoulhon07}.
Therefore, it is by definition a conserved quantity. However, when evaluated at a finite boundary for a system that gravitationally radiates,
it is seen to decrease with time as gravitational energy is carried out of the finite computational grid by gravitational radiation.
The mass that is evaluated using Eq.~\eqref{eq:ADMmass}
in such a scenario does not have strict physical meaning although it can be used in an approximate conservation equation primarily to monitor the accuracy of $3+1$ numerical simulations \cite{Shibata03a}. The fact that the quantity lacks a strict physical meaning is because the hypersurface is not spatially isolated at a time after gravitational radiation.
This is evident in that no matter how far one pushes the computational boundary spatially outward, one would encounter the gravitational radiation that is propagating to null infinity.
In other words, the ADM mass measurement does not by construction separate the gravitational energy of the physical system that we are interested in 
and that of gravitational radiation.
Even for Eq.~\eqref{eq:ADMmass} to hold approximately in the computational grid,
the boundary of the computational grid has to be asymptotically flat and thus free from the presence
of any gravitational radiation, the ambient spacetime has to be fully stationary and fulfil the right gauge conditions as mentioned earlier.
Therefore, in order to employ this equation even in an approximate sense to evaluate the gravitational energy of a physical system after gravitational radiation leaves it into the ambient spacetime, one has to do the measurement at a very late time of the critical set dynamics when one is sure that gravitational radiation is no longer present inside the computational grid and is no longer interacting with the computational grid boundary in any way. 
However, none of the initial data in numerical simulations can be set directly on the critical threshold in principle, implying that
they will either evolve into a neutron star-like object or collapse into a black hole at one point in time.
Also, due to the accumulation of numerical errors, in practice, we are not able to accurately simulate the critical set dynamics until very late time. Intrinsically as well, even though by definition stationary and
an initial data that dwells directly on the threshold will inevitably evolve into it, the critical set itself is not a TOV equilibrium configuration. A measurement that is based on a strict physical meaning however is the Bondi mass measurement which evaluates the gravitational mass of the system isolated at null infinity. This measurement puts any gravitational radiation to the past of the null surface of the evolving physical system as the radiation itself propagates to null infinity.

Considering the question of the parameter dependence of the critical index on the above-mentioned conserved geometric quantities,
we are basically asking a question of how many classes of neutron star semi-attractors exist according to the rest mass and ADM mass.
A natural approach to this question that we have employed is to vary the rest mass and when the rest mass is fixed, to vary the ADM mass of the initial data or the intrinsic gravitational energy of the physical system minus the energy that will be radiated away during its evolution to the critical set.
We recall that the neutron star-like initial data that we have constructed possess additional degrees of freedom in comparison to the neutron star initial data,
that enables us to fix the rest mass and vary their intrinsic gravitational energies. This intrinsic gravitational energy consists of the binding energy inherent in the Gaussian packet system and in principle will not be radiated away unless there is a physical mechanism that diffuses this energy, eg. neutrino emissions,
which is absent in our numerical simulations.
Therefore, when we vary the compactness of the Gaussian packets, we are intrinsically providing different gravitational energies to these systems which will be conserved in time.
This is evidenced in the fact that when we fix the boost velocity of the initial data sets and vary only the height of the Gaussian matter distribution, and correspondingly the compactness of the matter packets, by $\Delta\rho_c\sim 2\times 10^{-4}$, we see a change in the ADM mass by $~0.01$. This change is free from the factor of the mechanical energy inherent in the boost velocity that can be radiated away by the gravitational radiation - a physical system does indeed radiate more when it is moving faster.
If they are radiated away by some physical mechanism which we can choose to impose on our numerical simulations, or by numerical viscosity, and if we assume that the gravitational energies of the remnants which in our case is the semi-attractors, are all the same, our entire investigation of whether there exists a 2-parameter dependence of the critical index becomes a meaningless endeavor, as the motivation of the question depends on the premise that we are indeed able to manipulate the additional degrees of freedom in our initial data sets to vary the compactness and correspondingly the intrinsic gravitational energies of the physical system. In other words, the premise of our question in the first place is that the semi-attractors themselves \textit{can} have differing gravitational energies for the same rest mass.
Also, the critical index itself is a geometric quantity and is evaluated during the semi-attractor dynamics.
The observation that they indeed differ even when we adopt the above assumption that the gravitational energies of the semi-attractors are all the same, would result in the observation that semi-attractors with the same gravitational energies have differing critical indices, a confounding observation that not only is unlikely, but throws this entire question of 2-parameter dependence meaningless.
Considering realistic astrophysical systems where dissipative physical mechanisms which depletes the binding energies of the packets of matter do indeed occur,
it is still very unlikely that a fine-tuned balance exists for the gravitational dynamics, where the gravitational energies are all depleted by nonlinear physical mechanisms in just such a way that they all end up the same for all the remnants.
We cannot rule out such a scenario, which would be an interesting physical property for neutron star semi-attractors, but such a scenario would still violate the premise of the 2-parameter dependence question as mentioned above.
In addition, we have to recall that the neutron star semi-attractor is \textit{not} a TOV equilibrium configuration, which does not allow us to draw strict analogies with stable equilibrium TOV configurations which has a unique ADM mass for each of its rest masses.

\begin{figure}
\begin{center}
\includegraphics[scale=1.0]{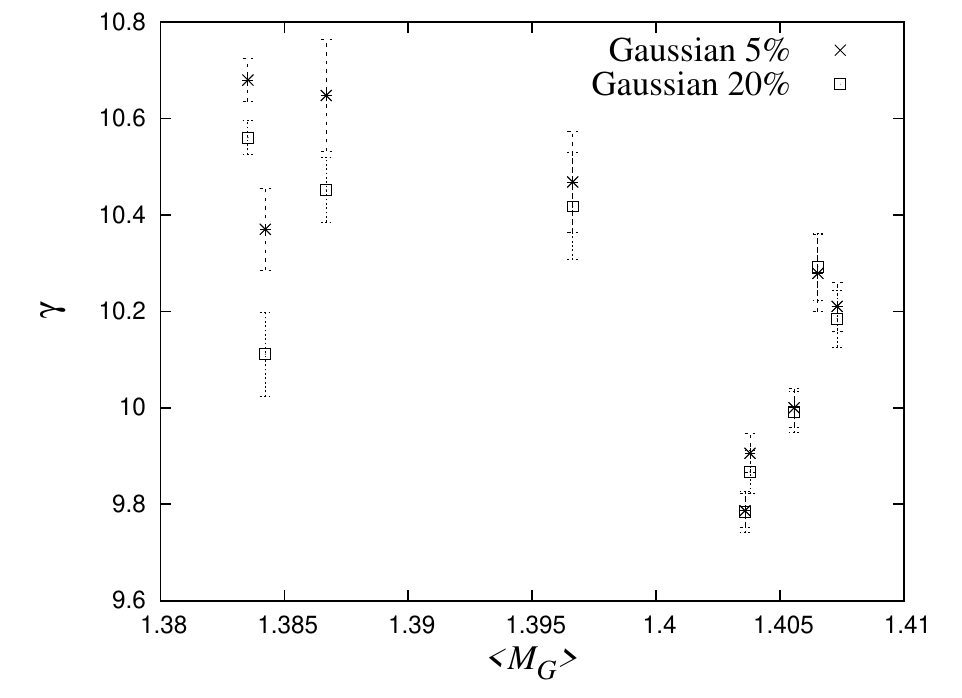}
\caption[Variation of critical index with respect to $\langle M_G\rangle$ for rest mass 1.6389.]{}
\label{fig:GAvVn}
\end{center}
\end{figure}

\begin{figure}
\begin{center}
\includegraphics[scale=1.0]{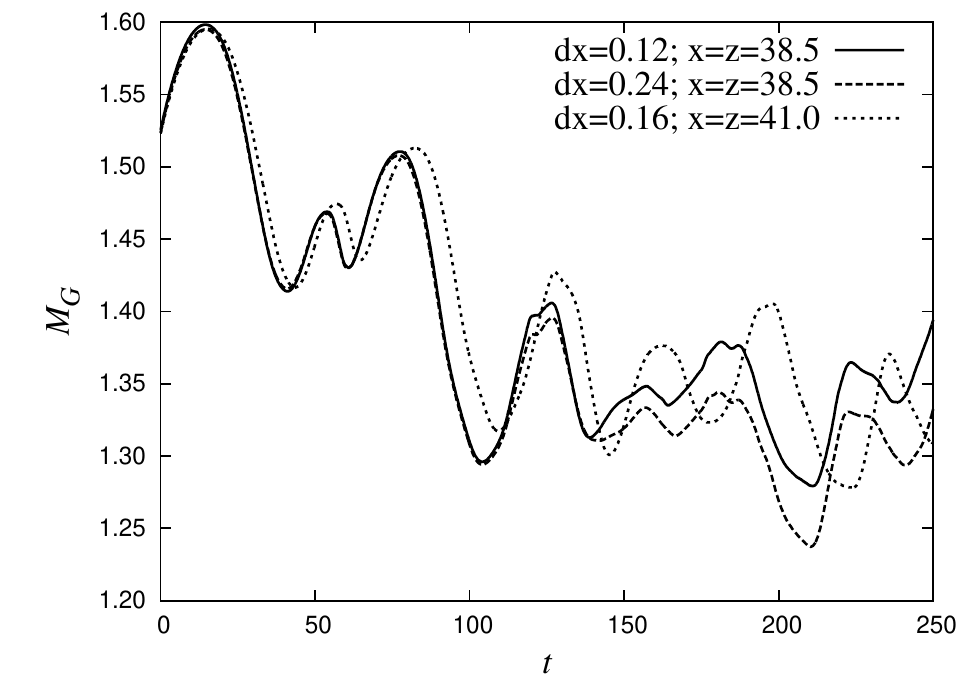}\caption[Evolution of $M_G$ using differing resolutions and boundary locations.]{}
\label{fig:ADMevol}
\end{center}\end{figure}

Fig.~\ref{fig:GAvVn} shows the dependence of the critical indices with respect to the averages of the gravitational masses, $\langle M_G\rangle$, measured by Eq.~\eqref{eq:ADMmass} on the finite computational grid over the period of two oscillations in the lapse function at the center of collision of each configuration along the critical surface in Fig.~\ref{fig:6fT}.
Fig.~\ref{fig:ADMevol} shows the evolution of this gravitational mass calculated using Eq.~\eqref{eq:ADMmass} on the finite computational grid, which we denote as just $M_G$, for a sample Gaussian packet system that evolves toward the critical set and dwells there until $t\sim 240$ for two differing resolutions and differing grid sizes. 
The observation that the time shift in the curves for the simulation with the boundary at $x=z=41.0$ from the one at $x=z=38.5$, namely $t\sim 3$ measured at the first dips of the curves, which is comparable to the spatial shift using $c=1$, and that the oscillations do not converge away with resolution,
shows that the oscillations in $M_G$ is caused by the interaction of gravitational radiation with the computational grid boundary.
As mentioned above, this renders the measurement itself not only numerically inaccurate but without even an approximate physical meaning.
Nonetheless, we still see a dependence of the critical index with respect to $\langle M_G\rangle$. As mentioned above, the fact that the critical indices do vary when we vary the compactness of the Gaussian packet systems,
will not change whether we plot them with respect to either the ADM masses or with $\langle M_G\rangle$, unless all the $\langle M_G\rangle$ values
- assuming that they are even accurate and possess an approximate physical meaning - end up being the same up to numerical error,
which is both physically unlikely and a violation of the premise of the question that we sought to answer.
Therefore, even though we do not rule out the possibility that the premise of our question is unfounded, ie. that indeed, neutron star semi-attractors \textit{cannot} in principle, even though they are \textit{not} in a TOV equilibrium state, intrinsically possess differing gravitational masses for the same rest mass, we conjecture from our observations (Fig.~\ref{fig:GAvV} and Fig.~\ref{fig:GAvVn}) that there is still a high likelihood that there exist classes of semi-attractors that can be labeled by both the rest mass and the gravitational mass of the neutron star-like system, due to the fact it would be a very rare although interesting occurence that there exists a fine-tuned balance in the gravitational dynamics such that gravitational energies of physical systems with intrinsically different binding energies are dissipated by nonlinear physical mechanisms in the matter to end up at the same value.
In addition, that there exist such classes of semi-attractors is also in line with our observation of two distinct phase thresholds for the critical surface in Fig.~\ref{fig:6gT} at the same rest mass, where the distinct phase thresholds represent two differing semi-attractors with the same rest mass but with differing ADM masses. 

\begin{figure}
\begin{center}
\includegraphics[scale=1.0]{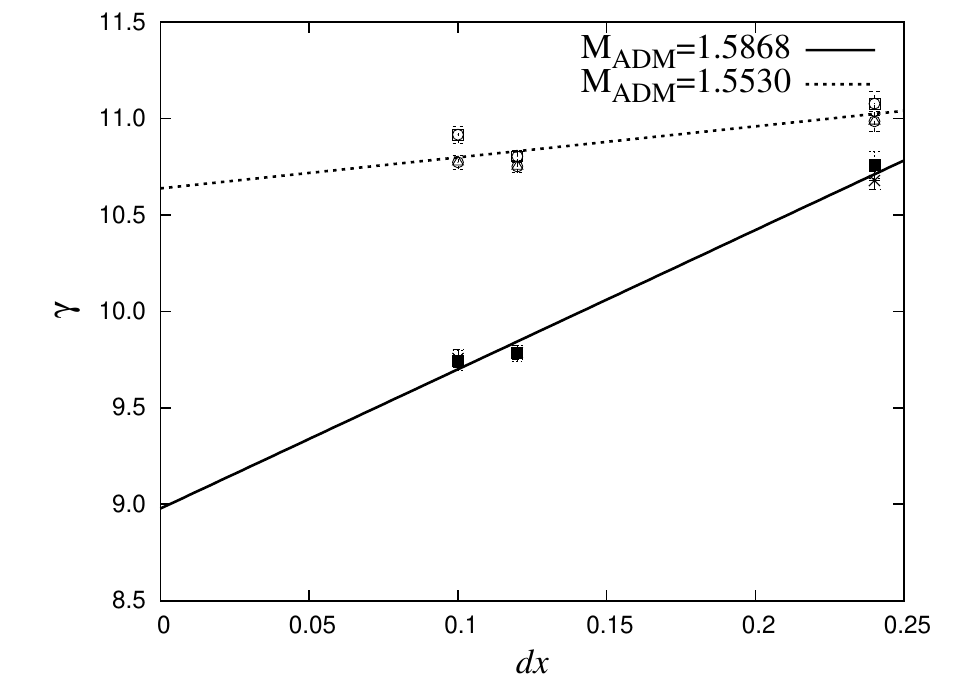}\caption[Convergence of critical indices for two configurations with differing compactness on the ADM mass-Gaussian height phase space.]{}
\label{fig:convGA}
\end{center}\end{figure}

\begin{figure}
\begin{center}
\includegraphics[scale=1.0]{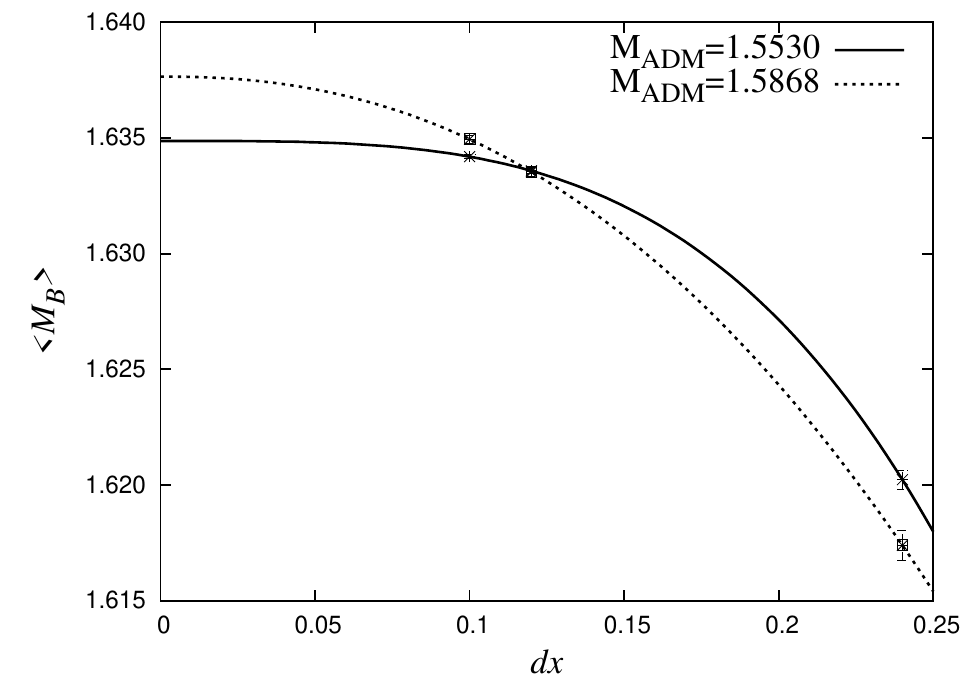}\caption[Convergence of $\langle M_B\rangle$ for two configurations with differing compactness on the ADM mass-Gaussian height phase space.]{}
\label{fig:convMb}
\end{center}\end{figure}

\begin{figure}
\begin{center}
\includegraphics[scale=1.0]{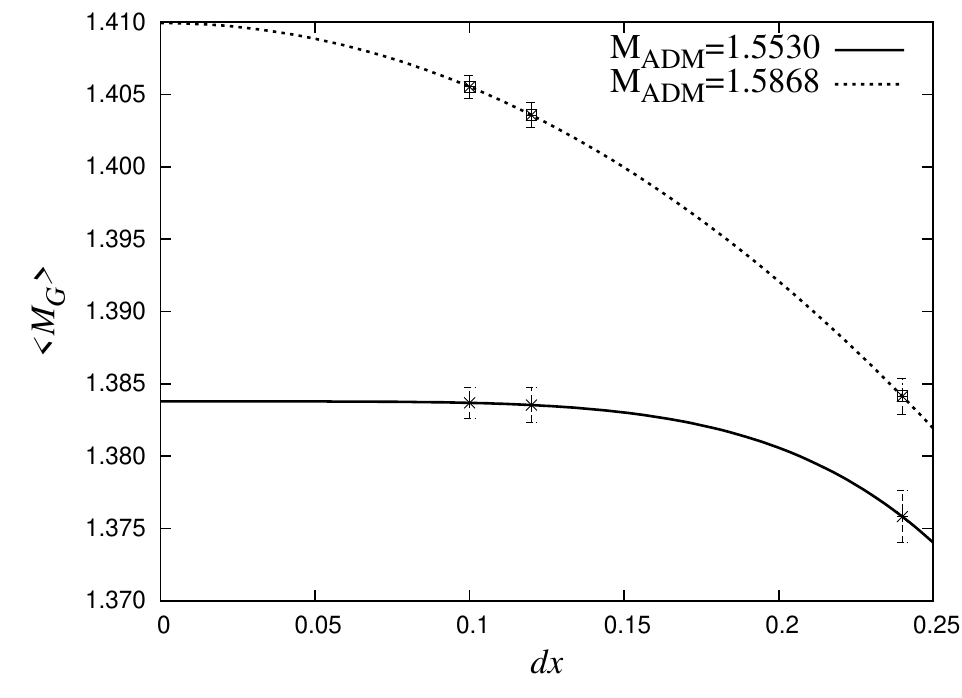}\caption[Convergence of $\langle M_G\rangle$ for two configurations with differing compactness on the ADM mass-Gaussian height phase space.]{}
\label{fig:convMa}
\end{center}\end{figure}

Fig.~\ref{fig:convGA} shows the convergence of the critical indices for two configurations on the ADM mass-Gaussian height phase space, ie. the configurations with $M_{ADM}=1.553$ and $M_{ADM}=1.5868$ respectively, whereas Fig.~\ref{fig:convMb} shows the convergence of the rest mass averaged over the period of the first two oscillations of the lapse function at the center of collision, which we will denote as $\langle M_B\rangle$, and Fig.~\ref{fig:convMa} the convergence of $\langle M_G\rangle$. 
We see that $\langle M_B\rangle$ and $\langle M_G\rangle$ converge on a higher order for the configuration with higher compactness and vice versa, ie. for the configuration with higher compactness, the $\langle M_B\rangle$ and $\langle M_G\rangle$ converge in a fourth order manner and fifth order manner respectively, but
for the less compact configuration, they both converge in a second order manner. 
However, the $\langle M_B\rangle$ for both configurations converge to $1.64$ up to the numerical error caused by the finite differencing, whilst their $\langle M_G\rangle$ values converge to different values on the order of $0.01$ respectively, ie. $1.38$ for the more compact configuration and $1.41$ for the less compact configuration, outside the error bars from the finite differencing.     
The convergence of the $\langle M_B\rangle$ to the same value is expected due to the fact that the rest mass of the initial data have been set to the same up to the eleventh decimal, and the fact that rest mass is a conserved quantity even when evaluated on a finite computational grid.
The variations in the twelfth and above decimals of the rest masses of the different initial data or in their $\langle M_B\rangle$ values in the limit of infinite resolution are completely random.
As for the critical indices, similar to $\langle M_G\rangle$, their values for both configurations are observed to converge to different values, where the more compact configuration exhibits first order convergence at a faster rate than the one for the less compact configuration, which is to be expected due to the different distances of the computational grid boundaries from the physical systems as mentioned previously.
These observations imply that in the limit of infinite resolution, where a large portion of numerical errors is eliminated, even though putting aside the ambiguous physical meaning of $\langle M_G\rangle$, the critical index depends on both the compactness, with its corresponding binding energy present in the physical system, as well as the rest mass of the system, which is in line with the conjecture we make that is mentioned earlier in this section.      
Since we have seen that convergence of the critical index is achieved at an even faster rate when the computational grid boundary is located further away from the physical system, when the boundary of the computational grid is pushed further out for each configuration, we expect to obtain similar observations that point to the above-mentioned conjecture.

   \chapter{Conclusion}

\section{Conclusions}

In this chapter, we review the results obtained in this dissertation in accordance with the objectives we had in mind.
In order to understand the time scales involved in non-rotating neutron star critical collapses, 
we first study the time scales of non-radiative pulsation modes of single static neutron stars.
We find that these time scales are an order of magnitude higher than the time scales involved in the non-rotating neutron star critical collapses observed
in simulations. We thus confirm that the pulsations exhibited by the quasi-stationary object during the critical collapse dynamics are not that of non-radiative
pulsation modes of single static neutron stars. This points to the fact that the quasi-stationary object is not in a state of TOV equilibrium. 

Via curve fitting and Fourier transform, we determine the time scale of attraction of the neutron star system toward the neutron star critical set.
This time scale is an order of magnitude smaller than that reported of neutron star systems undergoing slow cooling. We thus conjecture that dynamical neutron star systems
with significant amount of kinetic energy undergoing slow cooling via formation processes and radiation transport or systems undergoing slow accretion and 
a gradual loss of angular momentum have a possibility of undergoing critical collapse.

By constructing new neutron star-like initial data not restricted by the TOV equations and studying the behavior of their evolutions, 
we confirm that the neutron star critical set is a semi-attractor. We employ the freedom in the
gauge choice and the matter configuration to construct different such neutron star-like initial data
and by measuring their spacetime and matter 'flows', we observe that these initial data evolve towards the neutron star critical set. 
The critical indices extracted from these evolutions are the same up to numerical error as 
that obtained from neutron star simulations with the same rest mass and ADM mass. 
We thus reinforce the claim that the neutron star critical set is universal with respect to
initial data, going beyond the perturbative regime of the critical set. 

We further study the properties of the neutron star critical set dynamics itself. 
We observe that the oscillations of matter contained within different density thresholds 
of the object undergoing critical collapse exhibit phase differences 
across a certain density threshold, indicating the existence of 
matter fluxes across a certain density threshold within the object.
We also find that there exists an outer envelope within the object  
that damps out at a rate higher than the central region. 
In addition, the proper circumferences of different density thresholds within object are nearly constant.
To further shed light on whether the semi-attractor is a limit point or a limit cycle, 
we measure the power of the gravitational radiation emission throughout the critical collapse dynamics.
It is observed that the average of this power during the critical collapse process is very close to zero, 
suggesting that the oscillations of the critical collapse object may not damp out with time and thus that the semi-attractor
is a limit cycle. Combined with the possibility of realistic neutron star systems undergoing critical collapse based on the time scale study mentioned above, 
this characteristic of the neutron star critical solution,
which is characterized mainly by the existence of a universal unstable mode and its specific gravitational radiation signature,
may cause a queiscent effect on the gravitational radiation spectrum of realistic neutron star systems.

Using the additional degrees of freedom available to us in the new neutron star-like initial data we construct, 
we also investigate the properties of the characteristic time scale of the neutron star critical set,
which is the time scale of its unstable mode, describing the time scale of departure of a near-critical evolution from the critical set.
We find that this critical index depends almost linearly on the rest mass of the system.
We also observe that there is a possibility of a 2-parameter dependence of the critical index, where the parameters can be taken as the rest mass and the ADM mass of the system, 
both of which are conserved quantities. This indicates that there exist classes of neutron star semi-attractors that can be labeled by both the rest mass
and the gravitational mass of the system. This finding extends the observation that for a given rest mass, there exist two phase thresholds 
through one of which the system passes from the neutron star to the black hole phase at a lower ADM mass 
and through the other of which the system passes from the black hole phase back to the neutron phase at a higher ADM mass.
We note that the observation of the existence of two phase thresholds for a given rest mass 
is similar to that obtained from the neutron star initial data.

We then analyse the phase space properties of the neutron star critical set, by constructing phase spaces using
the parameters of the rest mass, ADM mass and central density of the new system.
For both the rest mass-ADM mass and ADM mass-central density phase spaces, we find turning point features with the 
aforementioned two-threshold feature. We note that the new system provides the additional degrees of freedom 
to construct the ADM mass-central density phase space. The extent of the critical surfaces in these phase spaces gives an
indication of the size of the attraction basin of the neutron star critical set.
For a certain rest mass, we determined the range of central densities and ADM masses of configurations that exhibit critical gravitational collapses,
which helps us draw an approximation of the range of realistic neutron star or neutron star-like configurations that can exhibit critical phenomena. 
  
We further explore the boundaries of the attraction basin
of the neutron star critical set by constructing a phase space using the 
separation distance and the boost velocity (which also corresponds to the ADM mass) of the new system. In decreasing the separation distance, we move
toward an initial configuration that consists of a single neutron star-like object with a varying implosion velocity.
Fixing the rest mass at a slightly lower value than the maximum rest mass allowable for a single equilibrium TOV configuration with $\Gamma=2$, 
we find that there exists a boundary whereby the threshold disappears when we increase the central density towards the maximum rest mass point. 

\section{Future work}

In order to further clarify the implications of the results of this dissertation on realistic astrophysical observations,
we will extend the analysis on neutron star systems that possess angular momentum.
To do this we intend to construct a similar neutron star-like initial data but with the additional incorporation of angular momentum, 
in both the irrotational and corotational orientations
in accordance with the spin orientations that have so far been observed in realistic neutron star systems.
This will also help us to see if an object that is simultaneously rotating and oscillating can undergo a dynamical process
whereby the critical set is characterized by a limit cycle, which poses an interesting scenario in the theory of general relativity itself.

   \lhead{}

   \lhead{}
\rhead{\textit{Appendix}}
\begin{appendix}

\section{The finite-differencing scheme}

The finite-differencing scheme is a numerical scheme where a grid is introduced as the domain of functions. Values of functions are evaluated on points on the grid 
thereby discretizing the functions according to the resolution of the grid. Using Taylor's expansion, an unknown function $u(x,y)$ on a 2-dimensional grid is 
evaluated as follows:
\begin{eqnarray}
\label{eq:A1}
u(x_0+\Delta x,y_0)=u(x_0,y_0)+\Delta x\frac{\partial u}{\partial x}(x,y_0)+\frac{(\Delta x)^2}{2!}\frac{\partial^2 u}{\partial x^2}(x,y_0)+ \nonumber \\
\frac{(\Delta x)^3}{3!}\frac{\partial^3 u}{\partial x^3}(x,y_0)+\frac{(\Delta x)^4}{4!}\frac{\partial^4 u}{\partial x^4}(x,y_0)+...
\end{eqnarray}
where:
\begin{equation}
\label{eq:A2}
u(x_0+\Delta x,y_0)-u(x_0-\Delta x,y_0)=2\Delta x\frac{\partial u}{\partial x}+\frac{(\Delta x)^3}{3}\frac{\partial^3 u}{\partial x^3}+\mathcal{O}[(\Delta x)^5].
\end{equation}
The partial derivative of $u(x,y)$ with respect to $x$ can thus be evaluated using a second-order central operator as:
\begin{equation}
\label{eq:A3}
\frac{\partial u}{\partial x}=\frac{u_{i+1,j}-u_{i-1,j}}{2\Delta x}+\mathcal{O}[(\Delta x)^2],
\end{equation}
where $i,j$ are the indices of the grid points in the $x,y$ directions respectively. In turn, the next higher derivative can be written as:
\begin{equation}
\label{eq:A4}
\frac{\partial^2 u}{\partial x^2}=\frac{u_{i+1}-2u_{i,j}+u_{i-1,j}}{(\Delta x)^2}+\mathcal{O}[(\Delta x)^2].
\end{equation}
Eq.s~\eqref{eq:A3} and ~\eqref{eq:A4} describe a second-order finite-differencing scheme, as the dominant term on the right is second-order with respect to the grid 
resolution $\Delta x$. In addition, the order of the dominant term is preserved as we go to higher orders of the derivative. The scheme is thus said to be nth-order 
following the order of the dominant term of partial derivatives with respect to the grid resolution. Convergence of finite-differencing schemes is reached when 
$\Delta x\rightarrow 0$. At this limit, functions become in principle infinitely accurate. Without prior 
knowledge of a function, we are still able to check how accurately it is evaluated by checking its convergence with respect to the grid resolution. In numerical general 
relativistic hydrodynamics codes implementing finite-difference schemes based on the 3+1 formalism, convergence is checked by performing a simulation using different grid 
resolutions and observing how the violations of the Hamiltonian Eq. (2.37) and momentum constraint Eq. (2.38) scale with respect to the grid resolution.   

\section{Push-forward of vectors and pull-back of forms}

The push-forward of a vector along another vector field congruence is defined as an operator that acts on the vector such that it produces a new vector whose Lie 
derivative along the vector field congruence is zero. Consider a vector $\mathbf{B}$ being pushed forward along the vector field 
congruence $\mathbf{A}$. Let $t$ be the affine parameter along $\mathbf{A}$. The Lie derivative of $\mathbf{B}$ along $\mathbf{A}$ is as follows:
\begin{equation}
\label{eq:B1}
\mathcal{L}_{\mathbf{A}}\mathbf{B}=[\mathbf{A},\mathbf{B}]=\lim_{t\rightarrow 0}\frac{\mathbf{B}-\tilde{\psi}_t\mathbf{B}}{t},
\end{equation}
where $\tilde{\psi}_t\mathbf{B}$ is the push-forward of $\mathbf{B}$ along $\mathbf{A}$. The push-forward operator thus transports the vector $\mathbf{B}$ along 
the vector congruence $\mathbf{A}$ such that the new vector $\tilde{\psi}_t\mathbf{B}$ forms a coordinate basis with the form dual to $\mathbf{A}$.
Given this definition, we see that the push-forward of a vector acting on a form is the same as the vector acting on the pull-back of the form. 
The push-forward of a vector can also be seen as an active transformation on the vector whereas the pull-back of a form can be seen as a passive transformation on 
the form while actively transforming the coordinate system in which the form is defined.

\section{Geometric units}

The gravitational constant $G$ has the units of $[L^3 T^{-2} M^{-1}]$, where $L$ denotes the dimension of length, $T$ denotes the dimension of time and $M$ denotes 
the dimension of mass. Using the mass of the sun in metric units as a standard and the value of the constant as well as the speed of light in the same metric units, 
the unit for $G$ can be set dimensionless as follows:
\begin{eqnarray}
\label{eq:C1}
L^3 T^{-2} M_{\odot}^{-1} & = & 6.67\times 10^{-11} \nonumber \\
(L/T)^2 L & = & 1.33\times 10^{20} \nonumber \\
L & = & 1.33\times 10^{20}/(3\times 10^8)^2 \nonumber \\
L & = & 1.476\times 10^3 m,
\end{eqnarray}
which denotes $1M_{\odot}$ of length as equivalent to $1.476\times 10^3 m$. From this relation, we can similarly write all other variables in terms of $M_{\odot}$, 
such as:
\begin{eqnarray}
1M_{\odot} of time \equiv 4.92\times 10^{-6}s \label{eq:C2} \\
1M_{\odot} of mass \equiv 1.989\times 10^{30} kg \label{eq:C3} \\
1M_{\odot} of energy \equiv 1.786\times 10^{67} J \label{eq:C4}.
\end{eqnarray}
  
\end{appendix}

  \end{main}

\end{document}